\documentclass{article}
\usepackage{amsmath, amsthm, amssymb}
\usepackage{graphicx}
\usepackage{verbatim}
\usepackage{natbib}
\usepackage{caption}
\usepackage{subcaption}
\usepackage{fancyvrb}
\usepackage{enumerate}
\usepackage{relsize}
\usepackage{multirow}
\usepackage[boxruled]{algorithm2e}
\usepackage[section]{placeins}

\usepackage{hyperref}
\usepackage[margin=1.3in]{geometry}
\hypersetup{colorlinks,citecolor=blue,urlcolor=blue,linkcolor=blue}

\usepackage{stefan_tex}

\newcommand\blfootnote[1]{%
  \begingroup
  \renewcommand\thefootnote{}\footnote{#1}%
  \addtocounter{footnote}{-1}%
  \endgroup
}

\theoremstyle{plain}
\newtheorem{prop}{Proposition}

\newtheorem{lemm}[prop]{Lemma}
\newtheorem{theo}[prop]{Theorem}

\theoremstyle{definition}

\newtheorem{assumption}{Assumption}

\theoremstyle{remark}

\newtheorem{rema}[prop]{Remark}

\graphicspath{{plots/}}

\author{
  Amy Guan$^{\dagger}$ \and
  Roshni Sahoo$^{*\dagger}$ \and 
  Joshua Salomon$^{\ddagger}$ \and
  Stefan Wager$^{\mathsection}$ \and
  Marissa Reitsma$^{\ddagger}$
}

\date{Stanford University}

\title{Data Fusion for High-Resolution Estimation}

\begin{document}

\maketitle

\begin{abstract}
High-resolution estimates\blfootnote{\noindent Draft version: \ifcase\month\or January\or February\or March\or April\or May\or June\or July\or August\or September\or October\or November\or December\fi \ \number \year. $^{*}$Corresponding author, email \url{rsahoo@stanford.edu}, $^{\dagger}$Department of Computer Science, $^{\ddagger}$Department of Health Policy, $^{\mathsection}$Graduate School of Business. We thank Harrison Li, Yajuan Si, seminar participants at ICHPS, JSM, ACIC for their helpful feedback. This research was supported by RS's and MR's Stanford Data Science Fellowships and Spectrum PHS Pilot Grant and by NSF grant SES-2242876. We are grateful for the Data Science Fellowship for creating the opportunity for this interdisciplinary research and illuminating discussions with Lihua Lei. Data for replication is available \href{https://drive.google.com/drive/folders/1ALGzHTkK1k4X5HJIPtzwmwJE1EApMMbk?usp=drive_link}{here}. Code for replication is available \href{https://github.com/roshni714/data_fusion}{here}.} of population health indicators are critical for precision public health. We propose a method for high-resolution estimation that fuses distinct data sources: an unbiased, low-resolution data source (e.g. aggregated administrative data) and a potentially biased, high-resolution data source (e.g. individual-level online survey responses). We assume that the potentially biased, high-resolution data source is generated from the population under a model of sampling bias where observables can have arbitrary impact on the probability of response but the difference in the log probabilities of response between units with the same observables is linear in the difference between sufficient statistics of their observables and outcomes. Our data fusion method learns a distribution that is closest (in the sense of KL divergence) to the online survey distribution and consistent with the aggregated administrative data and our model of sampling bias. This approach significantly reduces bias in high-resolution estimates compared to baselines that rely on a single data source alone on a testbed that includes repeated measurements of three indicators measured by both the (online) Household Pulse Survey and ground-truth data sources at two geographic resolutions over the same time period.
\end{abstract}


\begin{section}{Introduction}




Precision public health aims to deliver ``the right intervention to the right population at the right time'' \citep{khoury2015precision,roberts2024precision}. In order to achieve this goal, public health decision makers need accurate and subgroup-specific estimates of health indicators. Such estimates can improve the effectiveness and efficiency of public health initiatives by enabling more accurate monitoring of key indicators and emerging threats, and by informing the design, implementation, and evaluation of programs and policies.

Traditional representative in-person household surveys are often the gold standard for public health measurement. While these surveys are representative at a low-resolution (i.e., regional or national) level, their deployment costs and limited sample size preclude their use in high-resolution (i.e., state or county) estimation. On the other hand, online surveys are a promising tool for obtaining high-resolution health estimates \citep{geldsetzer2020use, us2021measuring, salomon2021us} because compared to traditional in-person surveys, online surveys are cheaper to administer and scale to much larger sample sizes \citep{blumberg2021national}. These surveys can capture subgroup heterogeneity that is not captured by traditional in-person surveys. However, their key limitation is that they suffer heavily from sampling bias, meaning that the population of survey respondents differs from the target population. High-resolution estimates from online surveys alone may exhibit large bias relative to target population values.

\begin{figure}
\includegraphics[width=\textwidth]{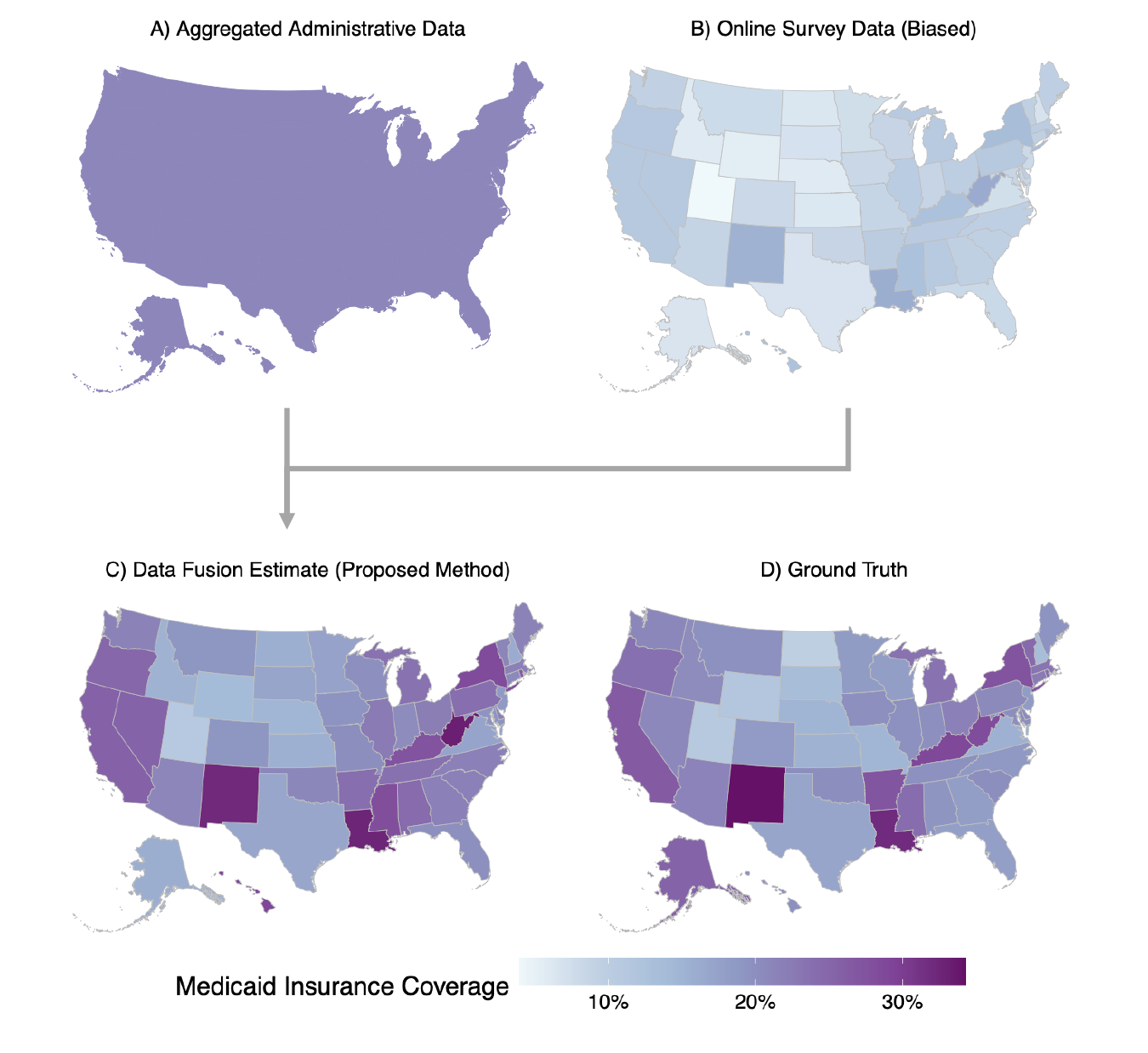}
\caption{We visualize predictions of Medicaid enrollment status obtained from different data sources. (a) We visualize a national-level estimate of the indicator from the American Community Survey, which is a traditional household survey that is assumed to be gold standard. (b) We visualize state-level estimates of the indicator from the Household Pulse Survey, an Experimental Data Product of the US Census Bureau that may suffer heavily from sampling bias. (c) State-level predictions from our data fusion approach, which leverages the data sources visualized in (a) and (b). (d) State-level ground truth obtained from the American Community Survey.}
\label{fig:motivation}
\end{figure}

Sampling bias in online surveys is typically addressed by applying covariate-based reweighting methods, such as inverse propensity weighting and post-stratification \citep{ groves2011survey, rosenbaum1983central}. These techniques permit unbiased estimation of the mean outcome if responses from units are ``missing at random" \citep{rubin1976inference}, meaning that an individual's probability of survey response depends only on observables. However, recent empirical analyses find that these techniques fail to sufficiently adjust for the bias in online surveys \citep{bradley2021unrepresentative, kessler2022estimated}, indicating that responses from units in online surveys are likely ``missing not at random." In other words, an individual's probability of survey response can depend on both observable and unobservable covariates. For instance, \citet{kessler2022estimated} find that an online survey called the Household Pulse Survey (HPS) yields implausibly high estimates of the prevalence of anxiety or depression compared to a telephone survey called the Behavioral Risk Factor Surveillance System (BRFSS). They observe that the HPS respondent population differs from the US adult population in geographic-demographic characteristics, and they hypothesize that the two populations differ in their unmeasured psychological characteristics, as well.

The main contribution of the present study is a data fusion method for high-resolution estimation that leverages two distinct data sources: an unbiased, low-resolution data source (e.g. aggregated administrative data or aggregated traditional in-person survey data) and a potentially biased, high-resolution data source (e.g. individual-level online survey responses). We refer to the former as ``aggregated administrative data" and the latter as an ``online survey" throughout; our results, however, apply to any setting where we seek to fuse two data sources with these general properties. To illustrate our approach, consider estimating state Medicaid enrollment with access to aggregated administrative data and online survey data in Figure \ref{fig:motivation}. The aggregated administrative data matches the mean outcome in the true population but does not capture subgroup heterogeneity; in Figure \ref{fig:motivation}(a), we visualize the national-level average Medicaid enrollment status in the United States. In contrast, the online survey predictions are biased relative to the ground truth but still capture meaningful subgroup heterogeneity that exists in the population; in Figure \ref{fig:motivation}(b), we observe that an online survey underestimates state-level Medicaid enrollment rates relative to the state-level ground truth (Figure \ref{fig:motivation}(d)). In Figure \ref{fig:motivation}(c), we demonstrate that our proposed
data fusion method improves the accuracy of high-resolution estimates of state-level Medicaid
enrollment rates.

Our method is enabled by the assumption that the online survey distribution is generated from the population distribution through a model of sampling bias where a unit's probability of survey response can depend on both observable characteristics, such as age, sex, and education level, as well as unobservable characteristics that may affect the health indicator, such as the extent to which a unit trusts health authorities. In particular, we permit observables to have arbitrary impact on the probability of response but require that the difference in the log probabilities of response among units with the same observables is linear in the difference between a known sufficient statistic function of their observables and outcomes. While structurally simple, this model can flexibly capture many forms of sampling bias, including the setting where the amount of sampling bias due to unobservables varies across units with different observables.

Given the nature of the sampling bias, the true population distribution is not identifiable from the available data. Nevertheless, we can partially identify the population distribution by the set of distributions that are consistent with our available information and assumptions. Then, we propose to select a candidate distribution from this set to generate high-resolution estimates. The partial identification set for the population distribution consists of distributions that (i) can generate the survey distribution under our proposed model of sampling bias and (ii) match the available moments of the aggregated administrative data. Among the partial identification set, we aim to learn the distribution that is most difficult to distinguish from the survey distribution, so that properties of the online survey distribution such as subgroup heterogeneity can be preserved. Section \ref{sec:setup} demonstrates that this candidate population distribution is the solution to a semiparametric moment matching problem. Section \ref{sec:alg_inference} outlines how to do statistical inference on a (high-resolution) subgroup mean outcome under the candidate distribution.

A secondary contribution of this work is the development of a testbed for evaluating the quality of high-resolution estimates obtained from online survey data. Evaluating estimates obtained from online surveys is often infeasible because ground-truth measurements of the outcomes in the population are unavailable. Nevertheless, we identify three different population health indicators for which there are repeated measurements from an online survey and a ground-truth data source for the same population over the same period of time (see Table \ref{tab:data_pairing}). Furthermore, these ground-truth data sources contain measurements of the indicators at the state-level and can be aggregated to form regional-level or national-level ground truth, as well. The indicators in our testbed include COVID-19 vaccination rates, SNAP (Supplemental Nutrition Assistance Program) enrollment rates, and Medicaid enrollment rates. The three indicators are measured by the online Household Pulse Survey (HPS) \citep{us2021measuring}. The ground-truth data sources include administrative data from the CDC and USDA and aggregated responses from the American Community Survey, a survey conducted by the Census Bureau that legally requires selected households to respond.

For the task of predicting state-level Medicaid enrollment, our method yields a 77\% decrease in population-weighted MAE of state-level estimates relative to a method that only utilizes the online survey data and a 31\% reduction in MAE compared to a method that only uses the aggregated administrative data. For the task of predicting state-level COVID-19 vaccination rates, our method yields an 80\% decrease in the population-weighted mean absolute error (MAE) of state-level estimates relative to a method that only utilizes the online survey data on average over time and 50\% reduction in MAE compared to a method that only utilizes the aggregated administrative data. For the task of predicting state-level SNAP enrollment, we find that online surveys already achieve low MAE (MAE 0.01) and our method does not degrade performance.


\begin{subsection}{Related Work}
Data fusion of high-resolution and low-resolution sources has previously received attention in economics \citep{hellerstein1999imposing,imbens1994combining}, statistics \citep{chatterjee2016constrained, zhai2022data,zhang2020generalized}, and computer science \citep{agostini2024bayesian,balachandar2023domain}, as well as in applied contexts in public health \citep{reitsma2024bias,schenker2007combining,ward2016redrawing}. Most related to our work, \citet{imbens1994combining} combine individual-level micro-data and aggregate-level macro-data to estimate regression coefficients of microeconomic models. However, in their setting, the population distribution and target parameters are identified by the micro-data alone, and the macro-data is used to improve efficiency of the estimate. In contrast, in our work, neither the online survey nor the aggregated administrative data identify the population distribution, but both sources contribute to partial identification of the population distribution.

Since we consider a setting where the population distribution is not identified by the online survey distribution, our work also contributes to the broad literature on learning under missing data \citep{rubin1976inference}. In particular, we work in the regime where data is ``missing not at random," where a unit's probability of survey response can depend on both observable and unobservable characteristics. Previous works that consider this setting define the target parameter of interest to be a functional of the (unidentifiable) population distribution. Taking a robust approach, \citet{aronow2013interval, manski2016credible, miratrix2018shape} posit worst-case bounds on the probability of sample selection and identify an uncertainty set of plausible population distributions that could have generated the observed data. Then, they obtain partial identification intervals for the target parameter by considering its maximum and minimum value over the uncertainty set. A salient conceptual difference between our work and partial identification approaches is that we commit to selecting a candidate distribution from the uncertainty set for high-resolution estimation.
In addition, we emphasize that the model of sampling bias that we consider is a conditional model that places a restriction on the population conditional distribution, rather than an unconditional model of sampling bias that would place a restriction on the population joint distribution. Conditional uncertainty sets have been considered in the literature on sensitivity analysis in causal inference \citep{jin2022sensitivity, dorn2023sharp, yadlowsky2022bounds} and in the literature on robust policy learning and decision making \citep{lei2023policy, sahoo2022learning, wang2024distributionally}. As discussed in \citet{sahoo2022learning}, conditional models of sample selection are often more suitable than unconditional models when the covariate shift is observable, as is the case in this work.

We demonstrate in Section \ref{sec:setup} that our model of sampling bias is equivalent to assuming that the population and online survey conditional distributions are related via an exponential tilting. Exponential tilting models have previously been used in sensitivity analyses in causal inference. \citet{birmingham2003pattern,andrea2001methods,scharfstein1999adjusting} use an exponential tilting to model nonignorable dropout in longitudinal studies and assess the sensitivity of the mean outcome over time under different tiltings. \citet{franks2020flexible,scharfstein2021semiparametric,zhou2023sensitivity} use an exponential tilting to assess unmeasured confounding in observational studies. Nevertheless, the aim of our work and that of these previous works are distinct. These works aim to assess the sensitivity of treatment effect estimates to an unmeasured confounder represented by an exponential tilting with access to expert-specified ranges for the severity of the tilting. In contrast, the main statistical problem that we focus on is equivalent to learning the closest exponential tilting of the survey distribution that is consistent with moment conditions.




Our work is also conceptually related to the literature on combining experimental and observational data to estimate treatment effects \citep{athey2025experimental, bareinboim2014transportability, colnet2024causal,dahabreh2023efficient, guo2021multi, kallus2018removing, lanners2025data,li2023efficient, li2024efficient,li2025data,lu2021you,pearl2011transportability, rudolph2017robust,stuart2015assessing, yang2020combining, yang2024data}. Much like how online surveys can suffer from sampling bias, observational data can suffer from unmeasured confounding. As a result, structural assumptions, which can be expressed in terms of restrictions on the class of the conditional mean outcome \citep{li2024efficient} or a parametric selection model \citep{li2025data}, are required to link the observational data with the experimental data. Nevertheless, in this literature, the target parameter, the treatment effect, is usually identifiable from the experimental data alone so the observational data are primarily used to improve the efficiency of the estimate.

\end{subsection}

\end{section}

\begin{section}{Data Fusion}
\label{sec:setup}
In this section, we outline our data fusion approach to high-resolution estimation. First, we define the statistical information provided by the individual-level online survey data source and the aggregated administrative data source. We place an assumption on the relationship between the survey and population distribution. Second, we introduce our data fusion approach; we select a candidate distribution from the partial identification set of the population distribution and provide a semiparametric characterization of this distribution. Lastly, we show that when we do not impose a particular model of sampling bias, our approach reduces to an information projection \citep{csiszar1975divergence}.

\begin{subsection}{Sampling Model}
\label{subsec:setup}
We consider a population of units $i=1, 2, \dots n$. Let $\mathcal{X} \subset \mathbb{R}^{d}$ be the covariate space and let $\mathcal{Y}\subset \mathbb{R}$ be the outcome space. Let $X_{i} \in \mathcal{X}$ denote observable covariates and $Y_{i} \in \mathcal{Y}$ denote the outcome of a unit $i$. We suppose that $X_{i}$ consists of observable individual-level characteristics, such as age, sex, and education level of unit $i$. We assume a survey that collects observable covariates $X_{i}$ and outcomes $Y_{i}$ is administered in the population but some units may not respond to the survey. Let $R_{i} \in \{0, 1\}$ be a binary indicator that denotes whether unit $i$ selects into responding to the survey.

Let $P$ be the distribution over $(X_{i}, Y_{i}, R_{i})$ for units in the population. Let $\mathcal{G}$ be a set of subgroups of the population. Let each subgroup $g \in \mathcal{G}$ have a corresponding subgroup population distribution $P_{g}$ where $P_{g}$ may differ from $P$ in its covariate distribution but shares the same conditional distribution, i.e. $P_{g, Y|X} = P_{Y|X}$. Then, we aim to estimate
\begin{equation}
\mu_{g}:= \EE[P_{g}]{Y} \quad \forall g \in \mathcal{G}.
\end{equation}

Let $S$ be the distribution over survey respondents $(X_{i}, Y_{i}) \mid R_{i} = 1$. Since some units may not respond to the survey, the survey suffers from a missing data problem. Following the taxonomy of missing data problems from \citet{rubin1976inference}, we recall that responses are missing completely at random (MCAR) if the probability of response is the same for all units, i.e. there exists a constant $\alpha_{0} \in [0, 1]$ such that
\begin{equation}
\label{eq:mcar}
\PP[P]{R_{i}=1 \mid X_{i}=x, Y_{i}=y} = \alpha_{0} \quad \forall x \in \mathcal{X}, y \in \mathcal{Y}.
\end{equation}
Under MCAR, the survey distribution $S$ is the same as the population distribution $P$. Responses are defined to be missing at random (MAR) if the probability of response is independent of outcomes given the observable covariates, i.e. there exists a function $\alpha: \mathcal{X} \rightarrow [0, 1]$ such that
\begin{equation}
\label{eq:mar}
\PP[P]{R_{i}=1 \mid X_{i} = x, Y_{i}= y} =  \alpha(x) \quad \forall x \in \mathcal{X}, y \in \mathcal{Y}.
\end{equation}
Under MAR, the survey distribution may have a different covariate distribution than the population distribution, i.e. $S_{X} \neq P_{X}$ but the two distributions have the same conditional distribution of $Y \mid X$. The MAR assumption underlies covariate-based reweighting and post-stratification techniques that are widely used in the literature on generalizability \citep{stuart2011use,tipton2013improving,tipton2014generalizable}.
Responses are defined to be missing not at random (MNAR) if the probability of response can depend on the outcome, i.e. there exists a function $\alpha: \mathcal{X} \times \mathcal{Y} \rightarrow [0, 1]$ such that
\begin{equation}
\label{eq:mnar}
\PP[P]{R_{i} = 1 \mid X_{i} = x, Y_{i} = y} = \alpha(x, y) \quad \forall x \in \mathcal{X}, y \in \mathcal{Y}.
\end{equation}
We note that \eqref{eq:mnar} is equivalent to assuming that the probability of response depends on unobservable characteristics that are not independent from the outcome. Under MNAR, the survey distribution $S$ may differ from $P$ in both its covariate distribution and conditional distribution. 

Motivated by recent empirical analyses of online surveys \citep{bradley2021unrepresentative, kessler2022changes}, we suppose the online survey distribution $S$ suffers from a MNAR data problem. However, we do assume some structure on the missingness mechanism. 
Namely, we assume that the difference in the log probability of response between units with the same observables is linear in the difference between some known sufficient statistics $\eta(x, y)$ of their covariates and outcome.
\begin{assumption}[Sampling Bias due to Unobservables]
\label{assumption:exp_tilting}
For some known $\eta: \mathcal{X} \times \mathcal{Y} \rightarrow \mathbb{R}^{J}$, there exists $\theta \in \mathbb{R}^{J}$ such that
\begin{equation} 
\label{eq:binary_selection_prob}
\log(\alpha(x, y)) - \log(\alpha(x, y')) = \theta^{T}(\eta(x, y') - \eta(x, y)). 
\end{equation}
\end{assumption}
Note that when $\eta(x, y)$ does not depend on $y$, Assumption \ref{assumption:exp_tilting} reduces to a MAR assumption; in this case, $S$ will deviate from $P$ in only its covariate distribution. In addition, when $\eta(x, y)$ does not depend on $x$, then the sampling bias due to the outcome is the same for all units. Since $\eta(x, y)$ can depend on both $x, y$, there can be heterogeneity in the amount of sampling bias due to unobservables across units with different covariates. To provide notation, we write that $P_{Y|X} \sim_{\eta} S_{Y|X}$ if the response probabilities satisfy Assumption \ref{assumption:exp_tilting}.

We show that Assumption \ref{assumption:exp_tilting} is equivalent to assuming that the population and online survey conditional distributions are related through an exponential tilting with sufficient statistics $\eta(x, y).$

\begin{lemm}
\label{lemm:exp_family}
Let $Q$ be a population distribution over $(X_i, Y_i, R_i)$, and let $S$ be a survey distribution over $(X_i, Y_i) \mid R_{i}=1$. Let $\alpha: \mathcal{X} \times \mathcal{Y} \rightarrow [0, 1]$ denote the probability of survey response as defined in \eqref{eq:mnar}. 
Suppose that $\alpha$ satisfies Assumption \ref{assumption:exp_tilting} for some $\eta: \mathcal{X} \times \mathcal{Y}\rightarrow \mathbb{R}^{J}$, i.e. $Q_{Y|X} \sim_{\eta} S_{Y|X}$. Then, there exists $\theta \in \mathbb{R}^{J}$ such that
\begin{equation}
\label{eq:exp_family}
dQ_{Y \mid X=x}(y) \propto dS_{Y \mid X=x}(y) \cdot \exp(\theta^{T}\eta(x, y)) \quad \forall x \in \mathcal{X}, y \in \mathcal{Y}
\end{equation}
and $\sup_{x \in \mathcal{X}} \frac{dS_{X}(x)}{dQ_{X}(x)} < \infty.$ Conversely, if \eqref{eq:exp_family} and $\sup_{x \in \mathcal{X}} \frac{dS_{X}(x)}{dQ_{X}(x)} < \infty$ hold, then $Q_{Y|X} \sim_{\eta} S_{Y|X}$  (Assumption \ref{assumption:exp_tilting} holds).
\end{lemm}

In addition to data from the online survey, we assume access to aggregated administrative data that provides low-resolution information on the population distribution in the form of moment conditions. For some $\gamma: \mathcal{X} \times \mathcal{Y} \rightarrow \mathbb{R}^{K}$, we observe
\begin{equation} \label{eq:moment} \EE[P]{\gamma(X, Y)} = \bar{\gamma}_{P}, \end{equation} where $\bar{\gamma}_{P} \in \mathbb{R}^{K}$. For example, we may have access to the mean outcome for different non-overlapping covariate-defined groups. Let $\mathcal{A}_{k} \subset \mathcal{X}$ for $k=1, 2, \dots K$ be disjoint subsets of the covariate space, we can set $\gamma_{k}(x, y) = y \cdot \mathbb{I}(x \in \mathcal{A}_{k})$ for $k=1, 2, \dots K$.

Throughout, we will also assume that the population covariate distribution $P_{X}$ is known and does not need to be estimated. Note that our data-generating process implies that the online survey covariate distribution $S_{X}$ is absolutely continuous with respect to the population covariate distribution $P_{X}$, i.e. $S_{X} \ll P_{X}$. The following assumption combined with this observation will imply that the population and online survey covariate distributions share the same support.
\begin{assumption}[Overlap]
\label{assumption:cov_shift}
There exists $\overline{r} > 0$ for which $\sup_{x \in \mathcal{X}} dP_{X}(x)/dS_{X}(x) \leq \overline{r}.$
\end{assumption}

Assumption \ref{assumption:exp_tilting}, Assumption \ref{assumption:cov_shift}, the moment condition in \eqref{eq:moment}, and access to the population covariate distribution $P_{X}$ immediately enable partial identification of the population distribution. The partial identification set for the population distribution $P$ consists of distributions $Q$ that satisfy $Q_{Y|X} \sim_{\eta} S_{Y|X},$ $\EE[Q]{\gamma(X, Y)} = \bar{\gamma}_{P},$ and $Q_{X} = P_{X}.$ 

We propose to select the distribution from this partial identification set that is most difficult to distinguish from the survey distribution $S$ because we aim to preserve properties of $S$, such as subgroup heterogeneity. The KL divergence from $Q$ to $S$, or $\text{KL}(Q||S)$, can be viewed as a measure of how easy it is to distinguish between the two distributions. In the context of a statistical test for the distribution from which i.i.d. samples are drawn, from $Q$ under the null or from $S$ under the alternative, the most powerful test is a threshold rule on the log of the likelihood ratio, which converges to $\text{KL}(Q||S)$ when the samples are drawn from $Q$.
As a result, to find the distribution that is the most difficult to statistically distinguish from the survey distribution, we minimize $\text{KL}(Q||S)$ over the partial identification set, i.e. we solve
\begin{equation}
\label{eq:kl_w_covariate}
\min_{Q} \left\{ \mathrm{KL}(Q || S) : Q_{Y|X} \sim_{\eta} S_{Y|X},\, \quad \EE[Q]{\gamma(X, Y)} = \bar{\gamma}_{P},\,\quad Q_{X} = P_{X} \right\}.
\end{equation}
In \eqref{eq:kl_w_covariate}, we are projecting the survey distribution $S$ onto the set of distributions that are consistent with our model of sampling bias and available data. This problem is closely related to an information projection \citep{csiszar1975divergence}, which corresponds to minimization of KL divergence in the first argument over a convex set of distributions. We will later show that the feasible set of \eqref{eq:kl_w_covariate} is nonconvex, complicating the optimization. 

We demonstrate that \eqref{eq:kl_w_covariate} can be written as a moment matching problem.
\begin{lemm}
\label{lemm:parametric_characterization}
Suppose that Assumption \ref{assumption:exp_tilting}, \ref{assumption:cov_shift} hold. Define the family of distributions $Q(\theta)$ where the conditional distribution $Q_{Y|X}(\theta)$ is given by \eqref{eq:exp_family} and the covariate distribution $Q_{X}(\theta) = P_{X}$. Any $Q$ that solves \eqref{eq:kl_w_covariate} is equal to some $Q(\theta_0)$ where $\theta_0 \in \mathbb{R}^{J}$ is a solution to
\begin{equation}
\label{eq:kl_param}
\min_{\theta \in \mathbb{R}^{J}} \{ \EE[P_{X}]{\mathrm{KL}(Q_{Y|X}(\theta)|| S_{Y|X})} : \EE[Q(\theta)]{\gamma(X, Y)} = \bar{\gamma}_{P}\}.
\end{equation}
\end{lemm}

\end{subsection}

\begin{subsection}{Model-Free Approach}
One can also ask about how our method for selecting the candidate distribution behaves when the researcher does not assume that Assumption \ref{assumption:exp_tilting} holds. We call this a ``model-free" approach because it does not require the researcher to commit to a particular model of sampling bias. Provided the same data as in \eqref{eq:kl_w_covariate}, the model-free approach aims to solve
\begin{equation}
\label{eq:kl_agnostic}
\min_{Q} \left\{ \mathrm{KL}(Q || S) : \EE[Q]{\gamma(X, Y)} = \bar{\gamma}_{P},\, \quad Q_{X} = P_{X}\right\}.
\end{equation}
This is closely related to the classical problem of entropy maximization subject to moment constraints \citep{jaynes1957information} and is an example of an information projection \citep{csiszar1975divergence}. In particular, we project the survey distribution $S$ onto the set of distributions that match the moment conditions, without imposing any model of sampling bias. Applying standard results from convex optimization yields that the optimal solution to \eqref{eq:kl_agnostic} is $\tilde{Q}$ such that
\begin{equation}   
\tilde{Q}_{X} = P_{X},\, \quad \label{eq:psi_tilting} d\tilde{Q}_{Y \mid X=x}(y) \propto dS_{Y \mid X=x}(y) \cdot \exp(\theta^{T}\gamma(x, y))\end{equation}
for some $\theta \in \mathbb{R}^{J}$ where $\EE[\tilde{Q}]{\gamma(X, Y)} = \bar{\gamma}_{P}.$ Leveraging properties of exponential families, the optimal natural parameter $\theta$ can be learned by moment matching, i.e. solving
\begin{equation}
\label{eq:model_free_alg}
\min_{\theta \in \mathbb{R}^{J}} \EE[P_{X}]{\tilde{A}_{X}(\theta)} - \bar{\gamma}_{P}^{T}\theta,
\end{equation}
where $\tilde{A}_{x}(\theta)$ denote the log-partition function of $d\tilde{Q}_{Y \mid X=x}$ defined as
\begin{equation} \tilde{A}_{x}(\theta) = \log \left( \int_{\mathcal{Y}} dS_{Y \mid X=x}(y) \exp(\theta^{T}\gamma(x, y))dy \right).\end{equation}

The solution in \eqref{eq:psi_tilting} is an exponential tilting where the sufficient statistics are given by the moment function $\gamma$. Interestingly, we find that the solution to the model-free problem \eqref{eq:kl_agnostic} is the same as the solution to model-based problem \eqref{eq:kl_w_covariate} when the tilting function $\eta$ is the same as the moment function $\gamma$. In other words, if we conjecture sampling bias along the same axes for which we observe moments of the population distribution then \eqref{eq:kl_agnostic} has the same solution as \eqref{eq:kl_w_covariate}. 

However, in many cases, the axes along which we observe target moments $\gamma$ may differ from the axes along which sampling bias occurs $\eta$. For instance, we may observe the average regional-level outcome from the Eastern region of the United States but conjecture that a constant amount of sampling bias due to unobservables affects survey responses from all states. In this case,  we observe $\EE[P]{\gamma(X, Y)} = \bar{\gamma}_{P}$ for $\gamma(x, y) = y \cdot \mathbb{I}(x \in \mathcal{A}_{\text{East}})$, where $\mathcal{A}_{\text{East}} \subset \mathcal{X}$ is the portion of the covariate space that corresponds to the Eastern United States but conjecture $\eta(x, y) = y$. The solution to \eqref{eq:kl_w_covariate} will tilt the online survey conditional distribution for all $x \in \mathcal{X}$, whereas the solution to \eqref{eq:kl_agnostic} will tilt the online survey conditional distribution for $x \in \mathcal{A}_{\text{East}}$ so that $\EE[Q]{\gamma(X, Y)} = \bar{\gamma}_{P}$ but will otherwise match the online survey conditional distribution for $x \notin \mathcal{A}_{\text{East}}$. Our empirical results in Section \ref{sec:exp} demonstrate this limitation of the model-free approach. The remainder of the paper will focus on the model-based approach.

\end{subsection}

\end{section}

\begin{section}{Estimation and Inference}
\label{sec:alg_inference}
We move from population-based identification results provided in Section \ref{sec:setup} to estimation and inference results that can be used with finite data. We assume access to $n$ i.i.d. samples from the online survey distribution $S$. Let $(X_{1}, Y_{1}), \dots (X_{n}, Y_{n}) \sim S$ denote i.i.d. samples from $S$. We also assume access to the population covariate distribution $P_{X}$.

A standard approach to statistical inference leverages Z-estimation. In Z-estimation, the target parameter of interest $\nu_{0}$ can be expressed as the solution to population estimating equations that provide identification of the target parameter, i.e. $\Psi(\nu) := \EE[S]{\psi(X_{i}, Y_{i}; \nu)} = 0.$ In this case, the solution to the empirical version of the estimating equations, i.e. $\Psi_{n}(\nu) := n^{-1} \sum_{i=1}^{n} \psi(X_{i}, Y_{i}; \nu) = 0$ yields a consistent estimator of the target parameter and its asymptotic distribution can be derived by linearizing the estimating equations. However, we cannot directly apply this strategy in our setting because the optimization problem studied in \eqref{eq:kl_param} is nonconvex, and its first-order conditions do not uniquely identify the global minimum of \eqref{eq:kl_param}. Directly solving an empirical version of the estimating equations may not yield a consistent estimate of the solution to \eqref{eq:kl_param}.

For this reason, our approach to statistical inference relies on one-step estimation \citep[Ch 5.6]{van2000asymptotic}. In one-step estimation, we first assume access to a preliminary, consistent estimator of $\nu_{0}$ called $\tilde{\nu}_{n}$. Then, we refine this estimator by taking one Newton step on the empirical moment conditions at the preliminary estimator. Let $\dot{\Psi}_{\nu}$ denote the Jacobian matrix of $\psi$ with respect to $\nu$. Let $\dot{\Psi}_{n}(\nu) := n^{-1} \sum_{i=1}^{n} \dot{\Psi}_{\nu}(X, Y; \nu).$ The one-step estimator $\hat{\nu}_{n}$ is defined as the solution to
\begin{equation}
\label{eq:one_step_simple}
\Psi_{n}(\tilde{\nu}_{n}) + \dot{\Psi}_{\nu, n}(\tilde{\nu}_{n}) \cdot (\nu - \tilde{\nu}_{n}) = 0.
\end{equation}
In the parametric regime, where the population estimating equations $\Psi$ only depend on the finite-dimensional target parameter $\nu$, the one-step estimator $\hat{\nu}_{n}$ is consistent for $\nu_{0}$ and is asymptotically normal, provided that the preliminary estimator $\tilde{\nu}_{n}$ converges to $\nu_{0}$ at a sufficiently fast rate (e.g. $n^{-1/4}$ rate). 

A complication that arises when applying this strategy to our setting is that the estimating equation $\Psi$ depends on nonparametric nuisance parameters in addition to the finite-dimensional target parameter $\nu$. Since the true values of these nuisance parameters may be complex, we aim to permit estimation via flexible machine learning methods (e.g. neural networks, random forests) to guarantee consistency. However, machine learning methods typically yield estimators that converge slowly (slower than $n^{-1/2}$ rate) to the true value, so the naive one-step estimator that relies on plug-in estimates of the nuisance parameters may converge at slower than a $n^{-1/2}$ rate. To develop a one-step estimator of $\nu_0$ that is consistent and asymptotically normal while allowing the nuisance parameters to be estimated nonparametrically, we use two strategies that are often employed in the literature on ``double/debiased machine learning'' \citep{chernozhukov2018double}. First, we construct population estimating equations $\Psi$ that satisfy orthogonality properties, which ensure that the estimating equations are not sensitive to first-order errors in the nuisance parameters. Second, we leverage cross-fitting to avoid bias due to overfitting. In cross-fitting, we estimate the nuisance parameters using separate samples than the ones used to construct the one-step estimator.

In the following subsections, we address the problem of statistical inference on the solution to \eqref{eq:kl_param} in the case where the outcome $Y$ is binary. We give the explicit form of the population estimating equation in Section \ref{subsec:pop_estimating_equations} and demonstrate that they have desirable orthogonality properties. Then, assuming access to a consistent, preliminary estimator, we give a procedure for constructing the one-step estimator of $\nu_{0}$ that leverages cross-fitting. We analyze the asymptotic variance of the one-step estimator and demonstrate how to build valid confidence intervals for $\nu_{0}.$ Finally, we give an algorithm for obtaining a preliminary estimator $\tilde{\nu}_{n}$ via sequential quadratic programming in Section \ref{sec:alg}. 

\begin{subsection}{Population Estimating Equations}
\label{subsec:pop_estimating_equations}
We highlight that the optimization problem in \eqref{eq:kl_param} is a semiparametric learning problem. We focus on the case where $Y$ is binary, i.e. $Y \in \{0, 1\}$. Let $\mu,r \in L^{2}(P_{X}, \mathcal{X})$, where $\mu: \mathcal{X} \rightarrow [0, 1]$ and $r: \mathcal{X} \rightarrow \mathbb{R}_{+}$. Let $S(\mu)$ denote the distribution over $(X, Y)$ where $\PP[S(\mu)]{Y=1 \mid X=x} = \mu$, and let $Q(\theta, \mu)$ denote the distribution over $(X, Y)$ where $dQ_{Y|X}(\theta, \mu) \propto dS_{Y|X}(\mu) \cdot \exp(\theta^{T}\eta(x, y))$. The optimization problem in \eqref{eq:kl_param} can be expressed as
\begin{equation}
\label{eq:theta_plug_in}
\min_{\theta \in \mathbb{R}^{J}} \Big\{ \EE[S]{r(X) \cdot  \text{KL}(Q_{Y|X}(\theta, \mu) || S_{Y|X}(\mu))}: \EE[S]{r(X) \cdot \EE[Q(\theta, \mu)]{\gamma(X, Y) \mid X}} = \bar{\gamma}_{P} \Big\}
\end{equation}
when the parameters $\mu, r$ are equal to the following functions
\begin{align*}
\mu_{0}(x) := \PP[S]{Y=1 \mid X=x},\, \quad r_{0}(x):= dP_{X}(x)/dS_{X}(x).
\end{align*} 
The optimization problem depends on a finite-dimensional parameter $\theta \in \mathbb{R}^{J}$, which is our target parameter, as well as potentially complex nuisance parameters $\mu_0, r_0 \in L^{2}(P_{X}, \mathcal{X})$. Note that the parameters $\mu_{0}, r_{0}$ are not known and must be estimated from data.

We define the population estimating equations. The population estimating equations depend on a finite-dimensional parameter $\nu:=(\theta, \lambda)$, where $\theta \in \mathbb{R}^{J}$ is the optimization variable in \eqref{eq:theta_plug_in} and $\lambda \in \mathbb{R}^{K}$ is the corresponding dual parameter, and nonparametric nuisance parameters $\mu, r$. The derivative of the Lagrangian of \eqref{eq:theta_plug_in} with respect to $\theta$ and the moment condition of \eqref{eq:theta_plug_in} appear in the population estimating equations. We define these below
\begin{align}
D_{L}(\nu; \mu) &:= \EE[P]{\Cov[Q(\theta, \mu)]{\eta(X, Y) \mid X} \theta + \Cov[Q(\theta, \mu)]{\eta(X, Y), \gamma(X, Y) \mid X}\lambda}, \\
M(\nu; \mu) &:= \EE[P]{\EE[Q(\theta, \mu)]{\gamma(X, Y) \mid X}} - \bar{\gamma}_{P}.
\end{align}

Our estimating equations also depend on residuals of nuisance parameter $\mu$, reweighted by the nuisance parameter $r$ and covariate-dependent adjustment functions $\delta_{D_{L}}(X; \nu, \mu), \delta_{M}(X; \nu, \mu)$. We define the adjustment functions by first defining a tilting function $w(X, Y; \nu, \mu)$ that captures how the conditional probabilities change under the exponential tilting:
\begin{equation}
w(X, Y; \nu, \mu) = \frac{\exp(\theta^{T}\eta(X, Y))}{\mu(X) \cdot \exp(\theta^{T}(\eta(X, 1) - \eta(X, 0))) + \exp(\theta^{T}\eta(X, 0))}.
\end{equation}
The covariate-dependent adjustments are given by
\begin{align}
\delta_{D_{L}}(X; \nu, \mu) &:=  (\rho(X, 1; \nu, \mu) - \rho(X, 0; \nu, \mu)) \cdot w(X, 1; \nu, \mu) \cdot w(X, 0; \nu, \mu), \\
\delta_{M}(X; \nu, \mu) &:= (\gamma(X, 1) - \gamma(X, 0)) \cdot w(X, 1; \nu, \mu) \cdot w(X, 0; \nu, \mu),
\end{align}
where 
\begin{align*}
\Delta_{g}(X, Y; \nu, \mu) &:= g(X, Y) - \EE[Q(\theta, \mu)]{g(X, Y) \mid X},\\
\rho(X, Y; \nu, \mu) &:= \Delta_{\eta}(X, Y; \nu, \mu)^{\otimes 2}\theta +  \Delta_{\eta}(X, Y; \nu, \mu) \otimes \Delta_{\gamma}(X, Y; \nu, \mu) \lambda.
\end{align*} Finally, our population estimating equations are given by $\Psi(\nu; \mu, r) = \EE[S]{\psi(X, Y; \nu, \mu, r)}$, where
\begin{equation}
\label{eq:pop_estimating_eq}
\psi(X, Y; \nu, \mu, r) := \begin{bmatrix} D_{L}(\nu; \mu) + r(X) \cdot \delta_{D_L}(X; \nu, \mu) \cdot (Y - \mu(X)) \\
M(\nu; \mu) + r(X) \cdot \delta_{M}(X; \nu, \mu) \cdot (Y - \mu(X))
\end{bmatrix}.
\end{equation}

Note that $\Psi(\nu; \mu_0, r_0) = 0$ corresponds to the first-order conditions of \eqref{eq:theta_plug_in} when $\mu= \mu_{0}$ and $r=r_{0}$. The first equation corresponds to setting the derivative of the Lagrangian equal to zero, and the second equation corresponds to the moment condition. In addition, we observe that $\psi$ is differentiable in $\nu$ and Gateaux differentiable in $\mu, r$. 

Since \eqref{eq:theta_plug_in} is a nonconvex problem, the first-order conditions do not uniquely identify a global minimizer. Nevertheless, if a global minimizer exists, it must satisfy the first-order conditions. We will assume that a global minimizer that satisfies the first-order conditions, specified above, exists and is locally identifiable.
\begin{assumption}[Existence of a Global Minimizer]
\label{assumption:existence_minimizer}
There exists a population global minimizer $\nu_0$ where $\Psi(\nu_0; \mu_{0}, r_{0}) = 0$ and $\dot{\Psi}(\nu_0; \mu_{0}, r_{0})$ has full rank.
\end{assumption}

We highlight the orthogonality properties of $\Psi$. We define $\dot{\Psi}_{\mu}$, the Gateaux derivative of $\Psi$ with respect to $\mu$ in the direction $\mu'$, where $\mu \in L^{2}(P_{X}, \mathcal{X})$,
\begin{align*}
\dot{\Psi}_{\mu}(\nu; \mu, r)[\mu' - \mu] = \frac{\partial}{\partial \epsilon} \Psi(\nu; \mu + \epsilon (\mu' - \mu), r) \Big|_{\epsilon=0}.
\end{align*}
Analogously, we define $\dot{\Psi}_{r}$, the Gateaux derivative of $\Psi$ with respect to $r$.
\begin{lemm}
\label{lemm:orthogonal}
The population estimating equations $\Psi$ satisfy the following orthogonality conditions:
\begin{align*}
\dot{\Psi}_{\mu}(\nu; \mu_0, r_0)[h] &= 0,\\
\dot{\Psi}_{r}(\nu; \mu_0, r_0)[h] &= 0
\end{align*}
for all $h \in L^{2}(P_{X}, \mathcal{X})$ and $\nu \in \mathbb{R}^{J + K}$.
\end{lemm}
Note that the notion of orthogonality described in Lemma \ref{lemm:orthogonal} that $\Psi$ satisfies is a stronger notion of orthogonality than Neyman orthogonality \citep{chernozhukov2018double}. Neyman orthogonality is a local notion of orthogonality that requires $\dot{\Psi}_{\mu}$ and $\dot{\Psi}_{r}$ to vanish in all directions $h$ when evaluated at the true value of the population minimizer $\nu_{0}$ and the nuisance components $(\mu_{0}, r_{0})$. In contrast, Lemma \ref{lemm:orthogonal} demonstrates that $\Psi$ satisfies a stronger condition that $\dot{\Psi}_{\mu}$ and $\dot{\Psi}_{r}$ vanish at any value of $\nu$, provided that the nuisance components are equal to their true values $\mu_{0}, r_{0}.$ This yields a simple analysis of the one-step estimator.

\begin{rema}
The estimating equations $\Psi$ are not the only way to represent the first-order conditions of \eqref{eq:theta_plug_in}. For example, the following estimating equations also correspond to the first-order conditions
\begin{equation}
\tilde{\Psi}(\nu; \mu) = \begin{bmatrix} D_{L}(\nu; \mu) \\ M(\nu; \mu) \end{bmatrix}.
\end{equation}
The estimating equations $\tilde{\Psi}$ only depend on the nuisance parameter $\mu$ and access to the population covariate distribution $P_{X}$. The first-order conditions of \eqref{eq:theta_plug_in} are captured by $\tilde{\Psi}(\nu; \mu_0) = 0.$ At first glance, one may prefer to work with $\tilde{\Psi}$ instead of $\Psi$ because it has a simpler form and does not depend on the nuisance parameter $r$ that corresponds to the covariate density ratio between the population and survey data distribution. However, the proof of Lemma \ref{lemm:orthogonal} highlights that $\tilde{\Psi}$ lacks the desirable orthogonality properties that $\Psi$ exhibits.
\end{rema}
\end{subsection}

\begin{subsection}{Inference via One-Step Estimation}
We use the population estimating equations derived in Section \ref{subsec:pop_estimating_equations} to define the one-step estimator. Recall that we have access to $(X_i, Y_i) \sim S$ i.i.d. for $i=1, 2, \dots n$. We construct the one-step estimator using cross-fitting. We randomly split the data into two folds $\mathcal{I}_{1},\, \mathcal{I}_{2}$. We evaluate the estimating equations on fold $\mathcal{I}$ as follows
\begin{align*}
\Psi_{n}^{\mathcal{I}}(\nu, \mu, r) &:= \frac{1}{|\mathcal{I}|} \sum_{i \in \mathcal{I}} \psi(X_i, Y_i; \nu, \mu, r), \\
\dot{\Psi}_{\nu, n}^{\mathcal{I}}(\nu, \mu, r) &:= \frac{1}{|\mathcal{I}|} \sum_{i \in \mathcal{I}} \dot{\psi}_{\nu}(X_i, Y_i; \nu, \mu, r),
\end{align*}
where $\psi_{\nu}$ is the Jacobian of $\psi$ with respect to $\nu$. Similarly, we denote the nuisance estimates that are fit using samples from fold $\mathcal{I}$ as $\hat{\mu}_{n}^{\mathcal{I}}, \hat{r}_{n}^{\mathcal{I}}$. Then, we define the cross-fit one-step estimator $\hat{\nu}_{n}$ as follows:
\begin{equation}
\label{eq:cross_fit_one_step}
\hat{\nu}_{n} = \frac{|\mathcal{I}_{1}|}{n} \hat{\nu}_{n}^{\mathcal{I}_{1}} + \frac{|\mathcal{I}_{2}|}{n} \hat{\nu}_{n}^{\mathcal{I}_{2}},
\end{equation}
where $\hat{\nu}_{n}^{\mathcal{I}_{1}}$ is given by the solution to
\begin{equation}
\label{eq:one_step}
 \dot{\Psi}_{\nu, n}^{\mathcal{I}_{1}}(\tilde{\nu}_{n}, \hat{\mu}_{n}^{\mathcal{I}_{2}}, \hat{r}_{n}^{\mathcal{I}_{2}}) \cdot (\nu - \tilde{\nu}_{n} ) +   \Psi_{n}^{\mathcal{I}_{1}}(\tilde{\nu}_{n}, \hat{\mu}_{n}^{\mathcal{I}_{2}}, \hat{r}_{n}^{\mathcal{I}_{2}}) = 0,
\end{equation}
and $\hat{\nu}^{\mathcal{I}_{2}}$ is defined analogously.

To develop guarantees on the one-step estimator, we require the following assumptions.
\begin{assumption}[Bounded Tilting and Moment Functions]
\label{assumption:bounded}
There exists $0< C, C' < \infty$ such that $\sup_{(x, y) \in \mathcal{X} \times \mathcal{Y}} \eta(x, y) \leq C  $ and $\sup_{(x, y) \in \mathcal{X} \times \mathcal{Y}} \psi(x, y) \leq C'.$
\end{assumption}
This assumption is needed for establishing that the derivatives of the estimating equations at the population minimizer are bounded.

\begin{assumption}[Consistency of Nuisance Estimates]
\label{assumption:nuisance}
The nuisance parameter estimators $\hat{\mu}^{\mathcal{I}_{2}}_{n}, \hat{r}^{\mathcal{I}_{2}}_{n}$ satisfy
\[n^{1/4} \cdot \sqrt{\frac{1}{|\mathcal{I}_{1}|} \sum_{i \in \mathcal{I}_{1}} (\hat{r}_{n}^{\mathcal{I}_{2}}(X_i) - r_{0}(X_i))^{2}}\xrightarrow{p} 0\] and \[n^{1/4} \cdot \sqrt{\frac{1}{|\mathcal{I}_{1}|}  \sum_{i \in \mathcal{I}_{1}} (\hat{\mu}^{\mathcal{I}_{2}}_{n}(X_i) - \mu_{0}(X_i))^{2}} \xrightarrow{p} 0,\]
and the analogous conditions hold for $\hat{\mu}^{\mathcal{I}_{1}}_{n}, \hat{r}^{\mathcal{I}_{1}}_{n}$.
\end{assumption}

\begin{assumption}[Consistency of Pilot Estimate]
\label{assumption:pilot}
The pilot estimate $\tilde{\nu}_{n}$ satisfies $n^{1/4} \cdot (\tilde{\nu}_{n} - \nu_{0}) \xrightarrow{p} 0$.
\end{assumption}

The following theorem establishes the asymptotic normality of the one-step estimator. The asymptotic variance of the one-step estimator depends on the following terms $\dot{\Psi}_{\nu, 0} := \dot{\Psi}_{\nu}(\nu_0; \mu_0, r_0)$ and $\Sigma_{0} := \EE[S]{\psi(X, Y; \nu_0, \mu_0, r_0) \psi(X, Y; \nu_0, \mu_0, r_0)^{T}}.$

\begin{theo}
\label{theo:one_step}
Let $\mathcal{X} \times \mathcal{Y}$ be a compact subset of $\mathbb{R}^{d} \times \mathbb{R}$. Suppose that Assumptions \ref{assumption:exp_tilting}, \ref{assumption:cov_shift}, \ref{assumption:existence_minimizer}, \ref{assumption:bounded},  \ref{assumption:nuisance}, \ref{assumption:pilot} hold. The one-step estimator satisfies $\sqrt{n}(\hat{\nu}_{n} - \nu_{0}) \xrightarrow{d} N(0, V)$, where
\begin{equation} 
\label{eq:variance} 
V := \dot{\Psi}_{\nu, 0}^{-1} \cdot \Sigma_{0} \cdot (\dot{\Psi}_{\nu, 0}^{-1})^{T}.
\end{equation}
\end{theo}

We propose to construct confidence intervals for $\nu_0$ by estimating the asymptotic variance $V$. Let $\Sigma_{n}^{\mathcal{I}}(\nu, \mu, r) := \frac{1}{|\mathcal{I}|} \sum_{i \in \mathcal{I}} \psi(X_i, Y_i; \nu, \mu, r) \cdot \psi(X_i, Y_i; \nu, \mu, r)^{T}.$

\begin{lemm}
\label{lemm:variance}
Let $\mathcal{X} \times \mathcal{Y}$ be a compact subset of $\mathbb{R}^{d} \times \mathbb{R}$. Suppose that Assumptions \ref{assumption:exp_tilting}, \ref{assumption:cov_shift}, \ref{assumption:existence_minimizer}, \ref{assumption:bounded}, \ref{assumption:nuisance}, \ref{assumption:pilot}, hold. Define the variance estimator
\begin{equation}
\label{eq:variance_est}
    V_{n} := \frac{|\mathcal{I}_{1}|}{n} V_{n}^{\mathcal{I}_{1}} + \frac{|\mathcal{I}_{2}|}{n} V_{n}^{\mathcal{I}_{2}},
\end{equation}
where
\[
    V_{n}^{\mathcal{I}_{1}} := (\dot{\Psi}_{n}^{\mathcal{I}_{1}}(\hat{\nu}_{n}, \hat{\mu}^{\mathcal{I}_{2}}_{n}, \hat{r}^{\mathcal{I}_{2}}_{n}))^{-1} \cdot \Sigma_{n}^{\mathcal{I}_{1}}(\hat{\nu}_{n}, \hat{\mu}_{n}^{\mathcal{I}_{2}}, \hat{r}_{n}^{\mathcal{I}_{2}}) \cdot ((\dot{\Psi}_{n}^{\mathcal{I}_{1}}(\hat{\nu}_{n}, \hat{\mu}_{n}^{\mathcal{I}_{2}}, \hat{r}_{n}^{\mathcal{I}_{2}} ))^{-1})^{T}\]
    and $V_{n}^{\mathcal{I}_{2}}$ is defined analogously. Then, $V_{n} \xrightarrow{p} V,$ where $V$ is defined in \eqref{eq:variance}.
\end{lemm}

We can apply Lemma \ref{lemm:variance} to obtain asymptotically-valid Gaussian confidence intervals for differentiable functionals of $\nu_{0}$, e.g. high-resolution mean outcomes. Define $\mu_{g}(\nu) := \EE[P_{g, X}]{\EE[Q_{Y|X}(\theta)]{Y \mid X}}$ to be the mean outcome of a subgroup $g$ under the conditional distribution $Q_{Y|X}(\theta)$. Note that $\mu_{g}(\nu_{0})$ is our target parameter. We can use the one-step estimator for $\nu_{0}$ to define an estimator for $\mu_{g}(\nu_{0})$ as follows
\begin{equation}
\label{eq:state_estimate}
\hat{\mu}_{g, \text{data-fusion}} = \mu_{g}(\hat{\nu}_{n}).
\end{equation}

\begin{lemm}
\label{lemm:delta}
Let $\mathcal{X} \times \mathcal{Y}$ be a compact subset of $\mathbb{R}^{d} \times \mathbb{R}$. Suppose that Assumptions \ref{assumption:exp_tilting}, \ref{assumption:cov_shift}, \ref{assumption:existence_minimizer}, \ref{assumption:bounded}, \ref{assumption:nuisance}, \ref{assumption:pilot}, hold. Let $\Phi$ denote the Normal cumulative distribution function. For any significance level $0 < \alpha < 1,$
\begin{equation} 
\label{eq:confidence_interval}
\lim_{n \rightarrow \infty} \PP[S]{\mu_{g}(\nu_0) \in \left(\mu_{g}(\hat{\nu}_{n}) \pm \Phi^{-1}(1 - \alpha/2) \cdot \sqrt{\frac{\nabla \mu_{g}(\hat{\nu}_{n})^{T} V_{n} \nabla \mu_{g}(\hat{\nu}_{n})}{n}}\right)} = 1 - \alpha.
\end{equation}
\end{lemm}

We present the overall procedure for constructing the one-step estimator and the confidence interval of $\mu_{g}(\nu_0)$ in Algorithm \ref{alg:one_step}. Note that the one-step estimation algorithm calls the algorithm for constructing the preliminary estimator, which is described in Section \ref{sec:alg}, as a subroutine.

\SetKwComment{Comment}{\#}{}
\SetKwInput{KwParams}{Parameters}

\begin{algorithm}
\small{
\DontPrintSemicolon
\caption{One-Step Estimation}\label{alg:one_step}
\KwData{Dataset $\mathcal{D} = \{(X_{i}, Y_{i}) \}_{i=1}^{n},\,$ Population Covariate Distribution $P_{X},\,$ Population Moment $\bar{\gamma}_{P}$, Population Subgroup Distributions $P_{g, X}$.
}
\KwParams{Significance Level $\alpha \in (0, 1)$.}
\KwResult{One-step Estimator $\hat{\nu}_{n}$, Confidence Interval $\mathrm{CI}_{\alpha}(\mu_{g}(\hat{\nu}_{n}))$}
\Comment{Construct preliminary estimator. See Algorithm \ref{alg:sqp}.}
$\tilde{\nu}_{n} \gets \texttt{getPreliminaryEstimator}(\mathcal{D}, P_{X}, \bar{\gamma}_{P})$ \;
\Comment{Randomly split the data into two folds.}
$\mathcal{I}_{1}, \mathcal{I}_{2} \gets \texttt{randomSplit}(\mathcal{D})$ \;
\For{$j \in \{1, 2\}$}{
\Comment{Fit nuisance parameters on opposite fold.}
$\hat{\mu}_{n}^{(- \mathcal{I}_{j})}, \hat{r}_{n}^{(- \mathcal{I}_{j})} \gets \texttt{getNuisanceParameters}(\mathcal{D}, P_{X}, \mathcal{I}_{j})$ \;
\Comment{Compute empirical estimating equations.}
$\{\psi(X_i, Y_i; \tilde{\nu}_{n}, \hat{\mu}_{n}^{(- \mathcal{I}_{j})}, \hat{r}_{n}^{(- \mathcal{I}_{j})})\}_{i \in \mathcal{I}_{j}} \gets \texttt{getEstimatingEquation}(\mathcal{D}, P_{X}, \mathcal{I}_{j}, \tilde{\nu}_{n}, \hat{\mu}_{n}^{(- \mathcal{I}_{j})}, \hat{r}_{n}^{(- \mathcal{I}_{j})}).$ \;
$\Psi_{n}^{\mathcal{I}_{j}}(\tilde{\nu}_{n}, \hat{\mu}_{n}^{(- \mathcal{I}_{j})}, \hat{r}_{n}^{(- \mathcal{I}_{j})}) \gets |\mathcal{I}_{j}|^{-1} \sum_{i \in \mathcal{I}_{j}} \psi(X_i, Y_i; \tilde{\nu}_{n}, \hat{\mu}_{n}^{(- \mathcal{I}_{j})}, \hat{r}_{n}^{(- \mathcal{I}_{j})})$\;
\Comment{Compute Jacobian via automatic differentiation.}
$\dot{\Psi}_{\nu, n}^{\mathcal{I}_{j}}(\tilde{\nu}_{n}, \hat{\mu}_{n}^{(- \mathcal{I}_{j})}, \hat{r}_{n}^{(- \mathcal{I}_{j})}) \gets \texttt{getJacobian}(\mathcal{D}, P_{X}, \mathcal{I}_{j}, \tilde{\nu}_{n}, \hat{\mu}_{n}^{(- \mathcal{I}_{j})}, \hat{r}_{n}^{(- \mathcal{I}_{j})}).$ \;
\Comment{Construct one-step estimator on current fold.} 
$\hat{\nu}_{n}^{\mathcal{I}_{j}} \gets \tilde{\nu}_{n} - (\dot{\Psi}_{\nu, n}^{\mathcal{I}_{j}}(\tilde{\nu}_{n}, \hat{\mu}_{n}^{(- \mathcal{I}_{j})}, \hat{r}_{n}^{(- \mathcal{I}_{j})}))^{-1}\Psi_{n}^{\mathcal{I}_{j}}(\tilde{\nu}_{n}, \hat{\mu}_{n}^{(- \mathcal{I}_{j})}, \hat{r}_{n}^{(- \mathcal{I}_{j})}).$\;}
\Comment{Compute cross-fit one-step estimator.}
$\hat{\nu}_{n} \gets n^{-1} (|\mathcal{I}_{1}| \cdot \hat{\nu}^{\mathcal{I}_{1}}_{n} + |\mathcal{I}_{2}| \cdot \hat{\nu}^{\mathcal{I}_{2}}_{n})$ \;
\For{$j \in \{1, 2\}$}{
\Comment{Estimate asymptotic variance on current fold.}
$\dot{\Psi}_{\nu, n}^{\mathcal{I}_{j}}(\hat{\nu}_{n}, \hat{\mu}_{n}^{(- \mathcal{I}_{j})}, \hat{r}_{n}^{(- \mathcal{I}_{j})}) \gets \texttt{getJacobian}(\mathcal{D}, P_{X}, \mathcal{I}_{j}, \hat{\nu}_{n}, \hat{\mu}_{n}^{(- \mathcal{I}_{j})}, \hat{r}_{n}^{(- \mathcal{I}_{j})}).$ \;
$\Sigma_{n}^{\mathcal{I}_{j}}(\hat{\nu}_{n}, \hat{\mu}_{n}^{(- \mathcal{I}_{j})}, \hat{r}_{n}^{(- \mathcal{I}_{j})}) \gets |\mathcal{I}_{j}|^{-1} \sum_{i \in \mathcal{I}_{j}} \psi(X_i, Y_i; \hat{\nu}_{n}, \hat{\mu}_{n}^{(- \mathcal{I}_{j})}, \hat{r}_{n}^{(- \mathcal{I}_{j})}) \psi(X_i, Y_i; \hat{\nu}_{n}, \hat{\mu}_{n}^{(- \mathcal{I}_{j})}, \hat{r}_{n}^{(- \mathcal{I}_{j})})^{T}$\;
$V_{n}^{\mathcal{I}_{j}} \gets (\dot{\Psi}_{\nu, n}^{\mathcal{I}_{j}}(\hat{\nu}_{n}, \hat{\mu}_{n}^{(- \mathcal{I}_{j})}, \hat{r}_{n}^{(- \mathcal{I}_{j})} ) )^{-1} \cdot \Sigma_{n}^{\mathcal{I}_{j}}(\hat{\nu}_{n}, \hat{\mu}_{n}^{(- \mathcal{I}_{j})}, \hat{r}_{n}^{(- \mathcal{I}_{j})}) \cdot ((\dot{\Psi}_{n}^{\mathcal{I}_{j}}(\hat{\nu}_{n},\hat{\mu}_{n}^{(- \mathcal{I}_{j})}, \hat{r}_{n}^{(- \mathcal{I}_{j})}))^{-1})^{T}$\;
}
\Comment{Compute asymptotic variance.}
$V_{n} \gets n^{-1} (|\mathcal{I}_{1}| \cdot V^{\mathcal{I}_{1}}_{n} + |\mathcal{I}_{2}| \cdot V^{\mathcal{I}_{2}}_{n})$ \;
\Comment{Construct confidence interval as in \eqref{eq:confidence_interval}.}
$\mathrm{CI}_{\alpha}(\mu_{g}(\hat{\nu}_{n}, n)) \gets \texttt{getConfidenceInterval}(P_{g, X}, \hat{\nu}_{n}, V_{n}, \alpha).$ \;
\Return{$\hat{\nu}_{n},\, \mathrm{CI}_{\alpha}(\mu_{g}(\hat{\nu}_{n}, n))$}
}
\end{algorithm}

\end{subsection}

\begin{subsection}{Initializing the one-step estimator}
\label{sec:alg}
We describe a practical algorithm for obtaining the preliminary estimator of the one-step estimation procedure. The main technical hurdle is that \eqref{eq:kl_param} is nonconvex. Our approach relies on sequential quadratic programming (SQP), which is an iterative technique for finding a local minimum of an optimization problem with objective and constraints that are twice differentiable but not necessarily convex \citep{boggs1995sequential,nocedal1999numerical}. Our SQP algorithm is locally convergent, so it will converge to the global minimum if initialized sufficiently close, but it is not guaranteed to converge to the global minimum from a remote starting point. 

We first describe the SQP algorithm in terms of population quantities and then provide a finite-sample algorithm in Algorithm \ref{alg:sqp}. Recall that $Q(\theta)$ is the family of distributions where the conditional distribution $Q_{Y|X}(\theta)$ is given by \eqref{eq:exp_family} and the covariate distribution $Q_{X}(\theta) = P_{X}$. Let $\bar{\gamma}(\theta) = \EE[Q(\theta)]{\gamma(X, Y)}.$ We can write the Lagrangian of \eqref{eq:kl_param} as follows
\begin{equation}
\label{eq:lagrangian}
L(\theta, \lambda) := \text{KL}(Q(\theta) || S) + \lambda^{T}(\bar{\gamma}(\theta) - \bar{\gamma}_{P}), \end{equation}
where $\lambda \in \mathbb{R}^{K}$ is the Lagrange multiplier. We express the derivatives of relevant terms as follows
\begin{align}
    B(\theta) &:= \nabla_{\theta} \bar{\gamma}(\theta),\, \label{eq:params1} \\
    g(\theta, \lambda) &:= \nabla_{\theta} L(\theta, \lambda), \label{eq:params2} \\
    H(\theta, \lambda) &:= \nabla^{2}_{\theta} L(\theta, \lambda), \label{eq:params3}
\end{align}
where $B(\theta) \in \mathbb{R}^{K \times J}$, $g(\theta, \lambda) \in \mathbb{R}^{1 \times J},$ and  $H(\theta, \lambda) \in \mathbb{R}^{J \times J}.$ 

The SQP algorithm begins with an approximate solution $(\theta_{(t)}, \lambda_{(t)}) \in \mathbb{R}^{J} \times \mathbb{R}^{K}.$ In each iteration of the SQP, we solve a quadratic program to find an update direction $\delta \in \mathbb{R}^{J}$ and its corresponding dual parameter $\rho \in \mathbb{R}^{K}$ at the current iterate and use them to define the next iterate $(\theta_{(t+1)}, \lambda_{(t + 1)})$. The objective of the quadratic program is a second-order approximation of the Lagrangian at the current iterate. The constraint of the quadratic program is a linearization of the constraint of \eqref{eq:kl_param} at the current iterate. The quadratic program solved in each iteration is given by
\begin{equation} 
\label{eq:iter_quad_program}
\min_{\delta \in \mathbb{R}^{J}} \left\{ \frac{1}{2} \delta^{T} H(\theta_{(t)}, \lambda_{(t)}) \delta + g(\theta_{(t)}, \lambda_{(t)}) \delta : B(\theta_{(t)}) \delta = \bar{\gamma}_{P} - \bar{\gamma}(\theta_{(t)}) \right\}. \end{equation}

Denote $\delta_{(t)}$ as the solution to \eqref{eq:iter_quad_program} and $\rho_{(t)}$ as the corresponding dual parameter of the quadratic program in \eqref{eq:iter_quad_program}. We obtain the subsequent iterate by
\begin{equation}
\label{eq:update}
\theta_{(t+1)} := \theta_{(t)} + \zeta_{\theta, (t)} \cdot \delta_{(t)}, \quad \lambda_{(t+1)} :=  \lambda_{(t)} + \zeta_{\lambda, (t)} \cdot (\rho_{(t)} - \lambda_{(t)})
\end{equation}
where $\zeta_{\theta, (t)}, \zeta_{\lambda, (t)} \in (0, 1]$ are step-size parameters.

\SetKwComment{Comment}{\#}{}

\begin{algorithm}
\DontPrintSemicolon
\caption{Preliminary Estimator Algorithm (with SQP Subroutine)}\label{alg:sqp}
\KwData{Dataset $\mathcal{D} = \{(X_{i}, Y_{i}) \}_{i=1}^{n},\,$ Population Covariate Distribution $P_{X}$, Population Moment $\bar{\gamma}_{P}$.}
\KwResult{Preliminary Estimator $\tilde{\nu}_{n}$}
\Comment{Estimate conditional distribution.}
$\widehat{dS}_{Y|X}(\cdot) \gets \texttt{getConditionalDistribution}(\mathcal{D})$ \;
\Comment{Run SQP Algorithm.}
\Comment{Initialize.}
$(\theta_{(0)}, \lambda_{(0)}) \gets 0, 0$ \;
$T=0$ \\
\For{$t=1,2, \dots$ }{
\Comment{Compute tilted conditional density.}
$\widehat{dQ}_{Y|X}(\theta_{(t)})(y) \propto \widehat{dS}_{Y|X}(y) \cdot \exp(\theta_{(t)}^{T}\eta(x, y)).$ \;
\Comment{Define QP parameters}
$\widehat{H}(\theta_{(t)}, \lambda_{(t)}),\, \widehat{B}(\theta_{(t)}),\, \widehat{g}(\theta_{(t)},\, \lambda_{(t)}),\, \widehat{\bar{\gamma}}(\theta_{(t)}) \gets \texttt{getQuadProgramParams}( P_{X}, \widehat{dQ}_{Y|X}(\theta)(\cdot), \theta_{(t)}, \lambda_{(t)}).$ \;
\Comment{Solve QP.}
$\delta_{(t)}, \rho_{(t)} \gets \texttt{QP}(\widehat{H}(\theta_{(t)}, \lambda_{(t)}),\, \widehat{B}(\theta_{(t)}),\, \widehat{g}(\theta_{(t)},\, \lambda_{(t)}),\, \widehat{\bar{\gamma}}(\theta_{(t)}),\, \bar{\gamma}_{P} )$  \;
\Comment{Update iterate.}
$\theta_{(t+1)} \gets \theta_{(t)} + \zeta_{\theta, (t)} \cdot \delta_{(t)}$ \\
$\lambda_{(t+1)} \gets \lambda_{(t)} + \zeta_{\lambda, (t)} \cdot (\rho_{(t)} - \lambda_{(t)}).$ \;
$T \gets T + 1$ \;
\Comment{End if $|\delta|$ is smaller than tolerance.}
\If{ $|\delta_{(t)}| < \epsilon $}{\KwSty{break}}
}
\Comment{Construct preliminary estimator from final iterate.}
$\tilde{\theta}_{n},\, \tilde{\lambda}_{n} \gets \theta_{(T)},\, \lambda_{(T)}$ \;
$\tilde{\nu}_{n} \leftarrow (\tilde{\theta}_{n},\, \tilde{\lambda}_{n}).$ \;
\Return{$\tilde{\nu}_{n}$}.
\end{algorithm}

We require closed-form expressions for the terms in \eqref{eq:params1}, \eqref{eq:params2}, \eqref{eq:params3}. In the following lemma, we derive expressions for these terms.

\begin{lemm}
\label{lemm:params}
Let $\theta \in \mathbb{R}^{J}$ and $\lambda \in \mathbb{R}^{K}.$ Let 
$I(\theta) := \nabla_{\theta} \EE[Q(\theta)]{\eta(X, Y)}$. The terms in \eqref{eq:params1}, \eqref{eq:params2}, \eqref{eq:params3}, and $I(\theta)$ are given by
\begin{align*}
I(\theta) &= \EE[P_{X}]{ \Cov[Q_{Y|X}(\theta)]{\eta(X, Y)|X}}, \\
B(\theta) &= \EE[P_{X}]{ \Cov[Q_{Y|X}(\theta)]{\gamma(X, Y), \eta(X, Y),|X}}, \\
g(\theta, \lambda) &= \theta^{T} I(\theta) + \lambda^{T} B(\theta), \\
H(\theta, \lambda) &= I(\theta) + \theta^{T} \EE[P_{X}]{ \EE[Q_{Y|X}(\theta)]{(\eta(X, Y) - \EE[Q_{Y|X}(\theta)]{\eta(X, Y)|X})^{\otimes 3} |X}} \\
&\indent+ \lambda^{T}\Big(\EE[P_{X}]{ \EE[Q_{Y|X}(\theta)]{\gamma(X, Y)\eta(X, Y)^{T} \otimes (\eta(X, Y) - \EE[Q_{Y|X}(\theta)]{\eta(X, Y) |X}) \mid X}} \\
&\indent\indent-  \EE[P_{X}]{ \EE[Q_{Y|X}(\theta)]{\gamma(X, Y) \mid X} \otimes \Cov[Q_{Y|X}(\theta)]{\eta(X, Y) \mid X}} \\
&\indent\indent-\EE[P_{X}]{\Cov[Q_{Y|X}(\theta)]{\gamma(X, Y), \eta(X, Y) \mid X} \otimes \EE[Q_{Y|X}(\theta)]{\eta(X, Y) \mid X}}\Big).
\end{align*}
\end{lemm}

To implement a finite-sample version of this SQP algorithm, we must build estimators of the quadratic program parameters. Note that by Lemma \ref{lemm:params}, all of these parameters can be expressed as expectations over $Q_{Y|X}(\theta)$ and $P_{X}$. While $P_{X}$ is known, $Q_{Y|X}(\theta)$ depends on an unknown quantity, the conditional online survey distribution $S_{Y|X}$. After developing an estimator for $S_{Y|X}$, the SQP algorithm can be implemented with the estimated quadratic program parameters. We set the preliminary estimator $\tilde{\nu}_{n}$ to the final iterate of the finite-sample SQP procedure. The full algorithm is provided in Algorithm \ref{alg:sqp}.

In the case where $Y$ is binary, $S_{Y|X}$ is fully characterized by the conditional mean function $\mu_{0}(x)=\EE[S]{Y=1 \mid X=x}$, so it suffices to estimate this quantity to develop an estimator of $Q_{Y|X}(\theta)$. Recall that in Section \ref{subsec:pop_estimating_equations}, we defined $Q_{Y|X}(\theta, \mu)$ as the distribution where $dQ_{Y|X}(\theta, \mu) \propto dS_{Y|X}(\mu) \cdot \exp(\theta^{T}\eta(x, y)).$ As a result, a plug-in estimator of $Q_{Y|X}(\theta)$ is given by $Q_{Y|X}(\theta, \hat{\mu})$, where $\hat{\mu}$ is an estimator of $\mu_{0}$. Then, the parameters in \eqref{eq:params1}, \eqref{eq:params2}, and \eqref{eq:params3} can be estimated using the formulas from Lemma \ref{lemm:params}, the plug-in estimator $Q_{Y|X}(\theta, \hat{\mu})$, and $P_{X}$. We compute conditional expectations using numerical integration with the plug-in estimator $Q_{Y|X}(\theta, \hat{\mu})$, and we compute expectations over $P_{X}$ via standard integration (though Monte Carlo integration can also be used for computational efficiency).

\end{subsection}
\end{section}


\begin{section}{High-Resolution Estimation of Public Health Indicators}
\label{sec:exp}
We evaluate our data fusion method on a testbed of public health indicators that are measured in both the online Household Pulse Survey \citep{acs} and a ground truth data source. We describe our data sources, empirical setup, and results.
\begin{subsection}{Data Sources and Uses}
\label{sec:data}
We create a testbed that pairs indicators from the online Household Pulse Survey \citep{acs} with ground-truth data from alternative sources.\footnote{Data for replication is available \href{https://drive.google.com/drive/folders/1ALGzHTkK1k4X5HJIPtzwmwJE1EApMMbk?usp=drive_link}{here}.}
\begin{table}[t]
    \centering
    \renewcommand{\arraystretch}{1.5} 
    \setlength{\tabcolsep}{6pt} 
    \begin{tabular}{|c|c|c|c|c|}
        \hline
        \multirow{3}{5em}{Health Indicator} & \multirow{3}{5em}{Online Survey} & \multirow{3}{5em}{Ground Truth (GT)} & \multirow{3}{5em}{GT Low-Resolution Level} & \multirow{3}{5em}{GT High-Resolution Level} \\
        & & & & \\
        & & & & \\
        \hline 
        COVID-19 Vaccination & HPS  & CDC    & Eastern Region & State  \\
        SNAP Enrollment     & HPS  & USDA   & Eastern Region & State  \\
        Medicaid Enrollment        & HPS  & ACS  & Eastern Region & State \\ 
        \hline
    \end{tabular}
    \caption{Our testbed spans three different population health indicators, pairing online survey data with three ground-truth data sources available at both low- and high-resolutions. Our learning procedure utilizes the online survey data and ground-truth data at the low-resolution level to provide high-resolution estimates. The ground-truth high-resolution data is used to evaluate these estimates.}
    \label{tab:data_pairing}
\end{table}
Relatively few survey indicators have ground-truth data available at a high-resolution level, which is required for evaluation of our learning procedure. We identified three indicators suitable for evaluation: COVID-19 vaccination status, SNAP enrollment, and Medicaid enrollment. The state-level ground-truth data from these indicators is aggregated as necessary to form state-level, regional-level, and national-level ground-truth data that are used for evaluation and learning.

Initially developed by the Census Bureau to capture near real-time impacts of the COVID-19 pandemic on multiple health, social, and economic outcomes, the online Household Pulse Survey collected responses in two week cycles from April 2020 until September 2024. The average sample size per cycle was 60,000. Compared to traditional population health surveys, such as BRFSS (annual state-level data with a sample size of 400,000) \citep{brfss_2021} and NHANES (biannual national-level data with a sample size of 12,000) \citep{nhanes}, HPS was a more timely source of state-level data. However, HPS suffered from greater sampling bias than traditional surveys, undermining its potential utility for high-resolution estimation \citep{bradley2021unrepresentative}.

COVID-19 vaccination status was a critical population health indicator in 2021, when vaccines became available and there was substantial geographic heterogeneity in uptake. This indicator was also used in a prior evaluation of the utility of large-scale online survey data  \citep{bradley2021unrepresentative}. Our indicator of COVID-19 vaccination status from HPS is based on the question ``Have you received a COVID-19 vaccine?", with ``Yes" coded as 1 and ``No" coded as 0. We used data on COVID-19 vaccination administration collected by the Centers for Disease Control and Prevention (CDC) as our ground-truth data \citep{covid_data}. These data report the percentage of people receiving at least one dose of the COVID-19 vaccine by day, at state-level and national-level. We collapsed daily ground-truth data to generate mean coverage targets that matched the HPS two-week cycle periods.

SNAP participation and Medicaid coverage status are important indicators of social determinants of health, including income, food security, and insurance status. We also hypothesize that these indicators may have different patterns of bias in HPS data compared to COVID-19 vaccination coverage. Our indicator of SNAP enrollment status from HPS is based on the question ``Do you or anyone in your household receive benefits from the Supplemental Nutrition Assistance Program (SNAP) or the Food Stamp Program?", with ``Yes" coded as 1 and ``No" coded as 0. We used data on SNAP enrollment collected by the United States Department of Agriculture (USDA) as our ground-truth data \citep{snap}. These data report the number of households enrolled in SNAP by month, at state-level. We generated mean SNAP ground targets that match HPS two-week cycle periods, by computing a weighted average of the monthly values, where the weights correspond to the number of days each two-week period overlaps with the given month.

Our indicator of Medicaid enrollment status from HPS is based on the question ``Are you currently covered by Medicaid, Medical Assistance, or any kind of government-assistance plan for those with low incomes or a disability?" with ``Yes" coded as 1 and ``No" coded as 0. We used data on Medicaid insurance coverage from the American Community Survey (ACS) as our ground-truth data \citep{acs}. These data report individuals with Medicaid insurance coverage by year for covariate-defined groups. We combined HPS data across cycles to match the annual ground-truth data.

Our learning procedure also leverages individual-level and state-level covariates. We included the following individual-level covariates from HPS: age group, sex, race/ethnicity, educational attainment, marital status, household income, and household size. Instead of using state-level fixed effects, we incorporated state-level covariates from the ACS in our procedure. This facilitated information sharing across similar states, characterized by socioeconomic features, including: educational attainment, income, employment status, housing status, marital status, veteran status, political party affiliation, and primary language.  
\end{subsection}
\begin{subsection}{Empirical Setup}
Both the online surveys and ground-truth data sources of our public health testbed are collected at repeated time intervals (see Section \ref{sec:data} for details), so we evaluate high-resolution estimates obtained from online surveys at multiple time steps $t=1, 2, \dots, T$. Let $S^{t}, P^{t}$ denote the online survey and population distribution at time step $t$, respectively. We assume the population covariate distribution $P_{X}^{t}$ is given by the 2020 Census and does not change with $t$. However, note that the survey distribution $S^{t}$ and the population conditional distribution $P_{Y|X}^{t}$ may evolve over time. 

At each time-step $t$, we observe $n_{t}$ i.i.d. samples from the survey distribution  $(X_i, Y_i) \sim S^{t}$, where $n_{t}$ ranges from 38,380 to 77,925 for a two-week time period and $n_{t}$ ranges from 659,641 to 1,187,526 for an annual time period. We also observe a moment condition that captures the average outcome in the Eastern region of the United States.\footnote{The Eastern region consists of states Alabama, Connecticut, Delaware, Florida, Georgia, Indiana, Kentucky, Maine, Maryland, Massachusetts, Michigan, New Hampshire, New Jersey, New York, North Carolina, Ohio, Pennsylvania, Rhode Island, South Carolina, Tennessee, Vermont, Virginia, and West Virginia.} Let $\mathcal{A}_{\text{East}} \subset \mathcal{X}$ denote a subset of the covariate space that corresponds to the Eastern region of the United States. Let $\psi(x, y) = y \cdot \mathbb{I}(x \in \mathcal{A}_{\text{East}})$. We observe 
\[\EE[P^{t}]{\psi(X, Y)} = \EE[P^{t}]{Y \cdot \mathbb{I}(X \in \mathcal{A}_{\text{East}})} = \bar{\psi}_{P^{t}}.\]

We evaluate the state-level estimates of health indicators obtained via different learning approaches. Let $\mathcal{G}$ be the set of US states. Given the data from time-step $t$, we compute high-resolution estimates $\{\hat{\mu}_{g}^{t}\}$ for $g \in \mathcal{G}$. Let $P_{g, X}$ denote the covariate distribution of state $g$, which can be obtained from the Census. To assess all methods, we compare the high-resolution estimates to the ground-truth high-resolution data $\{ \mu_{g}^{t}\}$ obtained from the ground-truth data source during the same time step (see Table \ref{tab:data_pairing}). We report a population-weighted MAE for each time step $t$. Let $p_{g}$ be the proportion of the US population from state $g$.
\begin{equation} \text{Population-Weighted MAE at Time-Step $t$} = \sum_{g \in \mathcal{G}} p_{g} \cdot |\hat{\mu}_{g}^{t} - \mu_{g}^{t}|.\end{equation}

We evaluate three baselines and our data fusion method.
\paragraph{Aggregated Administrative Data.} This baseline method produces state estimates by using only regional-level information provided by the moment conditions. We set
\[ \hat{\mu}^{t}_{g, \text{agg}} = \frac{\bar{\psi}_{P^{t}}}{\PP[P]{X \in \mathcal{A}_{\text{East}} }} \quad \forall g \in \mathcal{G}.\]
Note that this yields the same estimate for all states.
\paragraph{Biased High-Resolution Survey.} This baseline method produces state estimates by using only the high-resolution online survey data with standard covariate-based reweighting. We estimate the covariate density ratio via probabilistic classification \citep{menon2016linking,sugiyama2008direct}. Since all three outcomes we consider are binary, i.e. $\mathcal{Y} = \{0, 1\}$, we train a three-layer neural network with the covariate-reweighted logistic loss to estimate the conditional mean of the online survey data
\[ \mu_{S^{t}}(x) := \EE[S^{t}]{Y \mid X=x} = \PP[S^{t}]{Y=1 \mid X=x}.\]
Then, we aggregate this function over the group covariate distribution $P_{g}$ to produce high-resolution estimates, i.e.
\[ \hat{\mu}^{t}_{g, \text{biased}} = \EE[P_{g, X}]{ \hat{\mu}_{S^{t}}(X)} \quad \forall g \in \mathcal{G}.\]
This baseline corresponds to developing state estimates using only the potentially biased, high-resolution data source with standard covariate-based reweighting.
\paragraph{Model-Free.} The model-free approach  uses data from both the survey distribution and the moment condition at each time-step $t$ to solve \eqref{eq:kl_agnostic} and obtain a plausible population distribution $\tilde{Q}(\theta)$ with the form described in \eqref{eq:psi_tilting}. The choice of $\theta$ is learned by solving \eqref{eq:model_free_alg} via gradient descent. We compute state-level outcome estimates under this distribution
\[ \hat{\mu}^{t}_{g, \text{model-free}} = \EE[P_{g, X}]{\EE[\tilde{Q}^{t}_{Y|X}(\theta)]{Y \mid X}} \quad \forall g \in \mathcal{G}.\] 
This method requires that we estimate $\PP[S^{t}]{Y = 1 \mid X}$ from the online survey data. We use the same procedure for fitting this estimator as in the Biased High-Resolution Survey approach. 

\paragraph{Data Fusion.} Our data fusion approach produces state estimates by applying the method developed in Section \ref{sec:setup}. We use $J=1$ dimensional sufficient statistic $\eta(x, y) = y$, which is a simple model of response bias, where the amount of response bias due to unobservables is assumed to be the same for all units. 

The estimates $\hat{\mu}_{g, \text{data-fusion}}$ are obtained by leveraging the estimation and inference procedure from Section \ref{sec:alg_inference} to learn a plausible population distribution $Q^{t}(\theta)$ at each time-step $t$. Our procedure requires an estimate $\PP[S^{t}]{Y = 1 \mid X}$ from the online survey data. We obtain this estimate by fitting a three-layer neural network with the logistic loss via stochastic gradient descent on the online survey samples.

We compute state-level outcome estimates for $\hat{\mu}^{t}_{g, \text{data-fusion}}$ as defined in \eqref{eq:state_estimate}. We also compute confidence intervals on $\hat{\mu}^{t}_{g, \text{data-fusion}}$ as described in Section \ref{sec:alg_inference}.

\end{subsection}
\begin{subsection}{Results}
We evaluate the four approaches described above across all indicators in Table \ref{tab:data_pairing}. We observe that the methods perform consistently across all indicators, although the indicators exhibit different amounts of response bias. In this section, we visualize predictions from all methods for a single time period for each indicator, and we also visualize the population-weighted MAE of all methods for all time periods. Predictions for all time periods and indicators are provided in the appendix. 

\begin{figure}
\centering
\includegraphics[width=\textwidth]{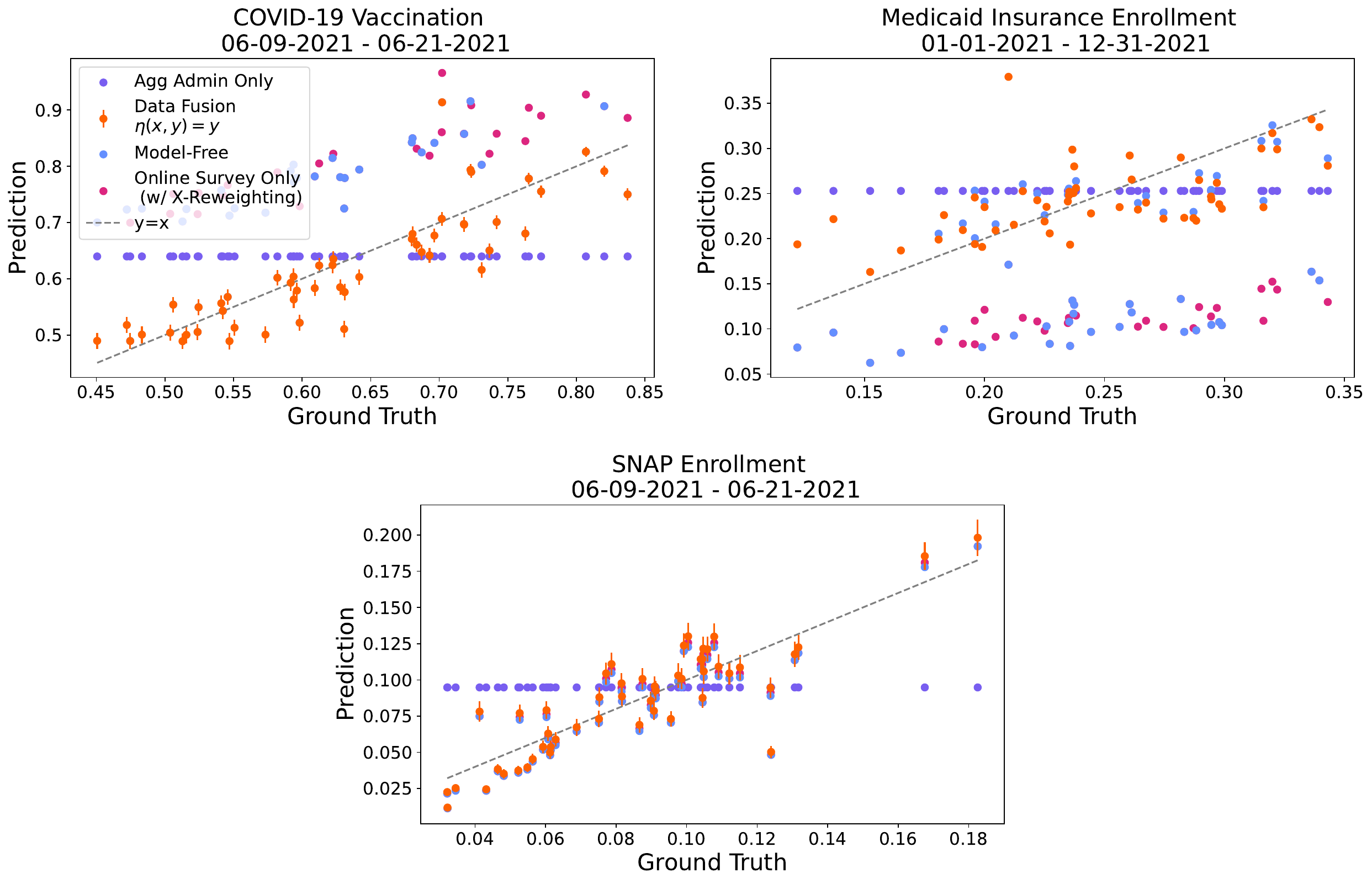}
\caption{This figure includes scatter-plots of state-level predictions under all methods vs. state ground truth for each indicator COVID-19 Vaccination (top left), Medicaid Enrollment (top right), and SNAP Enrollment (bottom). Each dot in a subplot corresponds to a prediction for a particular state. Predictions that lie on the $y=x$ line match the ground truth.}
\label{fig:pred}
\end{figure}

Figure \ref{fig:pred} visualizes the predictions of the different methods for all three indicators for a single time period plotted against the ground-truth estimates. Recall that the Biased High-Resolution Survey method has access to the online survey data, the Aggregated Administrative Data method has access to the Eastern regional mean outcome, and the Model-Free and Data Fusion methods have access to both the online survey data and Eastern regional mean outcome. 

The magenta series corresponds to the Biased High-Resolution Survey method. We find that the predictions of the Biased High-Resolution Survey method captures much of the heterogeneity across states that exists in the population distribution. Interestingly, we find that the Biased High-Resolution Survey predictions are inflated relative to the ground truth for COVID-19 vaccination and deflated relative to the ground truth for Medicaid enrollment, suggesting that units who are likely to respond to online surveys may also be more likely to get vaccinated but less likely to have Medicaid insurance. For the SNAP indicator, we find that the predictions of the Biased High-Resolution Survey method coincide with the predictions of the Data Fusion method and are quite close to the ground truth, suggesting that sampling bias in this indicator may be comparatively less severe. The purple series corresponds to the Aggregated Administrative Data method that relies only on Eastern regional mean outcome. This approach is unable to capture any of the heterogeneity across states. Finally, the orange series corresponds to our Data Fusion approach that relies on both the high-resolution online survey data and the Eastern regional mean outcome. We find that predictions from this approach are the most accurate, falling roughly on the $y=x$ line.

The Model-Free approach utilizes both the online survey data and the Eastern regional mean outcome. As described in Section \ref{sec:setup}, the key limitation of the model-free approach is that it only adjusts the survey conditional distribution for units belonging to the Eastern region and otherwise does not deviate from the survey conditional distribution.
\begin{figure}
\centering
\includegraphics[width=\textwidth]{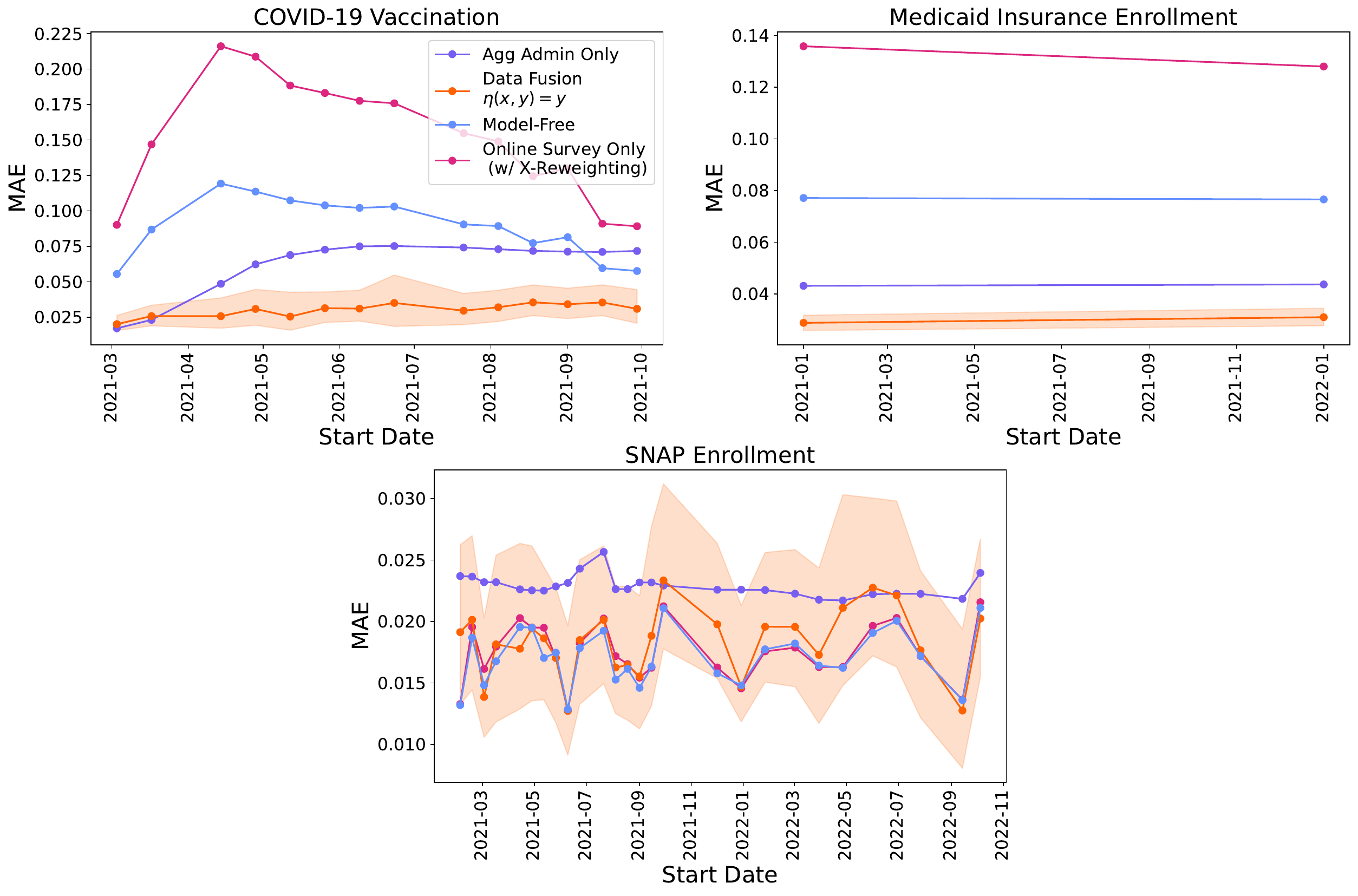}
\caption{We plot the population-weighted MAE of state level estimates of all three indicators over time. The leftmost panel corresponds to COVID-19 Vaccination, for which there are ground truth data at 2 week intervals from March to October 2021, the middle panel is Medicaid Enrollment, for which there is ground truth data at an annual time scale, and the rightmost panel is SNAP Enrollment, for which there is ground truth data at 2 week intervals from February 2021 - October 2022.}
\label{fig:mae}
\end{figure}

Figure \ref{fig:mae} visualizes the population-weighted MAE of all four methods and all indicators over time. In the top left panel on COVID-19 vaccination, we find that the MAE of the Biased High-Resolution Survey method is quite high, 0.15 on average over time. This result is consistent with the observation of \citet{bradley2021unrepresentative} that finds online surveys to be unrepresentative of the US adult population for indicators such as vaccination status. The MAE of the Aggregated Administrative Data method is 0.06 on average over time. The MAE of the Model-Free approach is 0.09 on average over time, outperforming the Biased High-Resolution Survey method but performing worse than the Aggregated Administrative Data. The MAE of the Data Fusion method is 0.03 on average over time, and it consistently performs on par or better than the baselines. In the top right panel, for Medicaid enrollment, we find that the MAE of the Biased High-Resolution Survey method is again quite high, 0.13 on average over time. The MAE of the Aggregated Administrative Data method is about 0.04. The MAE of the Model-Free approach is about 0.08, again providing an improvement over the Biased High-Resolution Survey method but performing worse than the Aggregated Administrative Data method. We find that the MAE of the Data Fusion method is 0.03, outperforming baselines. Lastly, in the bottom panel, we find that all methods perform similarly for the SNAP indicator.

\end{subsection}
\end{section}

\begin{section}{Discussion}
We develop a data fusion method that leverages the complementary strengths of high-resolution, but biased online survey data and low-resolution, but unbiased aggregate data to produce high-resolution estimates that outperform comparators derived from either data source alone. The key components of our framework include an exponential tilting to model a missing not at random data-generating process, extending similar models from sensitivity analysis in causal inference \citep{franks2020flexible,scharfstein1999adjusting,zhou2023sensitivity}, and moment matching using outcome data, inspired by covariate-balancing techniques for addressing shift due to observables \citep{zhao2019covariate, zubizarreta2015stable}. On a testbed of public health indicators, our method yields substantial reductions in the mean absolute error of high-resolution estimates of indicators that exhibit sampling bias but does not degrade the quality of these estimates for indicators that do not exhibit sampling bias.
 
Broadly, our work can contribute to the design of integrated multimodal public health surveillance systems and suggests that spreading investments across multiple complementary data sources could be more efficient than relying on a single imperfect source. Future work should consider the optimal design of such systems and also develop methods to support fusion of more than two sources. Continued innovation of statistical methods that leverage modern data and computation is needed to address increasingly complex contemporary public health challenges.
\end{section}


\bibliographystyle{plainnat} 
\bibliography{ref.bib}

\appendix

\begin{section}{Proofs}
\label{app:proof}
\begin{subsection}{Tensor Product Notation}
\label{sec:tensor_product}
We denote tensor products with $\otimes$ symbol. Let $A \in \mathbb{R}^{m \times n}$, $B \in \mathbb{R}^{m \times n \times p}$ and $v \in \mathbb{R}^{p}$. We define $v^{\otimes k} = v \otimes v \dots \otimes v$, where the right side is the product of $k$ terms. We define $v \otimes v$, $A \otimes v$, $v \otimes A$ as
\begin{align*}
    (v \otimes v)_{ij} = v_{i}v_{j},\, \quad (A \otimes v)_{ijk} = A_{ij} \cdot v_{k},\, \quad (v \otimes A)_{ijk} = v_{i} \cdot A_{jk}, 
\end{align*}
where $v \otimes v \in \mathbb{R}^{p \times p}$, $A \otimes v \in \mathbb{R}^{m \times n \times p}$, and $v \otimes A \in \mathbb{R}^{p \times m \times n}$. We can also define a tensor contraction as follows
\begin{align*}
    (B \cdot v)_{ij} &= \sum_{k=1}^{p} B_{ijk}v_{k},
\end{align*}
where $B \cdot v \in \mathbb{R}^{m \times n}.$ Lastly, define $\text{vec}(B)$ denote a vector of length $m \cdot n$, where the $i$-th index is $i$-th index corresponds to $\lfloor i/m \rfloor, i - m \cdot \lfloor i/m \rfloor$ entry of $B$.
\end{subsection}

\begin{subsection}{Proof of Lemma \ref{lemm:exp_family}}
First, we show that if Assumption \ref{assumption:exp_tilting} holds, then \eqref{eq:exp_family} for some $\theta \in \mathbb{R}^{J}.$

We use \eqref{eq:binary_selection_prob} to express $\PP[Q]{R=1 \mid X=x}.$
\begin{align*}
\PP[Q]{R=1 \mid X=x} &= \int_{\mathcal{Y}} dQ_{Y \mid X=x}(y') \cdot \PP[Q]{R=1 \mid X=x, Y=y'}dy' \\
&= \int_{\mathcal{Y}} \PP[Q]{R=1 \mid X=x, Y=y} \cdot dQ_{Y \mid X=x}(y') \cdot \exp(\theta^{T}(\eta(x, y) - \eta(x, y'))) dy' \\
&= \PP[Q]{R =1 \mid X=x, Y = y} \cdot \exp(\theta^{T}\eta(x, y)) \cdot \int_{\mathcal{Y}} dQ_{Y \mid X=x}(y') \cdot \exp(-\theta^{T}\eta(x, y')) dy'.
\end{align*}

Using Bayes' rule, we have that
\begin{align*}
\frac{dQ_{Y \mid X=x}(y)}{dS_{Y \mid X=x}(y)} &= \frac{\PP[Q]{R=1 \mid X=x}}{ \PP[Q]{R=1 \mid X=x, Y=y}} \\
&= \frac{\PP[Q]{R =1 \mid X=x, Y = y} \cdot \exp(\theta^{T}\eta(x, y)) \cdot \int_{\mathcal{Y}} dQ_{Y \mid X=x}(y') \cdot \exp(-\theta^{T}\eta(x, y')) dy'}{\PP[Q]{R=1 \mid X=x, Y=y}} \\
&= \exp(\theta^{T}\eta(x, y)) \cdot \int_{\mathcal{Y}} dQ_{Y \mid X=x}(y') \cdot \exp(-\theta^{T}\eta(x, y')) dy'.
\end{align*}

In summary,
\begin{equation}
\label{eq:ratio}
\frac{dQ_{Y \mid X=x}(y)}{dS_{Y \mid X=x}(y)} = \exp(\theta^{T}\eta(x, y)) \cdot \int_{\mathcal{Y}} dQ_{Y \mid X=x}(y') \cdot \exp(-\theta^{T}\eta(x, y')) dy'.
\end{equation}

Define the log-partition function
\begin{align*}
A_{x}(\theta) := \log \left(\int_{\mathcal{Y}} dS_{Y \mid X=x}(y) \exp(\theta^{T}\eta(x, y)) dy\right).
\end{align*}

Since $\frac{dQ_{Y \mid X=x}(y)}{dS_{Y \mid X=x}(y)}$ is a likelihood ratio, we must have that $\EE[S_{Y \mid X=x}]{\frac{dQ_{Y \mid X=x}(Y)}{dS_{Y \mid X=x}(Y)}} = 1.$ As a result,

\begin{align*}
\EE[S_{Y \mid X=x}]{\frac{dQ_{Y \mid X=x}(Y)}{dS_{Y \mid X=x}(Y)}} &= \int_{\mathcal{Y}} dS_{Y \mid X=x}(y) \exp(\theta^{T}\eta(x, y)) \cdot \int_{\mathcal{Y}} dQ_{Y \mid X=x}(y') \cdot \exp(-\theta^{T}\eta(x, y')) dy' dy \\
&= \int_{\mathcal{Y}} dS_{Y \mid X=x}(y) \exp(\theta^{T}\eta(x, y)) dy \cdot \int_{\mathcal{Y}} dQ_{Y \mid X=x}(y') \cdot \exp(-\theta^{T}\eta(x, y')) dy' \\
&= \exp(A_{x}(\theta)) \cdot \int_{\mathcal{Y}} dQ_{Y \mid X=x}(y') \cdot \exp(-\theta^{T}\eta(x, y')) dy' \\
&= 1.
\end{align*}

This yields that 
\begin{equation}
\label{eq:A_x_def}
\exp(-A_{x}(\theta)) = \int_{\mathcal{Y}} dQ_{Y \mid X=x}(y') \cdot \exp(-\theta^{T}\eta(x, y')) dy'.
\end{equation}
We can plug \eqref{eq:A_x_def} into \eqref{eq:ratio} to see that \eqref{eq:exp_family} holds.

Second, we show that if \eqref{eq:exp_family} holds, then this implies that Assumption \ref{assumption:exp_tilting} must hold. We define $Q$ to be a distribution over $(X, Y, R)$. We define that 
\begin{equation} 
\label{eq:defi}
\PP[Q]{R=1 \mid X=x, Y=y} := \frac{1}{N} \cdot \frac{dS_{X}(x)}{dQ_{X}(x)} \cdot \exp(- \theta^{T} \eta(x, y) + A_{x}(\theta)),
\end{equation}
where $N \leq \frac{1}{C} \cdot \sup_{x \in \mathcal{X}, y \in \mathcal{Y}} \exp(- \theta^{T} \eta(x, y) + A_{x}(\theta)).$ We must verify that \eqref{eq:binary_selection_prob} and that $Q_{X, Y \mid R=1} = S.$

We note that 
\begin{align*}
&\log\left(\int_{\mathcal{Y}} \exp(-\theta^{T}\eta(x, y)) dQ_{Y \mid X=x}(y) dy \right) \\
&= \log \left(\int_{\mathcal{Y}} \exp(-\theta^{T}\eta(x, y)) dS_{Y \mid X=x}(y) \exp(\theta^{T}\eta(x, y) - A_{x}(\theta)) dy  \right) \\
&= \log \left(\int_{\mathcal{Y}} dS_{Y \mid X=x}(y) \exp(-A_{x}(\theta)) dy \right) \\
&= - A_{x}(\theta).
\end{align*}
So, $-A_{x}(\theta)$ is the log-partition function of the exponential family with sufficient statistics $\eta(x, y)$, natural parameter $-\theta$, and base measure $dQ_{Y \mid X=x}.$ As a result,
\begin{align*}
\PP[Q]{R=1 \mid X=x} &=  \frac{1}{N} \cdot \frac{dS_{X}(x)}{dQ_{X}(x)} \cdot \EE[Q_{Y \mid X}]{\exp(- \theta^{T} \eta(X, Y) + A_{X}(\theta)) \mid X=x } \\
&= \frac{1}{N} \cdot \frac{dS_{X}(x)}{dQ_{X}(x)} \cdot \int_{y \in \mathcal{Y}} \exp(-\theta^{T}\eta(x, y) - (-A_{x}(\theta))) \cdot dQ_{Y \mid X=x}(y) \\
&= \frac{1}{N} \cdot \frac{dS_{X}(x)}{dQ_{X}(x)}.
\end{align*}
Thus, 
\begin{equation}
\label{eq:p_s_1_given_x}
\PP[Q]{R=1 \mid X=x} = \frac{1}{N} \cdot \frac{dS_{X}(x)}{dQ_{X}(x)}
\end{equation}

Plugging \eqref{eq:p_s_1_given_x} into the definition in \eqref{eq:defi}, we have that
\begin{equation}
\label{eq:rearrange}
\PP[Q]{R =1 \mid X=x, Y=y} = \PP[Q]{R =1 \mid X=x} \cdot \exp(- \theta^{T} \eta(x, y) + A_{x}(\theta)).
\end{equation}

Thus, we have that  Assumption \ref{assumption:exp_tilting} holds:
\begin{align*}
\frac{\alpha(x, y)}{\alpha(x, y')} &= \frac{\PP[Q]{R=1 \mid X=x, Y= y}}{\PP[Q]{R=1 \mid X=x, Y= y'}} \\
&= \frac{\PP[Q]{R =1 \mid X=x} \cdot \exp(- \theta^{T} \eta(x, y) + A_{x}(\theta))}{\PP[Q]{R =1 \mid X=x} \cdot \exp(- \theta^{T} \eta(x, y') + A_{x}(\theta))} \\
&= \exp(\theta^{T}(\eta(x, y') - \eta(x, y))).
\end{align*}


\end{subsection}

\begin{subsection}{Proof of Lemma \ref{lemm:parametric_characterization}}
By Lemma \ref{lemm:exp_family}, a distribution $Q$ that lies in the feasible set of \eqref{eq:kl_w_covariate} satisfies $Q_{Y|X} \sim_{\eta} S_{Y|X}$ must be a conditional exponential tilting of $S_{Y|X}$ as defined in \eqref{eq:exp_family}. Furthermore, if $Q$ lies in the feasible set of \eqref{eq:kl_w_covariate}, then $Q_{X} = P_{X}$. We note that the family of distributions that satisfies these properties is $Q(\theta)$. We observe that
\begin{align*}
\text{KL}(Q(\theta) || S) &= \text{KL}(Q_{X}(\theta) ||S_{X}) + \EE[P_{X}]{\text{KL}(Q_{Y|X}(\theta) || S_{Y|X})} \\
&= \text{KL}(P_{X} || S_{X}) + \EE[P_{X}]{\text{KL}(Q_{Y|X}(\theta) || S_{Y|X})} \\
&= C + \EE[P_{X}]{\text{KL}(Q_{Y|X}(\theta) || S_{Y|X})}
\end{align*} 
The last line follows from the fact that $S_{X} \ll P_{X}$ under our data-generating process and by Assumption \ref{assumption:cov_shift}, $P_{X} \ll S_{X}$, so the $\text{KL}(P_{X}||S_{X})$ must be bounded. This yields \eqref{eq:kl_param}.
\end{subsection}



\begin{subsection}{Proof of Lemma \ref{lemm:orthogonal}}

We can name the components of $\Psi$ by $M^{\perp}$ and $D_{L}^{\perp}$. These components both depend on a plug-in term $M, D_{L}$, respectively, and a correction term, $C_{M}, C_{D_{L}}$. We define the correction terms below.
\begin{align*}
C_{M}(\nu; \mu, r) &:= \EE[S]{r(X) \cdot \delta_{M}(X; \nu, \mu) \cdot (Y - \mu(X))} \\
C_{D_{L}}(\nu; \mu, r) &:= \EE[S]{r(X) \cdot \delta_{D_{L}}(X; \nu, \mu) \cdot (Y - \mu(X))}.
\end{align*}
Thus, the components of $\Psi$ are given by
\begin{align*}
M^{\perp}(\nu; \mu, r) &:= M(\nu; \mu) + C_{M}(\nu; \mu, r) \\
D_{L}^{\perp}(\nu; \mu, r) &:= D_{L}(\nu; \mu) + C_{D_{L}}(\nu; \mu, r).
\end{align*}
We establish the orthogonality properties of $\Psi$ by separately establishing the orthogonality properties of $M^{\perp}$ (Section \ref{sec:M_perp}) and $D_{L}^{\perp}$ (Section \ref{sec:D_L_perp}). We use the following lemma in the proof.

\begin{lemm}
\label{lemm:simple_deriv}
Let $G(\nu, \mu) := \EE[S]{g(X; \nu, \mu) \cdot (Y - \mu(X))}$. Then, 
\[\frac{\partial}{\partial \epsilon} G(\nu, \mu_{0} + \epsilon h) \Big|_{\epsilon=0} = -\EE[S]{g(X; \nu, \mu_{0}) \cdot h(X)}.\]
\end{lemm}

\begin{subsubsection}{Orthogonality of Second Estimating Equation}
\label{sec:M_perp}
We aim to show that the Gateaux derivatives of $M^{\perp}$ with respect to $\mu$ and $r$ in the direction $h$ vanish at $(\nu, \mu_{0}, r_{0})$ for any $\nu \in \mathbb{R}^{J + K}$  and any direction $h \in L^{2}(P_{X}, \mathcal{X})$. Note that
\begin{equation} 
\label{eq:M_function}
M(\nu, \mu) = \EE[P]{\gamma(X, 1)  \cdot \mu(X) \cdot w(X, 1; \nu, \mu) + \gamma(X, 0) \cdot (1 - \mu(X)) \cdot w(X, 0; \nu, \mu)} - \bar{\gamma}_{P}. 
\end{equation}

First, we compute the Gateaux derivative of $M^{\perp}$ with respect to $\mu$ at $(\mu_0, r_{0})$.
\begin{align*}
&\frac{\partial M^{\perp}(\nu, \mu_0 + \epsilon h, r_{0})}{\partial \epsilon} \Big|_{\epsilon=0} \\
&= \frac{\partial}{\partial \epsilon} M(\nu, \mu_0 + \epsilon h) \Big|_{\epsilon=0} + \frac{\partial C_{M}(\nu, \mu_{0} + \epsilon h, r_0)}{\partial \epsilon} \Big|_{\epsilon=0}   \\
&= \EE[P]{ (\gamma(X, 1) - \gamma(X, 0))  \cdot w(X, 1; \nu, \mu_0 + \epsilon h) \cdot w(X, 0; \nu, \mu_0 +\epsilon h) \cdot h(X) \Big|_{\epsilon=0} } \\
&\indent - \EE[S]{r_{0}(X) \cdot \delta_{M}(X; \nu, \mu_{0})\cdot h(X)} \\
&= \EE[P]{\delta_{M}(X; \nu, \mu_{0}) \cdot h(X)} - \EE[S]{r_{0}(X) \cdot \delta_{M}(X; \nu, \mu_{0})\cdot h(X)} \\
&= 0.
\end{align*}

The derivative of the first term follows from the expression for $M(\nu, \mu)$ in \eqref{eq:M_function} and 
\begin{equation}
\label{eq:derivative_measure}
\frac{\partial }{\partial \epsilon} \left( (\mu(x) + \epsilon h(x))  \cdot w(x, 1; \nu, \mu + \epsilon h)  \right) = w(x, 1; \nu, \mu + \epsilon h) \cdot w(x, 0; \nu, \mu + \epsilon h) \cdot h(x).
\end{equation} The derivative of the correction term follows from Lemma \ref{lemm:simple_deriv}. Thus, $M^{\perp}$ has the property that $\frac{\partial M^{\perp}(\nu, \mu_0 + \epsilon h, r_0)}{\partial \epsilon} \Big|_{\epsilon=0} = 0$ for all $\nu \in \mathbb{R}^{J+ K}$ and $h \in L^{2}(P_{X}, \mathcal{X})$, so it satisfies our desired orthogonality condition with respect to $\mu$.

Second, we compute the Gateaux derivative of $M^{\perp}$ with respect to $r$ at $(\mu_0, r_0)$. It suffices to examine the correction term
\begin{align*}
\frac{\partial}{\partial \epsilon} C_{M}(\nu, \mu_{0}, r_{0} + \epsilon h) \Big|_{\epsilon=0} &= \frac{\partial}{\partial \epsilon} \EE[S]{ (r_{0}(X) + \epsilon h(X))  \cdot \delta_{M}(X; \nu, \mu_{0}) \cdot (Y - \mu_{0}(X) )} \Big|_{\epsilon=0} \\
&= \EE[S]{\delta_{M}(X; \nu, \mu_{0}) \cdot (Y - \mu_{0}(X)) \cdot h(X)} \\
&= 0.
\end{align*}
The last equality holds because $\EE[S]{(Y - \mu_{0}(X)) \cdot g(X) \mid X} = \EE[S]{(Y - \EE[S]{Y \mid X}) \cdot g(X) \mid X} = 0.$
Thus, $M^{\perp}$ satisfies $\frac{\partial}{\partial \epsilon} M^{\perp}(\nu, \mu_0, r_0 + \epsilon h) \Big|_{\epsilon=0} = 0$ the desired orthogonality condition with respect to $r$.
\end{subsubsection}

\begin{subsubsection}{Orthogonality of First Estimating Equation}
\label{sec:D_L_perp}
We aim to show that the Gateaux derivatives of $D_{L}^{\perp}$ with respect to $\mu$ and $r$ in the direction $h$ vanish at $(\nu, \mu_{0}, r_{0})$ for any $\nu \in \mathbb{R}^{J + K}$  and any direction $h \in L^{2}(P_{X}, \mathcal{X})$.

We will use the notation that $m_{g}(X; \nu, \mu) := \EE[Q(\theta, \mu)]{g(X, Y) \mid X}.$

First, we examine the Gateaux derivative of $D_{L}^{\perp}$ with respect to $\mu$. We observe that by Lemma \ref{lemm:simple_deriv},
\begin{align*}
\frac{\partial C_{D_{L}}}{\partial \epsilon}(\nu, \mu_{0} + \epsilon h, r_0) \Big|_{\epsilon = 0} &= - \EE[S]{r_{0}(X) \cdot \delta_{D_{L}}(X; \nu, \mu_0) \cdot h(X)} \\
&= - \EE[P]{\delta_{D_{L}}(X; \nu, \mu_{0}) \cdot h(X)}.
\end{align*}

Second, we examine the Gateaux derivative of $D_{L}$. We can see that 
\begin{align*}
D_{L}(\nu; \mu) &= \EE[P]{\Delta_{\eta}(X, 1; \nu, \mu)^{\otimes 2} \cdot \theta \cdot \mu(X) \cdot w(X, 1; \nu, \mu)} \\
&\indent+ \EE[P]{\Delta_{\eta}(X, 0; \nu, \mu)^{\otimes 2} \cdot \theta \cdot  (1- \mu(X)) \cdot w(X, 0; \nu, \mu)} \\
&\indent+ \EE[P]{\Delta_{\eta}(X, 1; \nu, \mu) \otimes \Delta_{\gamma}(X, 1; \nu, \mu) \cdot \lambda \cdot \mu(X) \cdot w(X, 1; \nu; \mu)}\\
&\indent+ \EE[P]{\Delta_{\eta}(X, 0; \nu, \mu) \otimes \Delta_{\gamma}(X, 0; \nu, \mu) \cdot \lambda \cdot (1 - \mu(X)) \cdot w(X, 0; \nu; \mu)}.
\end{align*}

We name each of these terms $A_{1}, A_{0}, B_{1}, B_{0},$ respectively. We also define $A := A_{1} + A_{0}$ and $B := B_{1} + B_{0}.$

\begin{align*}
\frac{\partial A_{1}}{\partial \epsilon} (\nu, \mu_0 + \epsilon h) &= \EE[P]{\Delta_{\eta}(X, 1; \nu, \mu_0 + \epsilon h)^{\otimes 2} \cdot \theta \cdot \frac{\partial}{\partial \epsilon} \left( (\mu_0(X) + \epsilon h(X))  \cdot w(X, 1; \nu, \mu_0 + \epsilon h)  \right)} \\
&\indent+  \EE[P]{\Big( \frac{\partial}{\partial \epsilon} (\Delta_{\eta}(X, 1; \nu, \mu_0 + \epsilon h)^{\otimes 2}) \Big) \cdot \theta \cdot (\mu_0(X) + \epsilon h(X)) \cdot w(X, 1; \nu, \mu_0 + \epsilon h)} \\
&= \EE[P]{\Delta_{\eta}(X, 1; \nu, \mu_0 + \epsilon h)^{\otimes 2}  \cdot \theta \cdot w(X, 1; \nu, \mu_0 + \epsilon h) \cdot w(X, 0; \nu, \mu_0 + \epsilon h)  \cdot h(X) } \\
&\indent+  \EE[P]{ \Big( \frac{\partial}{\partial \epsilon} (\Delta_{\eta}(X, 1; \nu, \mu_0 + \epsilon h)^{\otimes 2}) \Big) \cdot \theta \cdot (\mu_0(X) + \epsilon h(X)) \cdot w(X, 1; \nu, \mu_0 + \epsilon h)}.
\end{align*}

The first equality follows from the product rule. The second equality is due to \eqref{eq:derivative_measure}. 

\begin{align*}
\frac{\partial A_{0}}{\partial \epsilon} ( \nu, \mu_0 + \epsilon h) &= \EE[P]{\Delta_{\eta}(X, 0; \nu, \mu_0 + \epsilon h)^{\otimes 2} \cdot \theta \cdot \frac{\partial}{\partial \epsilon} \left( (1 - \mu_0(X) - \epsilon h(X))  \cdot w(X, 0; \nu, \mu_0 + \epsilon h)  \right) } \\
&\indent+ \EE[P]{ \Big( \frac{\partial}{\partial \epsilon} (\Delta_{\eta}(X, 0; \nu, \mu_0 + \epsilon h)^{\otimes 2}) \Big) \cdot \theta \cdot (1 - \mu_0(X) - \epsilon h(X)) \cdot w(X, 0; \nu, \mu_0 + \epsilon h)} \\
&= -\EE[P]{\Delta_{\eta}(X, 0; \nu, \mu_0 + \epsilon h)^{\otimes 2} \cdot \theta \cdot w(X, 1; \nu, \mu_0 + \epsilon h) \cdot w(X, 0; \nu, \mu_0 + \epsilon h) \cdot h(X)  }\\
&\indent+ \EE[P]{\Big( \frac{\partial}{\partial \epsilon} (\Delta_{\eta}(X, 0; \nu, \mu_0 + \epsilon h)^{\otimes 2}) \Big) \cdot \theta \cdot (1 - \mu_0(X) - \epsilon h(X)) \cdot w(X, 0; \nu, \mu_0 + \epsilon h)}.
\end{align*}
We combine the derivatives of $A_{1}$ and $A_{0}$ to obtain the derivative of $A$.

\begin{align*}
&\frac{\partial A}{\partial \epsilon}(\nu, \mu_0 + \epsilon h) \\
&= \EE[P]{\left( \Delta_{\eta}(X, 1; \nu, \mu_0 + \epsilon h)^{\otimes 2} -\Delta_{\eta}(X, 0; \nu, \mu_0 + \epsilon h)^{\otimes 2} \right) \cdot w(X, 1; \nu, \mu_0 + \epsilon h) \cdot w(X, 0; \nu, \mu_0 + \epsilon h) \cdot \theta \cdot h(X)} \\
&\indent+ \EE[P]{\EE[Q(\theta, \mu_0 + \epsilon h)]{  \frac{\partial}{\partial \epsilon} (\Delta_{\eta}(X, Y; \nu, \mu_0 + \epsilon h)^{\otimes 2})  \mid X} \cdot \theta } \\
&= \EE[P]{\left(  \Delta_{\eta}(X, 1; \nu, \mu_0)^{\otimes 2} -\Delta_{\eta}(X, 0; \nu, \mu_0)^{\otimes 2} \right) \cdot w(X, 1; \nu, \mu_0) \cdot w(X, 0; \nu, \mu_0)  \cdot \theta \cdot h(X)}.
\end{align*}

The second term is equal to zero because 
\begin{align*}
&\EE[Q(\theta, \mu_0 + \epsilon h)]{ \frac{\partial}{\partial \epsilon} (\Delta_{\eta}(X, Y; \nu, \mu_0 + \epsilon h)^{\otimes 2})  \mid X} \\
&= \EE[Q(\theta, \mu_0 + \epsilon h)]{\Delta_{\eta}(X, Y; \nu, \mu_0 + \epsilon h) \otimes (-\frac{\partial}{\partial \epsilon} m_{\eta}(X; \nu, \mu_{0} + \epsilon h)) \mid X} \\
&\indent + \EE[Q(\theta, \mu_0 + \epsilon h)]{(-\frac{\partial}{\partial \epsilon} m_{\eta}(X; \nu, \mu_{0} + \epsilon h)) \otimes \Delta_{\eta}(X, Y; \nu, \mu_0 + \epsilon h) \mid X } \\
&= 0.
\end{align*}
The last line follows because $\EE[Q(\theta, \mu)]{\Delta_{\eta}(X, Y; \nu, \mu) \mid X} = 0.$

Now, we can compute the derivatives of $B_{1}, B_{0}$ as follows.
\begin{align*}
&\frac{\partial B_{1}}{\partial \epsilon} (\nu, \mu_0 + \epsilon h) \\
&= \EE[P]{\Delta_{\eta}(X, 1; \nu, \mu_0 + \epsilon h) \otimes \Delta_{\gamma}(X, 1; \nu, \mu_0 + \epsilon h) \cdot \lambda \cdot \frac{\partial}{\partial \epsilon} \left( (\mu_0(X) + \epsilon h(X))  \cdot w(X, 1; \nu, \mu_0 + \epsilon h)  \right)} \\
&\indent+ \EE[P]{\frac{\partial}{\partial \epsilon} \Big( \Delta_{\eta}(X, 1; \nu, \mu_0 + \epsilon h) \otimes \Delta_{\gamma}(X, 1; \nu, \mu_0 + \epsilon h) \Big)  \cdot \lambda  \cdot (\mu_0(X) + \epsilon h(X)) \cdot w(X, 1; \nu, \mu_0 + \epsilon h)} \\
&= \EE[P]{\Delta_{\eta}(X, 1; \nu, \mu_0 + \epsilon h) \otimes \Delta_{\gamma}(X, 1; \nu, \mu_0 + \epsilon h) \cdot w(X, 1; \nu, \mu_0 + \epsilon h) \cdot w(X, 0; \nu, \mu_0 + \epsilon h) \cdot h(X)} \\
&\indent+ \EE[P]{\frac{\partial}{\partial \epsilon} \Big( \Delta_{\eta}(X, 1; \nu, \mu_0 + \epsilon h) \otimes \Delta_{\gamma}(X, 1; \nu, \mu_0 + \epsilon h) \Big)  \cdot \lambda  \cdot (\mu_0(X) + \epsilon h(X)) \cdot w(X, 1; \nu, \mu_0 + \epsilon h)}.
\end{align*}

\begin{align*}
&\frac{\partial B_{0}}{\partial \epsilon} (\nu, \mu_0 + \epsilon h) \\
&= \EE[P]{\Delta_{\eta}(X, 0; \nu, \mu_0 + \epsilon h) \cdot \Delta_{\gamma}(X, 0; \nu, \mu_0 + \epsilon h) \cdot \lambda \cdot \frac{\partial}{\partial \epsilon} \left( (1 - \mu_0(X) - \epsilon h(X))  \cdot w(X, 0; \nu, \mu_0 + \epsilon h)  \right)} \\
&\indent+ \EE[P]{\frac{\partial}{\partial \epsilon} \Big( \Delta_{\eta}(X, 0; \nu, \mu_0 + \epsilon h) \cdot \Delta_{\gamma}(X, 0; \nu, \mu_0 + \epsilon h)  \Big) \cdot \lambda \cdot (1 - \mu_0(X) - \epsilon h(X)) \cdot w(X, 0; \nu, \mu_0 + \epsilon h)} \\
&= -\EE[P]{\Delta_{\eta}(X, 0; \nu, \mu_0 + \epsilon h) \cdot \Delta_{\gamma}(X, 0; \nu, \mu_0 + \epsilon h) \cdot w(X, 1; \nu, \mu_0 + \epsilon h) \cdot w(X, 0; \nu, \mu_0 + \epsilon h) \cdot \lambda \cdot h(X)} \\
&\indent+ \EE[P]{\frac{\partial}{\partial \epsilon} \Big( \Delta_{\eta}(X, 0; \nu, \mu_0 + \epsilon h) \cdot \Delta_{\gamma}(X, 0; \nu, \mu_0 + \epsilon h)  \Big) \cdot \lambda \cdot (1 - \mu_0(X) - \epsilon h(X)) \cdot w(X, 0; \nu, \mu_0 + \epsilon h)}.
\end{align*}

Define
\begin{equation}
\label{eq:D}
d(X; \nu, \mu) = \Delta_{\eta}(X, 1; \nu, \mu) \otimes \Delta_{\gamma}(X, 1; \nu, \mu) - \Delta_{\eta}(X, 0; \nu, \mu) \otimes \Delta_{\gamma}(X, 0; \nu, \mu).
\end{equation}

Thus, we can use the definition of $d$ from \eqref{eq:D} to see that
\begin{align*}
&\frac{\partial B}{\partial \epsilon} (\nu, \mu_0 + \epsilon h) \\
&= \EE[P]{ d(X; \nu, \mu_0 + \epsilon h) \cdot \lambda \cdot w(X, 1; \nu, \mu_0 + \epsilon h) \cdot w(X, 0; \nu, \mu_0 + \epsilon h) \cdot h(X)} \\
&\indent+ \EE[P]{\EE[Q(\theta, \mu_0 + \epsilon h)]{\Delta_{\eta}(X, Y; \nu, \mu_0 + \epsilon h)  \otimes (- \frac{\partial }{\partial \epsilon} m_{\gamma}(X; \nu, \mu_0 + \epsilon h)) \lambda \mid X}} \\
&\indent+ \EE[P]{\EE[Q(\theta_0, \mu_0 + \epsilon h)]{(- \frac{\partial }{\partial \epsilon} m_{\eta}(X; \nu, \mu_0 + \epsilon h)) \otimes \Delta_{\gamma}(X, Y; \nu, \mu_0 + \epsilon h) \lambda \mid X}} \\
&= \EE[P]{ d(X; \nu, \mu_0 + \epsilon h) \cdot \lambda \cdot w(X, 1; \nu, \mu_0 + \epsilon h) \cdot w(X, 0; \nu, \mu_0 + \epsilon h) \cdot h(X)}.
\end{align*}

The second equality holds because $\EE[Q(\theta, \mu)]{\Delta_{\eta}(X, Y; \nu, \mu) \mid X} = 0$ and $\EE[Q(\theta, \mu)]{\Delta_{\gamma}(X, Y; \nu, \mu) \mid X} = 0.$

Finally, we have that
\begin{align*}
&\frac{\partial D_{L}}{\partial \epsilon}(\nu, \mu_0 + \epsilon h) \Big|_{\epsilon=0} \\
&= \EE[P]{\left( \Delta_{\eta}(X, 1; \nu, \mu_0)^{\otimes 2} -\Delta_{\eta}(X, 0; \nu, \mu_0 )^{\otimes 2} \right) \theta \cdot w(X, 1; \nu, \mu_0) \cdot w(X, 0; \nu, \mu_0) \cdot h(X)} \\
&\indent + \EE[P]{d(X; \nu, \mu_0) \cdot w(X, 1; \nu, \mu_0) \cdot w(X, 0; \nu, \mu_0) \cdot \lambda  \cdot h(X)} \\
&= \EE[P]{\delta_{D_{L}}(X; \nu, \mu_0) \cdot h(X)}
\end{align*}

We observe that
\[ \frac{\partial D^{\perp}_{L}}{\partial \epsilon}(\nu, \mu_0 + \epsilon h, r_0) \Big|_{\epsilon=0} =  \frac{\partial D_{L}}{\partial \epsilon}(\nu, \mu_0 + \epsilon h) + \frac{\partial C_{D_{L}}}{\partial \epsilon}(\nu, \mu_0 + \epsilon h, r_0) \Big|_{\epsilon=0}= 0. \]
Thus, the Gateaux derivative of $D_{L}^{\perp}$ with respect to $\mu$ in the direction $h$ vanishes for all $h \in L^{2}(P_{X}, \mathcal{X})$ and $\nu \in \mathbb{R}^{J + K}.$

We examine the Gateaux derivative of $D_{L}^{\perp}$ with respect to $r$ in the direction $h$. It suffices to examine the correction term.
\begin{align*}
\frac{\partial}{\partial \epsilon} C_{D_{L}}(\nu, \mu_{0}, r_{0} + \epsilon h) \Big|_{\epsilon= 0} &= \EE[S]{\delta_{D_{L}}(X; \nu, \mu_0) \cdot (Y - \mu_{0}(X)) \cdot h(X)} \\
&= 0.
\end{align*}

Thus, we also conclude that the Gateaux derivative of $D_{L}^{\perp}$ with respect to $r$ in the direction $h$ vanishes for all $h$ in $L^{2}(P_{X}, \mathcal{X})$ and $\nu \in \mathbb{R}^{J + K}.$
\end{subsubsection}
\end{subsection}

\begin{subsection}{Proof of Theorem \ref{theo:one_step}}
We use the following lemma to simplify the one-step estimator.

\begin{lemm}
\label{lemm:asymptotic_equicontinuity}
Under Assumptions \ref{assumption:exp_tilting}, \ref{assumption:cov_shift}, \ref{assumption:existence_minimizer}, \ref{assumption:bounded}, \ref{assumption:nuisance}, we have that
\begin{align}
\Psi(\tilde{\nu}_{n}, \hat{\mu}_{n}^{\mathcal{I}_{2}}, \hat{r}_{n}^{\mathcal{I}_{2}}) =  \dot{\Psi}_{\nu}(\tilde{\nu}_{n}, \hat{\mu}_{n}^{\mathcal{I}_{2}}, \hat{r}_{n}^{\mathcal{I}_{2}}) \cdot (\tilde{\nu}_{n} - \nu_0) + o_{P}(n^{-1/2}). \label{eq:taylor_pop}
\end{align}
and 
\begin{align}
\Psi_{n}^{\mathcal{I}_{1}}(\tilde{\nu}_{n}, \hat{\mu}_{n}^{\mathcal{I}_{2}}, \hat{r}_{n}^{\mathcal{I}_{2}}) &=  \Psi(\tilde{\nu}_{n}, \hat{\mu}_{n}^{\mathcal{I}_{2}}, \hat{r}_{n}^{\mathcal{I}_{2}}) + \Psi_{n}^{\mathcal{I}_{1}}(\nu_{0}, \mu_{0}, r_{0}) - \Psi(\nu_{0}, \mu_{0}, r_{0}) + o_{P}(n^{-1/2}) \label{eq:asym1}\\
\dot{\Psi}_{\nu, n}^{\mathcal{I}_{1}}(\tilde{\nu}_{n}, \hat{\mu}_{n}^{\mathcal{I}_{2}}, \hat{r}_{n}^{\mathcal{I}_{2}}) &=  \dot{\Psi}_{\nu}(\tilde{\nu}_{n}, \hat{\mu}_{n}^{\mathcal{I}_{2}}, \hat{r}_{n}^{\mathcal{I}_{2}}) + \dot{\Psi}_{\nu, n}^{\mathcal{I}_{1}}(\nu_{0}, \mu_{0}, r_{0}) - \dot{\Psi}_{\nu}(\nu_{0}, \mu_{0}, r_{0}) + o_{P}(n^{-1/2}) \label{eq:asym2}.
\end{align}
\end{lemm}

We can apply \eqref{eq:asym1}, \eqref{eq:asym2} to \eqref{eq:one_step} to see that
\begin{align*}
&(\dot{\Psi}_{\nu}(\tilde{\nu}_{n}, \hat{\mu}_{n}^{\mathcal{I}_{2}}, \hat{r}_{n}^{\mathcal{I}_{2}}) + \dot{\Psi}_{\nu, n}^{\mathcal{I}_{1}}(\nu_0, \mu_{0}, r_{0}) - \dot{\Psi}_{\nu}(\nu_{0}, \mu_{0}, r_{0}) + o_{P}(n^{-1/2})) \cdot (\hat{\nu}_{n}^{\mathcal{I}_{1}} - \tilde{\nu}_{n})\\
&\indent+  (\Psi(\tilde{\nu}_{n}, \hat{\mu}_{n}^{\mathcal{I}_{2}}, \hat{r}_{n}^{\mathcal{I}_{2}}) + \Psi_{n}^{\mathcal{I}_{1}}(\nu_{0}, \mu_{0}, r_{0}) - \Psi(\nu_{0}, \mu_{0}, r_{0}) + o_{P}(n^{-1/2})) = 0.
\end{align*}
We observe that $\Psi(\tilde{\nu}_{n}, \hat{\mu}_{n}^{\mathcal{I}_{2}}, \hat{r}_{n}^{\mathcal{I}_{2}}) \xrightarrow{p} \Psi(\nu_{0}, \mu_{0}, r_{0})$ and $\dot{\Psi}_{\nu}(\tilde{\nu}_{n}, \hat{\mu}_{n}^{\mathcal{I}_{2}}, \hat{r}_{n}^{\mathcal{I}_{2}}) \xrightarrow{p} \dot{\Psi}_{\nu}(\nu_{0}, \mu_{0}, r_{0})$ by the Continuous Mapping Theorem. Since $\Psi(\nu_{0}, \mu_{0}, r_{0}) = 0$, then we can see $\hat{\nu}_{n}^{\mathcal{I}_{1}} \xrightarrow{p} \nu_{0}.$

Rearranging the above equation and noting that $\Psi(\nu_0, \mu_{0}, r_{0}) = 0$, we find that
\begin{align*}
&\dot{\Psi}_{\nu}(\tilde{\nu}_{n}, \hat{\mu}_{n}^{\mathcal{I}_{2}}, \hat{r}_{n}^{\mathcal{I}_{2}}) \cdot (\hat{\nu}_{n}^{\mathcal{I}_{1}} - \tilde{\nu}_{n})+  \Psi(\tilde{\nu}_{n}, \hat{\mu}_{n}^{\mathcal{I}_{2}}, \hat{r}_{n}^{\mathcal{I}_{2}}) + \Psi_{n}^{\mathcal{I}_{1}}(\nu_{0}, \mu_{0}, r_{0}) \\
&= - (\dot{\Psi}_{\nu, n}^{\mathcal{I}_{1}}(\nu_0, \mu_{0}, r_{0}) - \dot{\Psi}_{\nu}(\nu_{0}, \mu_{0}, r_{0})) \cdot (\hat{\nu}^{\mathcal{I}_{1}}_{n} - \tilde{\nu}_{n}) +  o_{P}(n^{-1/2}).
\end{align*}

We can plug \eqref{eq:taylor_pop} into the above equation to get that 
\begin{align*}
&\dot{\Psi}_{\nu}(\tilde{\nu}_{n}, \hat{\mu}_{n}^{\mathcal{I}_{2}}, \hat{r}_{n}^{\mathcal{I}_{2}}) \cdot (\hat{\nu}^{\mathcal{I}_{1}}_{n} - \nu_{0}) + \Psi_{n}^{\mathcal{I}_{1}}(\nu_{0}, \mu_{0}, r_{0}) \\
&= - (\dot{\Psi}_{\nu, n}^{\mathcal{I}_{1}}(\nu_{0}, \mu_{0}, r_{0}) - \dot{\Psi}_{\nu}(\nu_{0}, \mu_{0}, r_{0}))  \cdot (\hat{\nu}^{\mathcal{I}_{1}}_{n} - \tilde{\nu}_{n}) + o_{P}(n^{-1/2}).
\end{align*}
We note that $(\dot{\Psi}_{n}(\nu_{0}) - \dot{\Psi}(\nu_{0}))  \cdot (\hat{\nu}^{\mathcal{I}_{1}}_{n} - \tilde{\nu}_{n}) = o_{P}(n^{-1/2})$. This follows from the fact that $\hat{\nu}^{\mathcal{I}_{1}}_{n}$ and $\tilde{\nu}_{n}$ both estimate $\nu_{0}$ consistently, so $\hat{\nu}^{\mathcal{I}_{1}}_{n} - \tilde{\nu}_{n} = o_{P}(1)$, and also $\dot{\Psi}_{\nu, n}(\nu_{0}) - \dot{\Psi}_{\nu}(\nu_{0}) = O_{p}(n^{-1/2})$ by the Central Limit Theorem. Thus, the above equation can be simplified as
\begin{equation}
\label{eq:4}
\dot{\Psi}_{\nu}(\tilde{\nu}_{n}, \hat{\mu}_{n}^{\mathcal{I}_{2}}, \hat{r}_{n}^{\mathcal{I}_{2}}) \cdot (\hat{\nu}_{n}^{\mathcal{I}_{1}} - \nu_{0}) + \Psi_{n}^{\mathcal{I}_{1}}(\nu_{0}, \mu_{0}, r_{0}) =  o_{P}(n^{-1/2}).\end{equation}
Finally, we can conclude that
\begin{equation}
\label{eq:5}
(\hat{\nu}^{\mathcal{I}_{1}}_{n} - \nu_0) = - (\dot{\Psi}(\tilde{\nu}_{n}, \hat{\mu}_{n}^{\mathcal{I}_{2}}, \hat{r}_{n}^{\mathcal{I}_{2}}))^{-1} \Psi_{n}^{\mathcal{I}_{1}}(\nu_{0}, \mu_{0}, r_{0}) + o_{P}(n^{-1/2}).
\end{equation}
An identical proof applies to show that
\begin{equation}
\label{eq:6}
(\hat{\nu}^{\mathcal{I}_{2}}_{n} - \nu_{0}) = - (\dot{\Psi}(\tilde{\nu}_{n}, \hat{\mu}_{n}^{\mathcal{I}_{1}}, \hat{r}_{n}^{\mathcal{I}_{1}}))^{-1} \Psi_{n}^{\mathcal{I}_{2}}(\nu_{0}, \mu_{0}, r_{0}) + o_{P}(n^{-1/2}).
\end{equation}
Note that \eqref{eq:5} and \eqref{eq:6} imply that
\begin{align*}
&\sqrt{n}(\hat{\nu}_{n} - \nu_{0}) \\
&= \sqrt{n}\left( \frac{|\mathcal{I}_{1}|}{n} (\hat{\nu}^{\mathcal{I}_{1}}_{n} - \nu_0) + \frac{|\mathcal{I}_{2}|}{n} (\hat{\nu}_{n}^{\mathcal{I}_{2}} - \nu_0) \right) \\
&= \sqrt{n} \left(- \frac{|\mathcal{I}_{1}|}{n} \cdot (\dot{\Psi}(\tilde{\nu}_{n}, \hat{\mu}_{n}^{\mathcal{I}_{2}}, \hat{r}_{n}^{\mathcal{I}_{2}}))^{-1} \Psi_{n}^{\mathcal{I}_{1}}(\nu_{0}, \mu_{0}, r_{0})  - \frac{|\mathcal{I}_{2}|}{n} \cdot (\dot{\Psi}(\tilde{\nu}_{n}, \hat{\mu}_{n}^{\mathcal{I}_{1}}, \hat{r}_{n}^{\mathcal{I}_{1}}))^{-1} \Psi_{n}^{\mathcal{I}_{2}}(\nu_{0}, \mu_{0}, r_{0})\right) + o_{P}(1)\\
&= \sqrt{n} \left( -\dot{\Psi}(\tilde{\nu}_{n}, \hat{\mu}_{n}^{\mathcal{I}_{2}}, \hat{r}_{n}^{\mathcal{I}_{2}}))^{-1} \cdot \frac{1}{n} \sum_{i \in \mathcal{I}_{1}} \psi(X_i, Y_i; \nu_0, \mu_0, r_0)  - \dot{\Psi}(\tilde{\nu}_{n}, \hat{\mu}_{n}^{\mathcal{I}_{1}}, \hat{r}_{n}^{\mathcal{I}_{1}}))^{-1} \cdot \frac{1}{n} \sum_{i \in \mathcal{I}_{2}} \psi(X_i, Y_i; \nu_0, \mu_0, r_0) \right) + o_{P}(1) \\
&= \sqrt{n} \left( -\dot{\Psi}(\nu_0, \mu_0, r_0)^{-1} \cdot \frac{1}{n} \sum_{i=1}^{n} \psi(X_i, Y_i; \nu_0, \mu_0, r_0)  \right) + o_{P}(1)
\end{align*}
The last line above follows from consistency of $\dot{\Psi}(\tilde{\nu}_{n}, \hat{\mu}_{n}^{\mathcal{I}_{2}}, \hat{r}_{n}^{\mathcal{I}_{2}}) \xrightarrow{p} \dot{\Psi}(\nu_0, \mu_0, r_0)$ and $\dot{\Psi}(\tilde{\nu}_{n}, \hat{\mu}_{n}^{\mathcal{I}_{1}}, \hat{r}_{n}^{\mathcal{I}_{1}}) \xrightarrow{p} \dot{\Psi}(\nu_0, \mu_0, r_0)$, $\dot{\Psi}(\nu_0, \mu_0, r_0)$ is full rank under Assumption \ref{assumption:existence_minimizer}, and $\frac{1}{n} \sum_{i \in \mathcal{I}_{k}} \psi(X_i, Y_i; \nu_0, \mu_0, r_0) = O_{p}(n^{-1/2}).$ Thus, we can apply the Central Limit Theorem to see that  $\sqrt{n}(\hat{\nu}_{n} - \nu_{0}) \xrightarrow{d} N(0, V)$, where $V$ is defined in \eqref{eq:variance}.
\end{subsection}

\begin{subsection}{Proof of Lemma \ref{lemm:variance}}
We can show that $V_{n}^{\mathcal{I}_{1}}$ is a consistent estimator of the asymptotic variance $V$. Recall that by Theorem \ref{theo:one_step}, we have that $\hat{\nu}_{n} \rightarrow \nu_{0}$. 

We show that $\dot{\Psi}^{\mathcal{I}_{1}}_{\nu, n}(\hat{\nu}_{n}, \hat{\mu}_{n}^{\mathcal{I}_{2}}, \hat{r}_{n}^{\mathcal{I}_{2}}) \xrightarrow{p} \dot{\Psi}_{\nu, 0}$. We use the following decomposition.
\begin{align*}
\dot{\Psi}_{\nu, n}^{\mathcal{I}_{1}}(\hat{\nu}_{n}, \hat{\mu}_{n}^{\mathcal{I}_{2}}, \hat{r}_{n}^{\mathcal{I}_{2}}) - \dot{\Psi}_{\nu, 0} = (\dot{\Psi}_{\nu, n}^{\mathcal{I}_{1}}(\hat{\nu}_{n}, \hat{\mu}_{n}^{\mathcal{I}_{2}}, \hat{r}_{n}^{\mathcal{I}_{2}}) - \dot{\Psi}_{\nu, n}^{\mathcal{I}_{1}}(\nu_0, \hat{\mu}_{n}^{\mathcal{I}_{2}}, \hat{r}_{n}^{\mathcal{I}_{2}})) + (\dot{\Psi}_{\nu, n}^{\mathcal{I}_{1}}(\nu_0, \hat{\mu}_{n}^{\mathcal{I}_{2}}, \hat{r}_{n}^{\mathcal{I}_{2}}) -  \dot{\Psi}_{\nu, 0}).
\end{align*}
Let $A_{n}, B_{n}$ correspond to the first and second terms above. We aim to show that $A_{n} = o_{P}(1)$ and $B_{n} = o_{P}(1)$. We note that for fixed $(X, Y, \mu, r)$, $\dot{\psi}_{\nu}$ is Lipschitz continuous in $\nu$ in a neighborhood $B_{r}(\nu_0)$ about $\nu_0$, i.e.
\[  ||\dot{\psi}_{\nu}(X, Y; \nu, \mu, r) - \dot{\psi}_{\nu}(X, Y; \nu', \mu, r)|| \leq L(X, Y; \mu, r) \cdot ||\nu - \nu'|| \quad \forall \nu, \nu' \in B_{r}(\nu_0),\]
because $\dot{\Psi}$ is continuous in $\nu$ and $B_{r}(\nu_{0})$ is bounded. Then for $\nu \in B_{r}(\nu_0)$, we have that
\begin{align*}
||\dot{\Psi}_{\nu, n}^{\mathcal{I}_{1}}(\nu, \hat{\mu}_{n}^{\mathcal{I}_{2}}, \hat{r}_{n}^{\mathcal{I}_{2}}) - \dot{\Psi}_{\nu, n}^{\mathcal{I}_{1}}(\nu_0, \hat{\mu}_{n}^{\mathcal{I}_{2}}, \hat{r}_{n}^{\mathcal{I}_{2}})|| &= ||P_{n}^{\mathcal{I}_{1}} \left(\psi(\nu, \hat{\mu}_{n}^{\mathcal{I}_{2}}, \hat{r}_{n}^{\mathcal{I}_{2}}) - \psi(\nu_0, \hat{\mu}_{n}^{\mathcal{I}_{2}}, \hat{r}_{n}^{\mathcal{I}_{2}}) \right)|| \\
&\leq P_{n}^{\mathcal{I}_{1}} || \left(\psi(\nu, \hat{\mu}_{n}^{\mathcal{I}_{2}}, \hat{r}_{n}^{\mathcal{I}_{2}}) - \psi(\nu_0, \hat{\mu}_{n}^{\mathcal{I}_{2}}, \hat{r}_{n}^{\mathcal{I}_{2}}) \right)|| \\
&\leq P_{n}^{\mathcal{I}_{1}} L(X_i, Y_i; \hat{\mu}^{\mathcal{I}_{2}}_{n}, \hat{r}^{\mathcal{I}_{2}}_{n}) \cdot ||\nu - \nu_0|| \\
&\leq C \cdot ||\nu - \nu_{0}||.
\end{align*}
Since $\hat{\nu}_{n} \xrightarrow{p} \nu_0$, we have that $\PP[]{\hat{\nu}_{n} \in B_{r}(\nu_0)} \rightarrow 1$. Thus, $\PP[]{||\dot{\Psi}_{\nu, n}^{\mathcal{I}_{1}}(\nu, \hat{\mu}_{n}^{\mathcal{I}_{2}}, \hat{r}_{n}^{\mathcal{I}_{2}}) - \dot{\Psi}_{\nu, n}^{\mathcal{I}_{1}}(\nu_0, \hat{\mu}_{n}^{\mathcal{I}_{2}}, \hat{r}_{n}^{\mathcal{I}_{2}})|| > \epsilon} \rightarrow 0.$ 

Second, we analyze $B_{n}$. 
\begin{align*}
B_{n} &= (\dot{\Psi}_{\nu, n}^{\mathcal{I}_{1}}(\nu_0, \hat{\mu}_{n}^{\mathcal{I}_{2}}, \hat{r}_{n}^{\mathcal{I}_{2}}) - \dot{\Psi}_{\nu}(\nu_{0}, \hat{\mu}_{n}^{\mathcal{I}_{2}}, \hat{r}_{n}^{\mathcal{I}_{2}})) + (\dot{\Psi}_{\nu}(\nu_{0}, \hat{\mu}_{n}^{\mathcal{I}_{2}}, \hat{r}_{n}^{\mathcal{I}_{2}}) - \dot{\Psi}_{\nu, 0}).
\end{align*}
When we condition on $\mathcal{I}_{2}$, $\hat{\mu}^{\mathcal{I}_{2}}_{n}, \hat{r}^{\mathcal{I}_{2}}_{n}$ can be treated as deterministic functions, so the Law of Large Numbers applies to the first term. The second term is $o_{P}(1)$ due to the Continuous Mapping Theorem. 

We can apply the identical argument to show that $\Sigma^{\mathcal{I}_{1}}_{n}(\hat{\nu}_{n}, \hat{\mu}_{n}^{\mathcal{I}_{2}}, \hat{r}_{n}^{\mathcal{I}_{2}}) \xrightarrow{p} \Sigma_{0}$. Finally, we can use Slutsky's theorem to see that $V_{n}^{\mathcal{I}_{1}}$ must be a consistent estimator for $V.$ The same argument applies to show that $V_{n}^{\mathcal{I}_{2}}$ is a consistent estimator for $V$. Thus, $V_{n}$ must be a consistent estimator for $V$.

\end{subsection}

\begin{subsection}{Proof of Lemma \ref{lemm:delta}}
Using the delta method, we can see that
\begin{align*}
\sqrt{n}(\mu_{g}(\hat{\nu}_{n}) - \mu_{g}(\nu_0)) \xrightarrow{d} N(0, \nabla \mu_{g}(\nu_0)^{T} V \nabla \mu_{g}(\nu_0)). 
\end{align*}
We observe that by the Continuous Mapping Theorem $\nabla \mu_{g}(\hat{\nu}_{n}) \xrightarrow{p} \nabla \mu_{g}(\nu_0).$ Furthermore, from Lemma \ref{lemm:variance}, $V_{n} \xrightarrow{p} V.$ Using Slutsky's theorem, we can see that $\nabla \mu_{g}(\hat{\nu}_{n})^{T} V_{n} \nabla \mu_{g}(\hat{\nu}_{n}) \xrightarrow{p} \nabla \mu_{g}(\nu_0)^{T} V \nabla \mu_{g}(\nu_0)$, so this term is a consistent estimator for the asymptotic variance of $\mu_{g}(\hat{\nu}_{n})$. Thus, we can derive asymptotically-valid Gaussian confidence intervals for $\mu_{g}(\hat{\nu}_{n})$ as in \eqref{eq:confidence_interval}.
\end{subsection}

\begin{subsection}{Proof of Lemma \ref{lemm:params}}
We recall that by Lemma \ref{lemm:exp_family}, $Q(\theta)$ must have conditional distribution of the form \eqref{eq:exp_family}. From Theorem 2.4 of \citet{keener2010theoretical}, since \eqref{eq:exp_family} is $J$-parameter exponential family, we have that
\begin{align*}
\nabla A_{X}(\theta) &= \EE[Q_{Y|X}(\theta)]{\eta(X, Y) \mid X} \\
\nabla^{2} A_{X}(\theta) &= \Cov[Q_{Y|X}(\theta)]{\eta(X, Y) \mid X} \\
\nabla^{3} A_{X}(\theta) &= \EE[Q_{Y|X}(\theta)]{(\eta(X, Y) - \EE[Q_{Y|X}(\theta)]{\eta(X, Y)|X})^{\otimes 3} |X}.
\end{align*} 

First, we note that
\begin{align*}
I(\theta) &= \nabla_{\theta} \EE[Q(\theta)]{\eta(X, Y)} \\
&= \nabla_{\theta} \EE[P_{X}]{\EE[Q_{Y|X}(\theta)]{\eta(X, Y) \mid X}} \\
&= \EE[P_{X}]{\nabla_{\theta} \EE[Q_{Y|X}(\theta)]{\eta(X, Y) \mid X} } \\
&= \EE[P_{X}]{\nabla_{\theta}^{2} A_{X}(\theta)} \\
&= \EE[P_{X}]{ \Cov[Q_{Y \mid X}(\theta)]{\eta(X, Y) \mid X}}.
\end{align*}

Second, we note that 
\begin{align*}
B(\theta) &= \nabla_{\theta} \EE[Q(\theta)]{\gamma(X, Y)} \\
&= \EE[P_{X}]{\nabla_{\theta} \EE[Q_{Y|X}(\theta)]{\gamma(X, Y) \mid X}} \\
&= \EE[P_{X}]{\EE[S_{Y|X}]{\nabla_{\theta} (\gamma(X, Y) \cdot \exp(\theta^{T}\eta(X, Y) - A_{X}(\theta)))\mid X } } \\
&= \EE[P_{X}]{\EE[S_{Y|X}]{(\gamma(X, Y)\otimes \eta(X, Y) - \gamma(X, Y) \cdot \nabla_{\theta} A_{X}(\theta)) \cdot \exp(\theta^{T} \eta(X, Y) - A_{X}(\theta)) \mid X}} \\
&= \EE[P_{X}]{\EE[Q_{Y|X}(\theta)]{\gamma(X, Y) \otimes \eta(X, Y) \mid X} - \EE[Q_{Y|X}(\theta)]{\gamma(X, Y)  \mid X} \otimes \EE[Q_{Y|X}(\theta)]{\eta(X, Y) \mid X}} \\
&= \EE[P_{X}]{\Cov[Q_{Y|X}(\theta)]{\gamma(X, Y), \eta(X, Y) \mid X}}.
\end{align*}

Then, we can observe that
\begin{align*}
g(\theta, \lambda) &= \nabla_{\theta} L(\theta, \lambda)\\
&= \theta^{T} I(\theta) + \lambda^{T}B(\theta).
\end{align*}

Finally, we compute $H(\theta, \lambda)$. First, we derive
\begin{align*}
&\EE[P_{X}]{\nabla_{\theta} \EE[Q_{Y|X}(\theta)]{\gamma(X, Y) \otimes \eta(X, Y) \mid X}} \\
&= \EE[P_{X}]{\EE[Q_{Y|X}(\theta)]{\gamma(X, Y) \otimes \eta(X, Y) \otimes (\eta(X, Y) - \EE[Q_{Y|X}(\theta)]{\eta(X, Y) |X}) \mid X}}.
\end{align*}

Second, we derive
\begin{align*}
&\EE[P_{X}]{\nabla_{\theta}  \EE[Q_{Y|X}(\theta)]{\gamma(X, Y) \mid X} \odot \EE[Q_{Y|X}(\theta)]{\eta(X, Y) \mid X} } \\ 
&= \EE[P_{X}]{ \EE[Q_{Y|X}(\theta)]{\gamma(X, Y) \mid X} \otimes \Cov[Q_{Y|X}(\theta)]{\eta(X, Y) \mid X}} \\
&\indent+ \EE[P_{X}]{\Cov[Q_{Y|X}(\theta)]{\gamma(X, Y), \eta(X, Y) \mid X} \otimes \EE[Q_{Y|X}(\theta)]{\eta(X, Y) \mid X}}.
\end{align*}

Combining these results, we have that
\begin{align*}
&\nabla_{\theta} B(\theta) \\
&= \nabla_{\theta} \EE[P_{X}]{\Cov[Q_{Y|X}(\theta)]{\gamma(X, Y), \eta(X, Y) \mid X}} \\
&= \EE[P_{X}]{\nabla_{\theta} \Bigg(\EE[Q_{Y|X}(\theta)]{\gamma(X, Y) \otimes \eta(X, Y) \mid X} - \EE[Q_{Y|X}(\theta)]{\gamma(X, Y) \mid X} \odot \EE[Q_{Y|X}(\theta)]{\eta(X, Y) \mid X}\Bigg) } \\
&= \EE[P_{X}]{\EE[Q_{Y|X}(\theta)]{\gamma(X, Y)\otimes \eta(X, Y) \otimes (\eta(X, Y) - \EE[Q_{Y|X}(\theta)]{\eta(X, Y) |X}) \mid X}} \\
&\indent-  \EE[P_{X}]{ \EE[Q_{Y|X}(\theta)]{\gamma(X, Y) \mid X} \otimes \Cov[Q_{Y|X}(\theta)]{\eta(X, Y) \mid X}} \\
&\indent-\EE[P_{X}]{\Cov[Q_{Y|X}(\theta)]{\gamma(X, Y), \eta(X, Y) \mid X} \otimes \EE[Q_{Y|X}(\theta)]{\eta(X, Y) \mid X}}.
\end{align*}

\begin{align*}
H(\theta, \lambda) &= \nabla_{\theta}^{2} L(\theta, \lambda) \\
&= \nabla_{\theta} \theta^{T}I(\theta)  + \lambda^{T} B(\theta) \\
&=I(\theta) + \theta^{T} (\nabla_{\theta} I(\theta)) + \lambda^{T} \nabla_{\theta} B(\theta) \\
&= I(\theta) + \theta^{T} (\EE[P_{X}]{\nabla^{3} A_{X}(\theta)})  + \lambda^{T} \nabla_{\theta} B(\theta) \\
&= I(\theta) + \theta^{T} \EE[P_{X}]{\EE[Q_{Y|X}(\theta)]{(\eta(X, Y) - \EE[Q_{Y|X}(\theta)]{\eta(X, Y)|X})^{\otimes 3} |X}} \\
&\indent+ \lambda^{T}\Big(\EE[P_{X}]{\EE[Q_{Y|X}(\theta)]{\gamma(X, Y) \otimes \eta(X, Y) \otimes (\eta(X, Y) - \EE[Q_{Y|X}(\theta)]{\eta(X, Y) |X}) \mid X}} \\
&\indent\indent-  \EE[P_{X}]{ \EE[Q_{Y|X}(\theta)]{\gamma(X, Y) \mid X} \otimes \Cov[Q_{Y|X}(\theta)]{\eta(X, Y) \mid X}} \\
&\indent\indent-\EE[P_{X}]{\Cov[Q_{Y|X}(\theta)]{\gamma(X, Y), \eta(X, Y) \mid X} \otimes \EE[Q_{Y|X}(\theta)]{\eta(X, Y) \mid X}}\Big).
\end{align*}

\end{subsection}

\begin{subsection}{Proof of Lemma \ref{lemm:simple_deriv}}

We note that 
\begin{align*}
&\frac{\partial}{\partial \epsilon} G(\nu, \mu_{0} + \epsilon h) \Big|_{\epsilon=0} \\
&=  \EE[S]{\frac{\partial}{\partial \epsilon} g(X; \nu, \mu_{0} + \epsilon h) \cdot (Y - \mu_{0}(X) - \epsilon h(X)) \Big|_{\epsilon=0}} \\
&= -\EE[S]{ g(X; \nu, \mu_{0} + \epsilon h)\cdot h(X) \Big|_{\epsilon=0}}  + \EE[S]{ \frac{\partial}{\partial \epsilon} g(X; \nu, \mu_{0} + \epsilon h) \Big|_{\epsilon= 0} \cdot (Y - \mu_{0}(X)) }  \\
&= -\EE[S]{ g(X; \nu, \mu_{0}) \cdot h(X)}.
\end{align*}
The first equality holds due to the Dominated Convergence Theorem, allowing us to exchange differentiation and expectation. The second equality follows from the product rule. The last equality holds because $\EE[S]{(Y - \mu_{0}(X)) \cdot g(X) \mid X} = \EE[S]{(Y - \EE[S]{Y \mid X}) \cdot g(X) \mid X} = 0.$

\end{subsection}

\begin{subsection}{Proof of Lemma \ref{lemm:asymptotic_equicontinuity}}
First, we define additional notation. Since $\mu, r \in L^{2}(P_{X}, \mathcal{X})$, then the Gateaux derivatives of $\psi$ with respect to $\mu, r$ is a linear functional. As a result, there exists $\dot{\psi}_{\mu}$ and $\dot{\psi}_{r}$ so that the Gateaux derivative of $\Psi$ as follows
\begin{align}
\dot{\Psi}_{\mu}(\nu, \mu, r)[h] := \EE[S]{\dot{\psi}_{\mu}(X, Y; \nu, \mu, r) \cdot h(X)}, \label{eq:dot_psi_mu} \\
\dot{\Psi}_{r}(\nu, \mu, r)[h] := \EE[S]{\dot{\psi}_{r}(X, Y; \nu, \mu, r) \cdot h(X)}. \label{eq:dot_psi_r}
\end{align}
In addition, define the empirical Gateaux derivative evaluated on fold $\mathcal{I}$
\begin{align*}
\dot{\Psi}^{\mathcal{I}}_{\mu, n}(\nu, \mu, r)[h] := \frac{1}{|\mathcal{I}|} \sum_{i \in \mathcal{I}} \dot{\psi}_{\mu}(X_i, Y_i; \nu, \mu, r) \cdot h(X_i), \\
\dot{\Psi}^{\mathcal{I}}_{r, n}(\nu, \mu, r)[h] := \frac{1}{|\mathcal{I}|} \sum_{i \in \mathcal{I}} \dot{\psi}_{r}(X_i, Y_i; \nu, \mu, r) \cdot h(X_i).
\end{align*}
We also define the following population second derivatives.
\begin{align*}
\ddot{\Psi}_{\nu \mu}(\nu, \mu, r)[h] &:= \EE[S]{\ddot{\psi}_{\nu \mu}(X, Y; \nu, \mu, r) \cdot h(X)}, \\
\ddot{\Psi}_{\nu r}(\nu, \mu, r)[h] &:= \EE[S]{\ddot{\psi}_{\nu r}(X, Y; \nu, \mu, r) \cdot h(X)}.
\end{align*}

The empirical second derivatives evaluated on fold $\mathcal{I}$ are given by
\begin{align*}
\ddot{\Psi}^{\mathcal{I}}_{\nu \mu, n}(\nu, \mu, r)[h] &:= \frac{1}{|\mathcal{I}|} \sum_{i \in \mathcal{I}} \ddot{\psi}_{\nu \mu}(X_i, Y_i; \nu, \mu, r) \cdot h(X_i), \\
\ddot{\Psi}^{\mathcal{I}}_{\nu r, n}(\nu, \mu, r)[h] &:= \frac{1}{|\mathcal{I}|} \sum_{i \in \mathcal{I}} \ddot{\psi}_{\nu r}(X_i, Y_i; \nu, \mu, r) \cdot h(X_i).
\end{align*}

\begin{subsubsection}{First Claim}
Since $\psi$ is differentiable in $\nu$ and Gateaux differentiable in $\mu, r$, we can perform a Taylor expansion of $\Psi$ at $(\tilde{\nu}, \hat{\mu}^{\mathcal{I}_{2}}_{n}, \hat{r}^{\mathcal{I}_{2}}_{n})$. By applying Assumption \ref{assumption:pilot}, we have that 
\begin{align*}
\Psi(\nu_0, \hat{\mu}_{n}^{\mathcal{I}_{2}}, \hat{r}_{n}^{\mathcal{I}_{2}}) = \Psi(\tilde{\nu}_{n}, \hat{\mu}_{n}^{\mathcal{I}_{2}}, \hat{r}_{n}^{\mathcal{I}_{2}}) + \dot{\Psi}_{\nu}(\tilde{\nu}_{n}, \hat{\mu}_{n}^{\mathcal{I}_{2}}, \hat{r}_{n}^{\mathcal{I}_{2}}) \cdot (\nu_0 - \tilde{\nu}_{n}) + o_{P}(n^{-1/2}).
\end{align*}
So, we have that
\begin{align*}
\Psi(\tilde{\nu}_{n}, \hat{\mu}_{n}^{\mathcal{I}_{2}}, \hat{r}_{n}^{\mathcal{I}_{2}}) &= \Psi(\nu_0, \hat{\mu}_{n}^{\mathcal{I}_{2}}, \hat{r}_{n}^{\mathcal{I}_{2}}) + \dot{\Psi}_{\nu}(\tilde{\nu}_{n}, \hat{\mu}_{n}^{\mathcal{I}_{2}}, \hat{r}_{n}^{\mathcal{I}_{2}}) \cdot (\tilde{\nu}_{n} - \nu_{0}) + o_{P}(n^{-1/2}).
\end{align*}
Then, by applying Assumption \ref{assumption:nuisance}, we have that
\begin{align*}
\Psi(\nu_0, \hat{\mu}_{n}^{\mathcal{I}_{2}}, \hat{r}_{n}^{\mathcal{I}_{2}}) &= \Psi(\nu_0, \mu_0, r_0) + \dot{\Psi}_{\mu}(\nu_0, \mu_0, r_0)[\hat{\mu}_{n}^{\mathcal{I}_{2}} - \mu_0] + \dot{\Psi}_{r}(\nu_0, \mu_0, r_0)[\hat{r}_{n}^{\mathcal{I}_{2}} - r_{0}] \\
&\indent+ O_{P}(||(\hat{\mu}_{n}^{\mathcal{I}_{2}}, \hat{r}_{n}^{\mathcal{I}_{2}}) - (\mu_{0}, r_{0})||^{2}_{n}) \\
&= o_{P}(n^{-1/2}).
\end{align*}

Thus, we have that 
\[ \Psi(\tilde{\nu}_{n}, \hat{\mu}_{n}^{\mathcal{I}_{2}}, \hat{r}_{n}^{\mathcal{I}_{2}}) =  \dot{\Psi}_{\nu}(\tilde{\nu}_{n}, \hat{\mu}_{n}^{\mathcal{I}_{2}}, \hat{r}_{n}^{\mathcal{I}_{2}}) \cdot (\tilde{\nu}_{n} - \nu_0) + o_{P}(n^{-1/2}). \]
\end{subsubsection}

\begin{subsubsection}{Second Claim}
To prove the second claim, we use a Taylor expansion of $\Psi_{n}$ at $(\nu_{0}, \mu_{0}, r_{0})$ below. In the expansion below, we use Lemma \ref{lemm:orthogonal} and Assumptions \ref{assumption:pilot} and \ref{assumption:nuisance}.
\begin{align*}
\Psi_{n}^{\mathcal{I}_{1}}(\tilde{\nu}_{n}, \hat{\mu}^{\mathcal{I}_{2}}_{n}, \hat{r}^{\mathcal{I}_{2}}_{n}) &= \Psi_{n}^{\mathcal{I}_{1}}(\nu_{0}, \mu_{0}, r_{0}) + \dot{\Psi}_{\nu, n}^{\mathcal{I}_{1}}(\nu_{0}, \mu_{0}, r_{0}) \cdot (\tilde{\nu}_{n} - \nu_{0}) + \dot{\Psi}_{\mu, n}^{\mathcal{I}_{1}}(\nu_{0}, \mu_{0}, r_{0})[\hat{\mu}_{n}^{\mathcal{I}_{2}} - \mu_{0}] \\
&\indent + \dot{\Psi}_{r, n}^{\mathcal{I}_{1}}(\nu_{0}, \mu_{0}, r_{0})[\hat{r}_{n}^{\mathcal{I}_{2}} - r_{0}] + O_{P}(||( \tilde{\nu}_{n}, \hat{\mu}_{n}^{\mathcal{I}_{2}}, \hat{r}_{n}^{\mathcal{I}_{2}}) - (\nu_{0}, \mu_{0}, r_{0})||_{n}^{2}) \\
&= \Psi_{n}^{\mathcal{I}_{1}}(\nu_{0}, \mu_{0}, r_{0}) + \dot{\Psi}_{\nu, n}^{\mathcal{I}_{1}}(\nu_{0}, \mu_{0}, r_{0}) \cdot (\tilde{\nu}_{n} - \nu_{0}) + \dot{\Psi}_{\mu, n}^{\mathcal{I}_{1}}(\nu_{0}, \mu_{0}, r_{0})[\hat{\mu}_{n}^{\mathcal{I}_{2}} - \mu_{0}] \\
&\indent + \dot{\Psi}_{r, n}^{\mathcal{I}_{1}}(\nu_{0}, \mu_{0}, r_{0})[\hat{r}_{n}^{\mathcal{I}_{2}} - r_{0}] + o_{P}(n^{-1/2}) \\
&= \Psi_{n}^{\mathcal{I}_{1}}(\nu_{0}, \mu_{0}, r_{0})  + \Psi_{\nu, n}^{\mathcal{I}_{1}}(\nu_{0}, \mu_{0}, r_{0}) \cdot(\tilde{\nu}_{n} - \nu_{0}) \\
&\indent+ \dot{\Psi}_{\mu}(\nu_{0}, \mu_{0}, r_{0}) \cdot [\hat{\mu}_{n}^{\mathcal{I}_{2}} - \mu_{0}] + \dot{\Psi}_{r}(\nu_{0}, \mu_{0}, r_{0}) \cdot [\hat{r}_{n}^{\mathcal{I}_{2}} - r_{0}] \\
&\indent+ (\dot{\Psi}_{r, n}^{\mathcal{I}_{1}}(\nu_{0}, \mu_{0}, r_{0}) [\hat{r}_{n}^{\mathcal{I}_{2}} - r_{0}] - \dot{\Psi}_{r}(\nu_{0}, \mu_{0}, r_{0})[\hat{r}_{n}^{\mathcal{I}_{2}} - r_{0}]) \\
&\indent+ (\dot{\Psi}_{\mu, n}^{\mathcal{I}_{1}}(\nu_{0}, \mu_{0}, r_{0}) [\hat{\mu}_{n}^{\mathcal{I}_{2}} - \mu_{0}] - \dot{\Psi}_{\mu}(\nu_{0}, \mu_{0}, r_{0})[\hat{\mu}_{n}^{\mathcal{I}_{2}} - \mu_{0}]) + o_{P}(n^{-1/2})\\
&= \Psi_{n}^{\mathcal{I}_{1}}(\nu_{0}, \mu_{0}, r_{0}) + \dot{\Psi}_{\nu, n}^{\mathcal{I}_{1}}(\nu_{0}, \mu_{0}, r_{0}) \cdot (\tilde{\nu}_{n} - \nu_{0}) \\
&\indent+ (\dot{\Psi}_{r, n}^{\mathcal{I}_{1}}(\nu_{0}, \mu_{0}, r_{0}) [\hat{r}_{n}^{\mathcal{I}_{2}} - r_{0}] - \dot{\Psi}_{r}(\nu_{0}, \mu_{0}, r_{0})[\hat{r}_{n}^{\mathcal{I}_{2}} - r_{0}])  \\
&\indent+(\dot{\Psi}_{\mu, n}^{\mathcal{I}_{1}}(\nu_{0}, \mu_{0}, r_{0}) [\hat{\mu}_{n}^{\mathcal{I}_{2}} - \mu_{0}] - \dot{\Psi}_{\mu}(\nu_{0}, \mu_{0}, r_{0})[\hat{\mu}_{n}^{\mathcal{I}_{2}} - \mu_{0}])+ o_{P}(n^{-1/2}).
\end{align*}

We argue that $\dot{\Psi}_{\mu, n}^{\mathcal{I}_{1}}(\nu_{0}, \mu_{0}, r_{0}) [\hat{\mu}_{n}^{\mathcal{I}_{2}} - \mu_{0}] - \dot{\Psi}_{\mu}(\nu_{0}, \mu_{0}, r_{0})[\hat{\mu}_{n}^{\mathcal{I}_{2}} - \mu_{0}]$ are $o_{P}(n^{-1/2}).$

\begin{lemm}
\label{lemm:conditional_variance}
Let $\dot{\psi}_{\mu}$ given in \eqref{eq:dot_psi_mu}. 
\begin{enumerate}
    \item There exists $0 < C_{\mu} < \infty$ so that $\Var[S]{\dot{\psi}_{\mu, j}(X, Y; \nu_{0}, \mu_{0}, r_{0}) \mid X=x} < C_{\mu}$ for all $j \in [J+ K].$
    \item There exists $0 < C_{r} < \infty$ so that $\Var[S]{\dot{\psi}_{r, j}(X, Y; \nu_{0}, \mu_{0}, r_{0}) \mid X=x} < C_{r}$ for all $j \in [J+ K].$
    \item There exists $0 < C' < \infty$ so that 
    \begin{align*}
    \Var[S]{\ddot{\psi}_{\nu \mu, ij}(X, Y; \nu_0, \mu_0, r_0) \mid X=x} < C'\\
    \Var[S]{\ddot{\psi}_{\nu r, ij}(X, Y; \nu_0, \mu_0, r_0) \mid X=x} < C'
    \end{align*} for all $i, j \in [J+ K] \times [J + K].$
\end{enumerate}
\end{lemm}

Let $A_{n}:= \dot{\Psi}_{\mu, n}^{\mathcal{I}_{1}}(\nu_{0}, \mu_{0}, r_{0}) [\hat{\mu}_{n}^{\mathcal{I}_{2}} - \mu_{0}] - \dot{\Psi}_{\mu}(\nu_{0}, \mu_{0}, r_{0})[\hat{\mu}_{n} - \mu_{0}]$. We argue that $n\cdot ||A_{n}||^{2} = o_{P}(1)$, which implies that $A_{n} = o_{P}(n^{-1/2})$. First, we note that Lemma \ref{lemm:orthogonal} gives that $\dot{\Psi}_{\mu}(\nu; \mu_{0}, r_{0}) [\hat{\mu}_{n} - \mu] = 0$ for all $\nu \in \mathbb{R}^{J + K}$, which implies that $\EE[S]{\dot{\psi}_{\mu}(X_i, Y_i; \nu_{0}, \mu_0, r_0) \cdot (\hat{\mu}_{n}^{\mathcal{I}_{2}}(X_i) - \mu(X_i)) \mid X_i} = 0$.

\begin{align*}
A_{n} &= \dot{\Psi}_{\mu, n}(\nu_{0}, \mu_{0}, r_{0}) [\hat{\mu}_{n}^{\mathcal{I}_{2}} - \mu_{0}] - \dot{\Psi}_{\mu}(\nu_{0}, \mu_{0}, r_{0})[\hat{\mu}_{n}^{\mathcal{I}_{2}} - \mu_{0}]\\
&= \frac{1}{|\mathcal{I}_{1}|} \sum_{i \in \mathcal{I}_{1}} \dot{\psi}_{\mu}(X_i, Y_i; \nu_{0}, \mu_0, r_0) \cdot (\hat{\mu}^{\mathcal{I}_{2}}_{n}(X_i) - \mu(X_i)) - \EE[S]{\dot{\psi}_{\mu}(X_i, Y_i; \nu_0, \mu_0, r_{0}) \cdot (\hat{\mu}_{n}^{\mathcal{I}_{2}}(X_i) - \mu(X_i)) \mid X_i}) \\
&= \frac{1}{|\mathcal{I}_{1}|} \sum_{i \in \mathcal{I}_{1}} \dot{\psi}_{\mu}(X_i, Y_i; \nu_{0}, \mu_0, r_0) \cdot (\hat{\mu}^{\mathcal{I}_{2}}_{n}(X_i) - \mu(X_i)).
\end{align*}
If we can show that $\EE[S]{n||A_{n}||^{2} \mid \mathcal{I}_{2}, \{X_{i}\}} = o_{P}(1),$ then $n||A_{n}||^{2} = o_{P}(1)$ by Lemma I.5 of \citet{jin2024tailored} and $A_{n} = o_{P}(n^{-1/2}).$ It remains to show $\EE[S]{n||A_{n}||^{2} \mid \mathcal{I}_{2}, \{X_{i}\}} = o_{P}(1)$. In the following derivation, we note that $\hat{\mu}^{\mathcal{I}_{2}}_{n}, \hat{r}^{\mathcal{I}_{2}}_{n}$ are fit using samples from $\mathcal{I}_{2}$, so they can be treated as deterministic quantities conditional on $\mathcal{I}_{2}.$

\begin{align*}
\EE[S]{ ||A_{n}||^{2} \mid \mathcal{I}_{2}, \{X_{i} \}} &= \sum_{j=1}^{J+K} \EE[S]{ A_{j, n}^{2} \mid \mathcal{I}_{2}, \{X_{i} \}} \\
&= \sum_{j=1}^{J+K}  \EE[S]{\Big( \frac{1}{|\mathcal{I}_{1}|} \sum_{i \in \mathcal{I}_{1}} \dot{\psi}_{\mu, j}(X_i, Y_i; \nu_{0}, \mu_{0}, r_{0}) \cdot (\hat{\mu}_{n}^{\mathcal{I}_{2}}(X_i) - \mu(X_i)) \Big)^{2}  \mid \mathcal{I}_{2}, \{X_i\}} \\ 
&= \sum_{j=1}^{J+K} \Var[S]{ \frac{1}{|\mathcal{I}_{1}|} \sum_{i \in \mathcal{I}_{1}} \dot{\psi}_{\mu, j}(X_i, Y_i; \nu_{0}, \mu_{0}, r_{0}) \cdot (\hat{\mu}_{n}^{\mathcal{I}_{2}}(X_i) - \mu(X_i))  \mid \mathcal{I}_{2}, \{X_i\}} \\
&= \sum_{j=1}^{J+K}  \sum_{i \in \mathcal{I}_{1}} \Var[S]{ \frac{1}{|\mathcal{I}_{1}|} \dot{\psi}_{\mu, j}(X_i, Y_i; \nu_{0}, \mu_{0}, r_{0}) \cdot (\hat{\mu}^{\mathcal{I}_{2}}_{n}(X_i) - \mu(X_i))  \mid \mathcal{I}_{2}, \{X_i\}} \\
&= \sum_{j=1}^{J+K}  \frac{1}{|\mathcal{I}_{1}|^{2}} \cdot \sum_{i \in \mathcal{I}_{1}} (\hat{\mu}^{\mathcal{I}_{1}}_{n}(X_i) - \mu(X_i))^{2} \cdot \Var[S]{\dot{\psi}_{\mu, j}(X_i, Y_i; \nu_0, \mu_0, r_0) \mid \mathcal{I}_{2}, \{X_i\}} \\
&= \sum_{j=1}^{J+K}  \frac{1}{|\mathcal{I}_{1}|} \cdot \Big( \frac{1}{|\mathcal{I}_{1}|} \sum_{i \in \mathcal{I}_{1}} (\hat{\mu}(X_i) - \mu(X_i))^{2} \Big) \cdot C_{\mu} \\
&= o_{P}(n^{-3/2}).
\end{align*}

We can apply an analogous argument to show that $(\dot{\Psi}^{\mathcal{I}_{1}}_{r, n}(\nu_{0}, \mu_{0}, r_{0}) [\hat{r}^{\mathcal{I}_{2}}_{n} - r_{0}] - \dot{\Psi}_{r}(\tilde{\nu}_{0}, \mu_{0}, r_{0})[\hat{r}_{n}^{\mathcal{I}_{2}} - r_{0}]) = o_{P}(n^{-1/2}).$

Thus, we have that
\begin{align*}
\Psi_{n}^{\mathcal{I}_{1}}(\tilde{\nu}_{n}, \hat{\mu}^{\mathcal{I}_{2}}_{n}, \hat{r}^{\mathcal{I}_{2}}_{n}) &= \Psi_{n}^{\mathcal{I}_{1}}(\nu_{0}, \mu_{0}, r_{0}) + \dot{\Psi}_{\nu, n}^{\mathcal{I}_{1}}(\nu_{0}, \mu_{0}, r_{0}) \cdot (\tilde{\nu}_{n} - \nu_{0}) + o_{P}(n^{-1/2}).
\end{align*}

We can combine the results from the Taylor expansion of $\Psi$ and $\Psi_{n}$ about $(\nu_{0}, \mu_{0}, r_{0})$ below.
\begin{align*}
&\Psi_{n}^{\mathcal{I}_{1}}(\tilde{\nu}_{n}, \hat{\mu}^{\mathcal{I}_{2}}_{n}, \hat{r}^{\mathcal{I}_{2}}_{n}) - \Psi(\tilde{\nu}_{n}, \hat{\mu}_{n}^{\mathcal{I}_{2}}, \hat{r}_{n}^{\mathcal{I}_{2}}) - \Psi_{n}^{\mathcal{I}_{1}}(\nu_{0}, \mu_{0}, r_{0})  + \Psi(\nu_{0}, \mu_{0}, r_{0}) \\
&= \dot{\Psi}_{\nu, n}^{\mathcal{I}_{1}}(\nu_{0}, \mu_{0}, r_{0}) \cdot (\tilde{\nu}_{n} - \nu_{0}) - \dot{\Psi}_{\nu}(\nu_{0}, \mu_{0}, r_{0}) \cdot (\tilde{\nu}_{n} - \nu_{0})  + o_{P}(n^{-1/2}) \\
&= (\dot{\Psi}_{\nu, n}^{\mathcal{I}_{1}}(\nu_{0}, \mu_{0}, r_{0}) - \Psi_{\nu}(\nu_{0}, \mu_{0}, r_{0})) \cdot (\tilde{\nu}_{n} - \nu_{0}) + o_{P}(n^{-1/2}) \\
&= o_{P}(n^{-1/2})
\end{align*}

We have that $\dot{\Psi}_{\nu, n}^{\mathcal{I}_{1}}(\nu_{0}, \mu_{0}, r_{0}) - \dot{\Psi}_{\nu}(\nu_{0}, \mu_{0}, r_{0}) = O_{p}(n^{-1/2})$ by the Central Limit Theorem and that $\tilde{\nu}_{n} - \nu_{0} = o_{P}(1)$, which yields the last equality above.

\end{subsubsection}

\begin{subsubsection}{Third Claim}

In this result, we leverage properties of the second derivatives of the estimating equations. Since the orthogonality properties from Lemma \ref{lemm:orthogonal} holds uniformly for all $\nu$, we have that $\dot{\Psi}_{\mu}(\nu, \mu_{0}, r_{0})[h] = 0$ and $\dot{\Psi}_{r}(\nu, \mu_{0}, r_{0})[h] = 0$ for all $\nu$. This implies that  
\begin{align*}
\dot{\Psi}_{\mu \nu}(\nu, \mu_{0}, r_{0})[\hat{\mu}_{n} - \mu_{0}] &= 0 \\
\dot{\Psi}_{r \nu}(\nu, \mu_{0}, r_{0})[\hat{r}_{n} - r_{0}] &= 0.
\end{align*}
Since we can interchange the order of differentiation, we also have that
\begin{align}
\dot{\Psi}_{\nu \mu}(\nu, \mu_{0}, r_{0})[\hat{\mu}_{n} - \mu_{0}] &= 0 \label{eq:deriv1}\\
\dot{\Psi}_{\nu r}(\nu, \mu_{0}, r_{0})[\hat{r}_{n} - r_{0}] &= 0. \label{eq:deriv2}
\end{align}

We use the above properties to simplify a Taylor expansion of $\dot{\Psi}_{\nu}$ at $(\nu_0, \mu_0, r_0)$.

\begin{align*}
\dot{\Psi}_{\nu}(\tilde{\nu}_{n}, \hat{\mu}^{\mathcal{I}_{2}}_{n}, \hat{r}^{\mathcal{I}_{2}}_{n}) &= \dot{\Psi}_{\nu}(\nu_{0}, \mu_{0}, r_{0}) + \ddot{\Psi}_{\nu\nu}(\nu_{0}, \mu_{0}, r_{0}) \cdot (\tilde{\nu}_{n} - \nu_{0}) + \ddot{\Psi}_{\mu \nu }(\nu_{0}, \mu_{0}, r_{0})[\hat{\mu}_{n}^{\mathcal{I}_{2}} - \mu_{0}] \\
&\indent + \ddot{\Psi}_{r\nu}(\nu_0, \mu_0, r_0)[\hat{r}_{n}^{\mathcal{I}_{2}} - r_{0}] + O_{P}(||( \tilde{\nu}_{n}, \hat{\mu}_{n}^{\mathcal{I}_{2}}, \hat{r}_{n}^{\mathcal{I}_{2}}) - (\nu_{0}, \mu_{0}, r_{0})||^{2}_{n}) \\
&= \dot{\Psi}_{\nu}(\nu_{0}, \mu_{0}, r_{0}) + \ddot{\Psi}_{\nu\nu}(\nu_{0}, \mu_{0}, r_{0}) \cdot (\tilde{\nu}_{n} - \nu_{0}) + o_{P}(n^{-1/2}).
\end{align*}

We can also perform a Taylor expansion of $\dot{\Psi}^{\mathcal{I}_{1}}_{\nu, n}$ at $(\nu_{0}, \mu_{0}, r_{0}).$
\begin{align*}
\dot{\Psi}_{\nu, n}^{\mathcal{I}_{1}}(\tilde{\nu}_{n}, \hat{\mu}_{n}, \hat{r}_{n}) &= \dot{\Psi}^{\mathcal{I}_{1}}_{\nu, n}(\nu_{0}, \mu_{0}, r_{0}) + \ddot{\Psi}_{\nu\nu, n}^{\mathcal{I}_{1}}(\nu_{0}, \mu_{0}, r_{0}) \cdot(\tilde{\nu}_{n} - \nu_{0}) + \ddot{\Psi}^{\mathcal{I}_{1}}_{\nu\mu, n}(\nu_{0}, \mu_{0}, r_{0})[\hat{\mu}_{n}^{\mathcal{I}_{2}} - \mu_{0}] \\
&\indent + \ddot{\Psi}^{\mathcal{I}_{1}}_{\nu r, n}(\nu_{0}, \mu_{0}, r_{0})[\hat{r}_{n}^{\mathcal{I}_{2}} - r_{0}] + O_{P}(||(\hat{\mu}_{n}^{\mathcal{I}_{2}}, \hat{r}_{n}^{\mathcal{I}_{2}}) - (\mu_{0}, r_{0})||^{2}_{n}) \\
&= \dot{\Psi}^{\mathcal{I}_{1}}_{\nu, n}(\nu_{0}, \mu_{0}, r_{0}) + \ddot{\Psi}_{\nu\nu, n}^{\mathcal{I}_{1}}(\nu_{0}, \mu_{0}, r_{0}) \cdot(\tilde{\nu}_{n} - \nu_{0}) + \ddot{\Psi}^{\mathcal{I}_{1}}_{\nu\mu, n}(\nu_{0}, \mu_{0}, r_{0})[\hat{\mu}_{n}^{\mathcal{I}_{2}} - \mu_{0}] \\
&\indent + \ddot{\Psi}^{\mathcal{I}_{1}}_{\nu r, n}(\nu_{0}, \mu_{0}, r_{0})[\hat{r}_{n}^{\mathcal{I}_{2}} - r_{0}] + o_{P}(n^{-1/2}) \\
&= \dot{\Psi}^{\mathcal{I}_{1}}_{\nu, n}(\nu_{0}, \mu_{0}, r_{0}) + \ddot{\Psi}_{\nu\nu, n}^{\mathcal{I}_{1}}(\nu_{0}, \mu_{0}, r_{0}) \cdot (\tilde{\nu}_{n} - \nu_{0}) \\
&\indent+ \ddot{\Psi}_{\nu\mu}(\nu_{0}, \mu_{0}, r_{0})[\hat{\mu}^{\mathcal{I}_{2}}_{n} - \mu_{0}] + \ddot{\Psi}_{\nu r}(\nu_{0}, \mu_{0}, r_{0})[\hat{r}^{\mathcal{I}_{2}}_{n} - r_{0}] \\
&\indent + ( \ddot{\Psi}_{\nu\mu, n}^{\mathcal{I}_{1}}(\nu_{0}, \mu_{0}, r_{0})[\hat{\mu}_{n}^{\mathcal{I}_{2}} - \mu_{0}] - \ddot{\Psi}_{\nu\mu}(\nu_{0}, \mu_{0}, r_{0})[\hat{\mu}^{\mathcal{I}_{2}}_{n} - \mu_{0}] ) \\
&\indent + (\ddot{\Psi}_{\nu r, n}^{\mathcal{I}_{1}}(\nu_{0}, \mu_{0}, r_{0})[\hat{r}_{n} - r_{0}] - \ddot{\Psi}_{\nu r}(\nu_{0}, \mu_{0}, r_{0})[\hat{r}_{n} - r_{0}]) + o_{P}(n^{-1/2}) \\
&= \dot{\Psi}_{\nu, n}^{\mathcal{I}_{1}}(\nu_{0}, \mu_{0}, r_{0}) + \ddot{\Psi}_{\nu\nu, n}^{\mathcal{I}_{1}}(\nu_{0}, \mu_{0}, r_{0}) \cdot (\tilde{\nu}_{n} - \nu_{0}) \\
&\indent + ( \ddot{\Psi}_{\nu\mu, n}^{\mathcal{I}_{1}}(\nu_{0}, \mu_{0}, r_{0})[\hat{\mu}_{n}^{\mathcal{I}_{2}} - \mu_{0}] - \ddot{\Psi}_{\nu\mu}(\nu_{0}, \mu_{0}, r_{0})[\hat{\mu}_{n}^{\mathcal{I}_{2}} - \mu_{0}] )\\
&\indent + (\ddot{\Psi}_{\nu r, n}^{\mathcal{I}_{1}}(\nu_{0}, \mu_{0}, r_{0})[\hat{r}_{n}^{\mathcal{I}_{2}} - r_{0}] - \ddot{\Psi}_{\nu r}(\nu_{0}, \mu_{0}, r_{0})[\hat{r}_{n}^{\mathcal{I}_{2}} - r_{0}]) + o_{P}(n^{-1/2}).
\end{align*}

The first equality follows from the fact that $\dot{\Psi}_{\nu}$ is Gateaux differentiable in $\mu, r$. The second equality follows from consistency of the nuisance estimates (Assumption \ref{assumption:nuisance}). The third equality follows from adding and subtracting the population Jacobians. The last equality follows from \eqref{eq:deriv1} and \eqref{eq:deriv2}, which holds due to the properties of the second derivatives demonstrated above.

Note that $B_{n} := \dot{\Psi}^{\mathcal{I}_{1}}_{\nu \mu, n}(\nu_{0}, \mu_{0}, r_{0}) [\hat{\mu}^{\mathcal{I}_{2}}_{n} - \mu_{0}] - \dot{\Psi}_{\nu \mu}(\nu_{0}, \mu_{0}, r_{0})[\hat{\mu}_{n}^{\mathcal{I}_{2}} - \mu_{0}]$. Note that $B_{n} \in \mathbb{R}^{(J + K) \times (J + K)}$. Let $B_{ij, n}$ denote the $(i, j)$-th index of $B_{n}$. Similar to before, we will show that $\EE[]{nB_{ij, n}^{2} \mid \mathcal{I}_{2}, \{X_{i}\}} = o_{P}(1),$ then $nB_{ij, n}^{2} = o_{P}(1)$ by Lemma I.5 of \citet{jin2024tailored} and $B_{ij, n} = o_{P}(n^{-1/2}).$ It remains to show $\EE[]{nB_{ij,n}^{2} \mid \mathcal{I}_{2}, \{X_{i}\}} = o_{P}(1),$
\begin{align*}
\EE[]{ B_{ij,n}^{2} \mid \mathcal{I}_{2}, \{X_{k} \}} &= \EE[]{\Big( \frac{1}{|\mathcal{I}_{1}|} \sum_{k \in \mathcal{I}_{1}} \dot{\psi}_{\nu\mu, ij}(X_k, Y_k; \nu_{0}, \mu_{0}, r_{0}) \cdot (\hat{\mu}_{n}^{\mathcal{I}_{2}}(X_k) - \mu(X_k)) \Big)^{2}  \mid \mathcal{I}_{2}, \{X_i\}} \\  
&=\Var[]{ \frac{1}{|\mathcal{I}_{1}|} \sum_{k \in \mathcal{I}_{1}} \dot{\psi}_{\nu\mu, ij}(X_k, Y_k; \nu_{0}, \mu_{0}, r_{0}) \cdot (\hat{\mu}_{n}(X_k) - \mu(X_k))  \mid \mathcal{I}_{2}, \{X_k\}} \\
&= \sum_{i \in \mathcal{I}_{1}} \Var[]{ \frac{1}{|\mathcal{I}_{1}|} \dot{\psi}_{\nu \mu}(X_k, Y_k; \nu_{0}, \mu_{0}, r_{0}) \cdot (\hat{\mu}_{n}(X_k) - \mu(X_k))  \mid \mathcal{I}_{2}, \{X_k\}} \\
&= \frac{1}{|\mathcal{I}_{1}|^{2}} \cdot \sum_{i \in \mathcal{I}_{1}} (\hat{\mu}(X_k) - \mu(X_k))^{2} \cdot \Var[]{\dot{\psi}_{\nu \mu, ij}(X_k, Y_k; \nu_0, \mu_0, r_0) \mid \mathcal{I}_{2}, \{X_k\}} \\
&= \frac{1}{|\mathcal{I}_{1}|} \cdot \Big( \frac{1}{|\mathcal{I}_{1}|} \sum_{k \in \mathcal{I}_{1}} (\hat{\mu}(X_k) - \mu(X_k))^{2} \Big) \cdot C \\
&= o_{P}(n^{-3/2}).
\end{align*}

We can apply an analogous argument to show that $(\dot{\Psi}^{\mathcal{I}_{1}}_{\nu r, n}(\nu_{0}, \mu_{0}, r_{0}) [\hat{r}^{\mathcal{I}_{2}}_{n} - r_{0}] - \dot{\Psi}_{\nu r}(\tilde{\nu}_{n}^{\mathcal{I}_{2}}, \mu_{0}, r_{0})[\hat{r}_{n} - r_{0}]) = o_{P}(n^{-1/2}).$

As a result, we have that
\begin{align*}
\dot{\Psi}^{\mathcal{I}_{1}}_{\nu, n}(\tilde{\nu}_{n}, \hat{\mu}_{n}^{\mathcal{I}_{2}}, \hat{r}_{n}^{\mathcal{I}_{2}}) &= \dot{\Psi}^{\mathcal{I}_{1}}_{\nu, n}(\nu_{0}, \mu_{0}, r_{0})  + \ddot{\Psi}^{\mathcal{I}_{1}}_{\nu\nu, n} (\nu_{0}, \mu_{0}, r_{0}) \cdot (\tilde{\nu}_{n} - \nu_{0}) + o_{P}(n^{-1/2}).
\end{align*}

We can combine the results from the Taylor expansion of $\dot{\Psi}_{\nu}$ and $\dot{\Psi}_{\nu, n}$ about $(\nu_{0}, \mu_{0}, r_{0})$ below.
\begin{align*}
&\dot{\Psi}_{\nu, n}^{\mathcal{I}_{1}}(\tilde{\nu}_{n}, \hat{\mu}^{\mathcal{I}_{2}}_{n}, \hat{r}^{\mathcal{I}_{2}}_{n}) - \dot{\Psi}_{\nu}(\tilde{\nu}_{n}, \hat{\mu}_{n}^{\mathcal{I}_{2}}, \hat{r}_{n}^{\mathcal{I}_{2}}) - \dot{\Psi}_{\nu, n}^{\mathcal{I}_{1}}(\nu_{0}, \mu_{0}, r_{0})  + \dot{\Psi}_{\nu}(\nu_{0}, \mu_{0}, r_{0}) \\
&= \ddot{\Psi}_{\nu\nu, n}^{\mathcal{I}_{1}}(\nu_{0}, \mu_{0}, r_{0}) \cdot (\tilde{\nu}_{n} - \nu_{0}) - \ddot{\Psi}_{\nu \nu}(\nu_{0}, \mu_{0}, r_{0}) \cdot (\tilde{\nu}_{n} - \nu_{0})  + o_{P}(n^{-1/2}) \\
&= (\ddot{\Psi}_{\nu\nu, n}^{\mathcal{I}_{1}}(\nu_{0}, \mu_{0}, r_{0}) - \ddot{\Psi}_{\nu \nu}(\nu_{0}, \mu_{0}, r_{0})) \cdot (\tilde{\nu}_{n} - \nu_{0}) + o_{P}(n^{-1/2}) \\
&= o_{P}(n^{-1/2})
\end{align*}

We have that $\ddot{\Psi}_{\nu \nu , n}^{\mathcal{I}_{1}}(\nu_{0}, \mu_{0}, r_{0}) - \ddot{\Psi}_{\nu \nu}(\nu_{0}, \mu_{0}, r_{0}) = O_{p}(n^{-1/2})$ by the Central Limit Theorem and that $\tilde{\nu}_{n} - \nu_{0} = o_{P}(1)$, which yields the last equality above.

\end{subsubsection}

\end{subsection}

\begin{subsection}{Proof of Lemma \ref{lemm:conditional_variance}}

\begin{subsubsection}{First Claim}
\label{sec:claim1}
First, we show that $\Var[S]{\dot{\psi}_{\mu, j}(X, Y; \nu_0, \mu_0, r_0) \mid X}$ is uniformly bounded. Recall the definition of $\psi$ from \eqref{eq:pop_estimating_eq}. There exists functions $u_{D_{L}}: \mathcal{X} \rightarrow \mathbb{R}^{J}, u_{M}: \mathcal{X} \rightarrow \mathbb{R}^{K}$ such that 
\begin{align*}
\frac{\partial}{\partial \epsilon } \delta_{D_{L}}(x; \nu_{0}, \mu_{0} + \epsilon h) \Big|_{\epsilon = 0} &= u_{D_{L}}(x; \nu_0, \mu_0) \cdot h(x),  \\
\frac{\partial}{\partial \epsilon } \delta_{M}(x; \nu_{0}, \mu_{0} + \epsilon h) \Big|_{\epsilon = 0} &= u_{M}(x; \nu_0, \mu_0) \cdot h(x).
\end{align*}

Note that
\begin{align*}
\dot{\psi}_{\mu}(X, Y; \nu_0, \mu_0, r_0) &= \begin{bmatrix} f_{\mu, D_{L}}(X; \nu_{0}, \mu_{0}, r_{0}) + r_{0}(X) \cdot u_{D_{L}}(X; \nu_0, \mu_0) \cdot (Y - \mu_{0}(X)) \\
f_{\mu, M}(X; \nu_{0}, \mu_{0}, r_{0}) + r_{0}(X) \cdot u_{M}(X; \nu_0, \mu_0) \cdot (Y - \mu_{0}(X))
\end{bmatrix},
\end{align*}
where $f_{\mu, D_{L}}, f_{\mu, M}, u_{D_{L}}, u_{M}$ are vector-valued functions. For $j= 1, 2, \dots J$, we have that
\begin{align*}
\Var[S]{\dot{\psi}_{\mu, j}(X, Y; \nu_0, \mu_0, r_0) \mid X} &\leq r_{0}(X)^{2} \cdot u_{D_{L}, j}(X; \nu_0, \mu_0)^{2} \cdot \Var[S]{Y - \mu_{0}(X) \mid X} \\
&\leq \overline{r}^{2} \cdot u_{D_{L}, j}(X; \nu_0, \mu_0)^{2} \cdot C.
\end{align*}
We observe that $f_{\mu, D_{L}}(X)$ is bounded and only depends on $X$ so it does not affect whether the conditional variance is bounded. In addition, by Assumption \ref{assumption:cov_shift}, $\sup_{x \in \mathcal{X}} r_{0}(x) \leq \overline{r}$. Finally $\Var[S]{Y - \mu_{0}(X) \mid X}$ is bounded by a constant $0 < C < \infty$ because $Y - \mu_{0}(X)$ is a bounded random variable.

Similarly, we observe that the analogous properties hold the conditional variances of the $j= J+1, \dots J+K$ entries. We have that
\begin{align*}
\Var[S]{\dot{\psi}_{\mu, j}(X, Y; \nu_0, \mu_0, r_0) \mid X} &\leq r_{0}(X)^{2} \cdot u_{M, j}(X; \nu_0, \mu_0)^{2} \cdot \Var[S]{Y - \mu_{0}(X) \mid X} \\
&\leq \overline{r}^{2} \cdot u_{M, j}(X; \nu_0, \mu_0)^{2} \cdot C.
\end{align*}

As a result, in order to show that the conditional variances are bounded it suffices to show that $\sup_{x \in \mathcal{X}} |u_{D_{L}}(x; \nu_0, \mu_0)|^{2}$ is bounded and $\sup_{x \in \mathcal{X}} |u_{M}(x; \nu_0, \mu_0)|^{2}$ is bounded. Let 
\[ \alpha(X; \nu, \mu) :=  \frac{\exp(\theta^{T}(\eta(X, 1) - \eta(X, 0)))}{ \mu(X)  \cdot \exp(\theta^{T}(\eta(X, 1) - \eta(X, 0))) + \exp(\theta^{T}\eta(X, 0))}.\]

We observe that
\begin{align*}
&\frac{\partial}{\partial \epsilon } \delta_{D_{L}}(X; \nu_{0}, \mu_{0} + \epsilon h) \Big|_{\epsilon = 0} \\
&= \frac{\partial }{\partial \epsilon } \Big( (\rho(X, 1; \nu_0, \mu_0 + \epsilon h) - \rho(X, 0; \nu_0, \mu_0 + \epsilon h)) \cdot w(X, 1; \nu_0, \mu_0 + \epsilon h) \cdot w(X, 0; \nu_0, \mu_0 + \epsilon h) \Big) \Big|_{\epsilon = 0} \\
&= \frac{\partial }{\partial \epsilon } \Big( \rho(X, 1; \nu_0, \mu_0 + \epsilon h) - \rho(X, 0; \nu_0, \mu_0 + \epsilon h) \Big) \Big|_{\epsilon = 0} \cdot w(X, 1; \nu_0, \mu_0) \cdot w(X, 0; \nu_0, \mu_0)  \\
&\indent+ (\rho(X, 1; \nu_0, \mu_0) - \rho(X, 0; \nu_0, \mu_0)) \cdot \frac{\partial}{\partial \epsilon} \Big( w(X, 1; \nu_0, \mu_0 + \epsilon h) \cdot w(X, 0; \nu_0, \mu_0 + \epsilon h) \Big) \Big|_{\epsilon = 0} \\
&= \Big[-2 \cdot [(\eta(X, 1) - \eta(X, 0))^{\otimes 2} \theta_{0} +  (\eta(X, 1) - \eta(X, 0)) \otimes (\gamma(X, 1) - \gamma(X, 0)) \lambda_{0}] \cdot w(X, 1; \mu_0, \nu_0)^{2} \cdot w(X, 0; \mu_0, \nu_0)^{2} \\
&\indent -2 \cdot (\rho(X, 1; \nu_{0}, \mu_{0}) - \rho(X, 0; \nu_{0}, \mu_{0})) \cdot  \alpha(X; \nu_{0}, \mu_{0}) \cdot w(X, 1; \nu_0, \mu_0 ) \cdot w(X, 0; \nu_0, \mu_0) \Big] \cdot h(X).
\end{align*}
Thus,
\begin{align*}
&u_{D_{L}}(X) \\
&= -2 \cdot [(\eta(X, 1) - \eta(X, 0))^{\otimes 2} \theta_{0} +  (\eta(X, 1) - \eta(X, 0)) \otimes (\gamma(X, 1) - \gamma(X, 0)) \lambda_{0}] \cdot w(X, 1; \mu_0, \nu_0)^{2} \cdot w(X, 0; \mu_0, \nu_0)^{2} \\
&\indent -2 \cdot (\rho(X, 1; \nu_{0}, \mu_{0}) - \rho(X, 0; \nu_{0}, \mu_{0})) \cdot  \alpha(X; \nu_{0}, \mu_{0}) \cdot w(X, 1; \nu_0, \mu_0 ) \cdot w(X, 0; \nu_0, \mu_0).
\end{align*}

By Assumption \ref{assumption:bounded}, $\eta(x, y), \gamma(x, y)$ are uniformly bounded. Thus, due to the compactness of $\mathcal{X}$, we can observe that $|u_{D_{L}}(x; \nu_0, \mu_0)|^{2}$ is also uniformly bounded by a constant.

Similarly, we observe that
\begin{align*}
&\frac{\partial}{\partial \epsilon } \delta_{M}(X; \nu_{0}, \mu_{0} + \epsilon h) \Big|_{\epsilon = 0} \\
&= (\gamma(X, 1) - \gamma(X, 0)) \cdot \frac{\partial}{\partial \epsilon} \Big( w(X, 1; \nu_0, \mu_0 + \epsilon h) \cdot w(X, 0; \nu_0, \mu_0 + \epsilon h) \Big) \Big|_{\epsilon = 0} \\
&= \Big[ -2 \cdot (\gamma(X, 1) - \gamma(X, 0)) \cdot  \alpha(X; \nu_{0}, \mu_{0}) \cdot w(X, 1; \nu_0, \mu_0 ) \cdot w(X, 0; \nu_0, \mu_0) \Big] \cdot h(X).
\end{align*}
Thus,
\begin{align*}
u_{M}(X; \nu_0, \mu_0) = -2 \cdot (\gamma(X, 1) - \gamma(X, 0)) \cdot  \alpha(X; \nu_{0}, \mu_{0}) \cdot w(X, 1; \nu_0, \mu_0 ) \cdot w(X, 0; \nu_0, \mu_0).
\end{align*}
We can also conclude here that $\sup_{x \in \mathcal{X}} |u_{M}(x; \nu_0, \mu_0)|^{2}$ must be bounded by a constant.
\end{subsubsection}

\begin{subsubsection}{Second Claim}
We show that $\Var[S]{\dot{\psi}_{r, j}(X, Y; \nu_0, \mu_0, r_0) \mid X}$ is uniformly bounded. Recall the definition of $\psi$ from \eqref{eq:pop_estimating_eq}. We observe that
\begin{align*}
\dot{\psi}_{r}(X, Y; \nu_0, \mu_0, r_0) := \begin{bmatrix} \delta_{D_{L}}(X; \nu_0, \mu_0) \cdot (Y - \mu_0(X)) \\ \delta_{M}(X; \nu_0, \mu_0) \cdot (Y - \mu_0(X))\end{bmatrix}.
\end{align*}
For $j= 1, 2, \dots J$, we have that
\begin{align*}
\Var[S]{\dot{\psi}_{r, j}(X, Y; \nu_0, \mu_0, r_0) \mid X}  \leq |\delta_{D_{L}}(X; \nu_0, \mu_0)|^{2} \cdot \Var[S]{Y - \mu_0(X) \mid X}.
\end{align*}
By Assumption \ref{assumption:bounded}, $\eta(x, y), \gamma(x, y)$ are uniformly bounded. Thus, due to the compactness of $\mathcal{X}$, we can observe that $|\delta_{D_{L}}(X; \nu_0, \mu_0)|^{2}$ is bounded by a constant. As above, the conditional variance of $Y - \mu_0(X)$ is also bounded because $Y- \mu_0(X)$ is a bounded random variable.

Similarly, for $j=J+1, \dots J+K$, we have that
\begin{align*}
\Var[S]{\dot{\psi}_{r, j}(X, Y; \nu_0, \mu_0, r_0) \mid X}  \leq |\delta_{M}(X; \nu_0, \mu_0)|^{2} \cdot \Var[S]{Y - \mu_0(X) \mid X}.
\end{align*}
We also have that $|\delta_{M}(X; \nu_0, \mu_0)|^{2}$ is bounded by a constant. Thus, the conditional variance must be uniformly bounded.
\end{subsubsection}

\begin{subsubsection}{Third Claim}
We note that
\[ \dot{\psi}_{\nu}(X, Y; \nu, \mu, r) = \begin{bmatrix} \dot{\psi}_{\theta}(X, Y; \nu, \mu, r) \\ \dot{\psi}_{\lambda}(X, Y; \nu, \mu, r) \end{bmatrix}. \]
Our first goal is to bound the conditional variance $\Var[S]{\ddot{\psi}_{\theta \mu, ij}(X, Y; \nu_0, \mu_0, r_0)}$. There exists functions $u_{D_{L}}: \mathcal{X} \rightarrow \mathbb{R}^{J \times J}, u_{M}: \mathcal{X} \rightarrow \mathbb{R}^{J \times K}$ such that 
\begin{align*}
\frac{\partial}{\partial \epsilon } \frac{\partial}{\partial \theta} \delta_{D_{L}}(x; \nu_{0}, \mu_{0} + \epsilon h) \Big|_{\epsilon = 0} &= u_{D_{L}}(x; \nu_0, \mu_0) \odot h(x),  \\
\frac{\partial}{\partial \epsilon } \frac{\partial}{\partial \theta} \delta_{M}(x; \nu_{0}, \mu_{0} + \epsilon h) \Big|_{\epsilon = 0} &= u_{M}(x; \nu_0, \mu_0) \odot h(x).
\end{align*}

As in the proof of the first claim (Section \ref{sec:claim1}), we note that $\ddot{\psi}_{\theta \mu}$ can be decomposed as follows
\begin{align*}
\ddot{\psi}_{\theta \mu}(X, Y; \nu_0, \mu_0, r_0) &= \begin{bmatrix} f_{D_{L}}(X; \nu_0, \mu_0, r_0) + r_0(X) \cdot u_{D_{L}}(X; \nu_0, \mu_0) \cdot (Y - \mu_0(X))  \\ 
f_{M}(X; \nu_0, \mu_0, r_0) + r_0(X) \cdot u_{M}(X; \nu_0, \mu_0) \cdot (Y - \mu_0(X)) \\
\end{bmatrix},
\end{align*}
where the first term in the decomposition does not affect the conditional variance of $\dot{\psi}_{\theta \mu}$ because it only depends on $X$. Then, bounding the conditional variance reduces to bounding $\sup_{x \in \mathcal{X}} |u_{D_{L}}(x; \nu_0, \mu_0)|^{2}, \sup_{x \in \mathcal{X}} |u_{M}(x; \nu_0, \mu_0)|^{2}$. In order to show that these terms are bounded, we show that $\sup_{x \in \mathcal{X}} |u_{D_{L}}(x)|^{2}$ can be written as a product of bounded functions. Define the following
\begin{align*}
d_{\eta} &:= \eta(X, 1) - \eta(X, 0), \\
d_{\gamma} &= \gamma(X, 1) - \gamma(X, 0), \\
q(X, 1; \nu, \mu) &:= \mu(X) \cdot w(X, 1; \nu, \mu), \\
q(X, 0; \nu, \mu) &:= (1 - \mu(X)) \cdot w(X, 0; \nu, \mu).
\end{align*}

Note that
\begin{align*}
\rho(X, 1; \nu, \mu) - \rho(X, 0; \nu, \mu) &= - (q(X, 1; \nu, \mu)^{2} -  q(X, 0; \nu, \mu)^{2}) \cdot (d_{\eta}^{\otimes 2} \cdot \theta + d_{\eta} \otimes d_{\gamma} \cdot \lambda).
\end{align*}
We observe that
\begin{align*}
&\frac{\partial}{\partial \epsilon} q(X, 1; \nu_0, \mu_0 + \epsilon h) \Big|_{\epsilon = 0} \\
&=\Big(w(X, 1; \nu_0, \mu_0) - \mu_0(X) \cdot w(X, 1; \nu_0, \mu_0) \cdot w(X, 0; \nu_0, \mu_0) \cdot (w(X, 1; \nu_0, \mu_0) - w(X, 0; \nu_0, \mu_0))\Big) \cdot h(X).
\end{align*}
The first term in the product above is bounded under Assumption \ref{assumption:bounded}. Similarly, we can write that $\frac{\partial}{\partial \epsilon} \rho(X, 1; \nu_0, \mu_0 + \epsilon h) - \rho(X, 0; \nu_0, \mu_0 + \epsilon h) \Big|_{\epsilon = 0} = g_{1}(X; \nu_0, \mu_0) \cdot h(X)$ where $\sup_{x \in \mathcal{X}} |g_{1}(X; \nu_0, \mu_0)|^{2}$ is uniformly bounded.

We also note that
\begin{align*}
&\frac{\partial}{\partial \theta} \Big( \rho(X, 1; \nu, \mu) - \rho(X, 0; \nu, \mu) \Big) \\
&= - (q(X, 1; \nu, \mu)^{2} -  q(X, 0; \nu, \mu)^{2}) \cdot d_{\eta}^{\otimes 2} - \frac{\partial}{\partial \theta} (q(X, 1; \nu, \mu)^{2} -  q(X, 0; \nu, \mu)^{2}) \cdot d_{\eta}^{\otimes 2} \theta \\
&= - (q(X, 1; \theta, \mu)^{2} -  q(X, 0; \theta, \mu)^{2}) \cdot d_{\eta}^{\otimes 2} \\
&\indent - 2 (w(X, 1; \nu, \mu) \cdot w(X, 0; \nu, \mu)) \cdot ( \mu(X)^{2} \cdot w(X, 1; \nu, \mu) + (1 - \mu(X))^{2} \cdot w(X, 0; \nu, \mu)) \cdot d_{\eta} \otimes (d_{\eta}^{\otimes 2} \cdot \theta + d_{\eta} \otimes d_{\gamma} \cdot \lambda).
\end{align*}
Again, applying the chain rule yields that $\frac{\partial^{2}}{\partial \epsilon \partial \theta} \Big( \rho(X, 1; \nu, \mu) - \rho(X, 0; \nu, \mu) \Big) \Big|_{\epsilon = 0} = g_{2}(X; \nu_0, \mu_0) \odot h(X)$, where $\sup_{x \in \mathcal{X}} |g_{2}(x; \nu_0, \mu_0)|^{2}$ is uniformly bounded. We observe that
\begin{align*}
&\frac{\partial}{\partial \theta} \delta_{D_{L}}(X; \nu, \mu) = \frac{\partial}{\partial \theta} \Big( \rho(X, 1; \nu, \mu) - \rho(X, 0; \nu, \mu) \Big) \cdot w(X,1; \nu, \mu) \cdot w(X, 0; \nu, \mu) \\
&\indent- (\rho(X, 1; \nu, \mu) - \rho(X, 0; \nu, \mu)) \cdot w(X,1; \nu, \mu) \cdot w(X, 0; \nu, \mu) \cdot \alpha(X; \nu, \mu).
\end{align*}
Thus, $\frac{\partial}{\partial \theta} \delta_{D_{L}}(X; \nu, \mu)$ also yields that $u_{D_{L}}(x)$ can also be uniformly bounded. Next, we observe that 
\begin{align*}
&\frac{\partial^{2}}{\partial \epsilon \partial \theta} \delta_{M}(X; \nu_0, \mu_0 + \epsilon h) \Big|_{\epsilon = 0} \\
&= - d_{\gamma} \otimes d_{\eta} \cdot \frac{\partial}{\partial \epsilon} w(X, 1; \nu_0, \mu_0 + \epsilon h) \cdot w(X, 0; \nu_0, \mu_0 + \epsilon h) \cdot \alpha(X; \nu_0, \mu_0 + \epsilon h) \Big|_{\epsilon = 0} \\
&= - d_{\gamma} \otimes d_{\eta} \cdot w(X, 1; \nu_0, \mu_0) \cdot w(X, 0; \nu_0, \mu_0) \cdot \alpha(X; \nu_0, \mu_0) \\
&\indent \cdot (1 - 6 w(X, 1; \nu_0, \mu_0) \cdot w(X, 0; \nu_0, \mu_0)) \cdot h(X).
\end{align*}
This implies that 
\begin{align*}
u_{M}(x) &=  - d_{\gamma} \otimes d_{\eta} \cdot w(x, 1; \nu_0, \mu_0) \cdot w(x, 0; \nu_0, \mu_0) \cdot \alpha(x; \nu_0, \mu_0) \\
&\indent \cdot (1 - 6 w(x, 1; \nu_0, \mu_0) \cdot w(x, 0; \nu_0, \mu_0)).
\end{align*}

Thus, $\sup_{x \in \mathcal{X}} |u_{M}(x)|^{2}$ is uniformly bounded. We can apply an identical proof strategy to show that the conditional variance of $\ddot{\psi}_{\theta r}(X, Y; \nu_0, \mu_0, r_0)$, $\ddot{\psi}_{\lambda \mu}(X, Y; \nu_0, \mu_0, r_0),$ and $\ddot{\psi}_{\lambda r}(X, Y; \nu_0, \mu_0, r_0)$ are all bounded.
\end{subsubsection}
\end{subsection}
\end{section}

\begin{section}{Additional Exhibits}
\label{app:additional_exhibits}
We visualize predictions of our data fusion approach and all baselines for all indicators and all time-steps.

\begin{figure}[b]
\caption{COVID-19 Vaccination predictions over time.}
\begin{subfigure}{0.48\textwidth}
    \includegraphics[width=\linewidth]{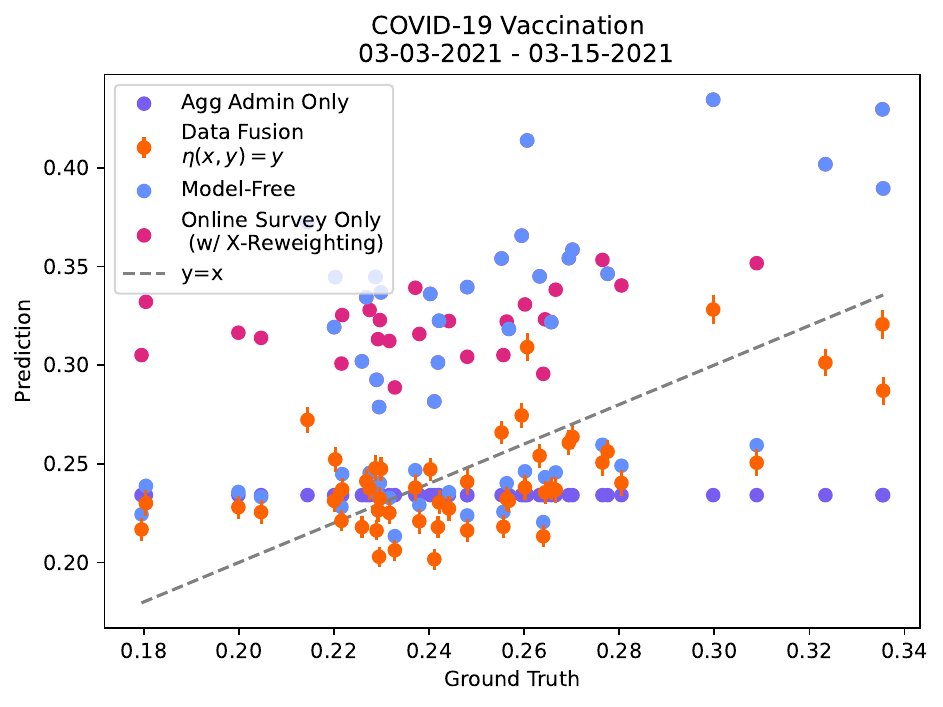}
\end{subfigure}%
\begin{subfigure}{0.48\textwidth}
    \includegraphics[width=\linewidth]{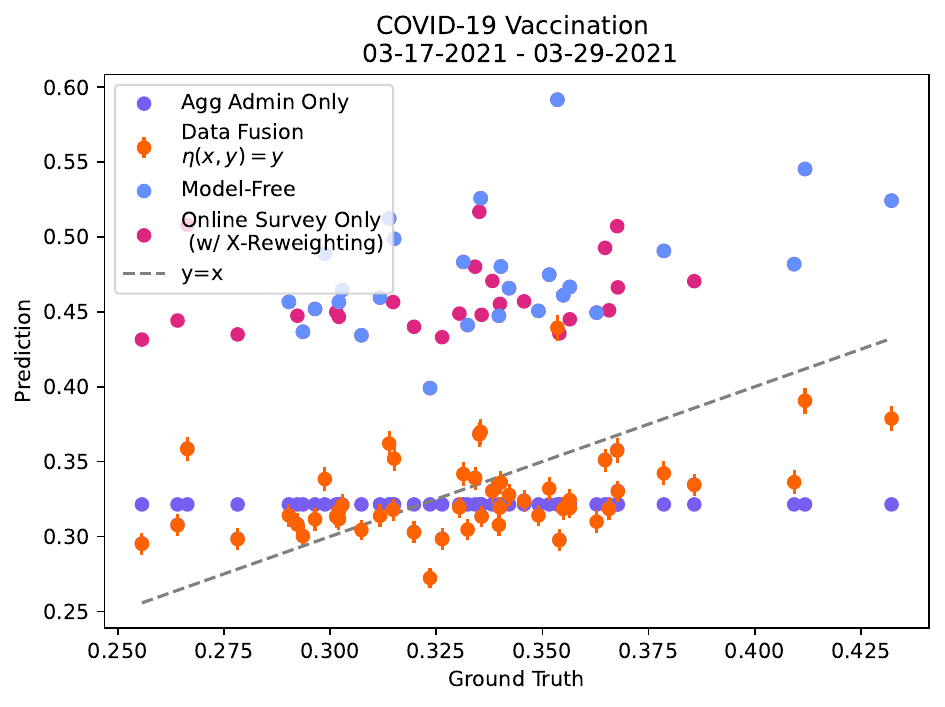}
\end{subfigure}
\begin{subfigure}{0.48\textwidth}
    \includegraphics[width=\linewidth]{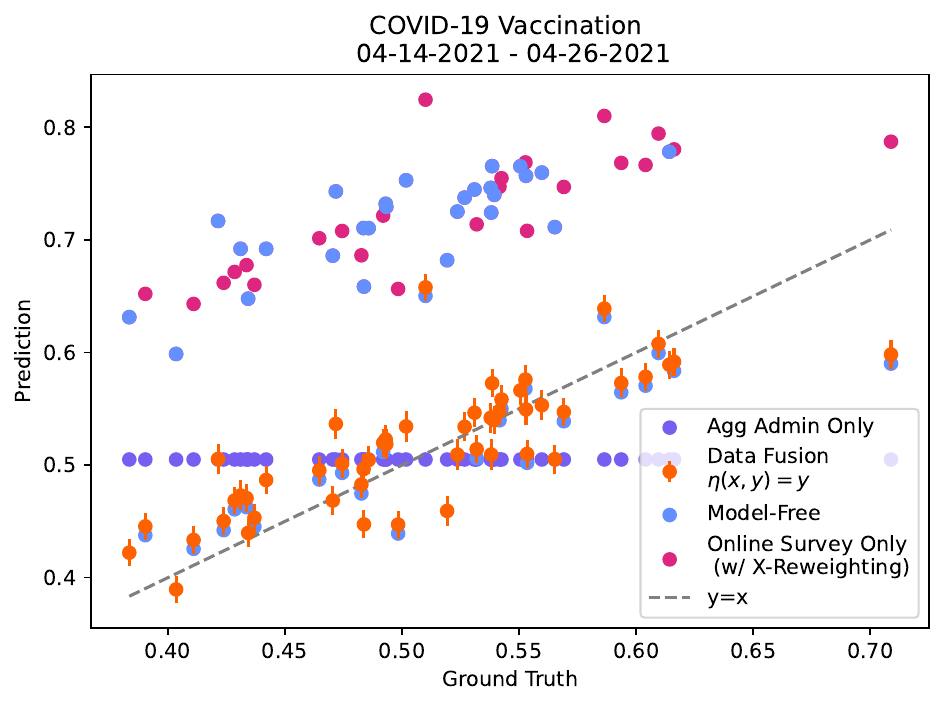}
\end{subfigure}%
\begin{subfigure}{0.48\textwidth}
    \includegraphics[width=\linewidth]{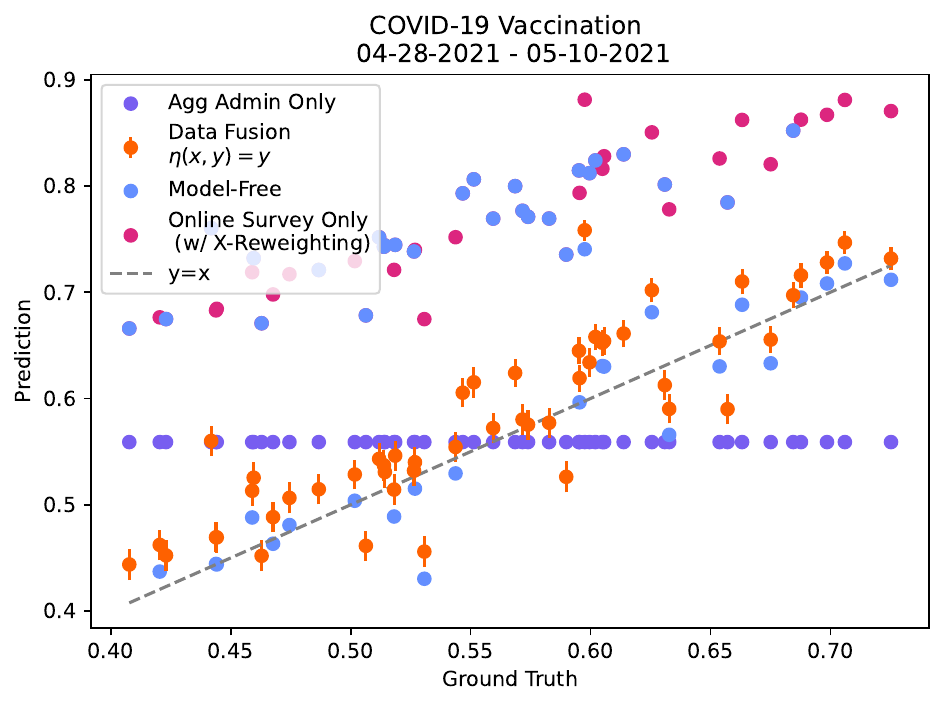}
\end{subfigure}
\begin{subfigure}{0.48\textwidth}
    \includegraphics[width=\linewidth]{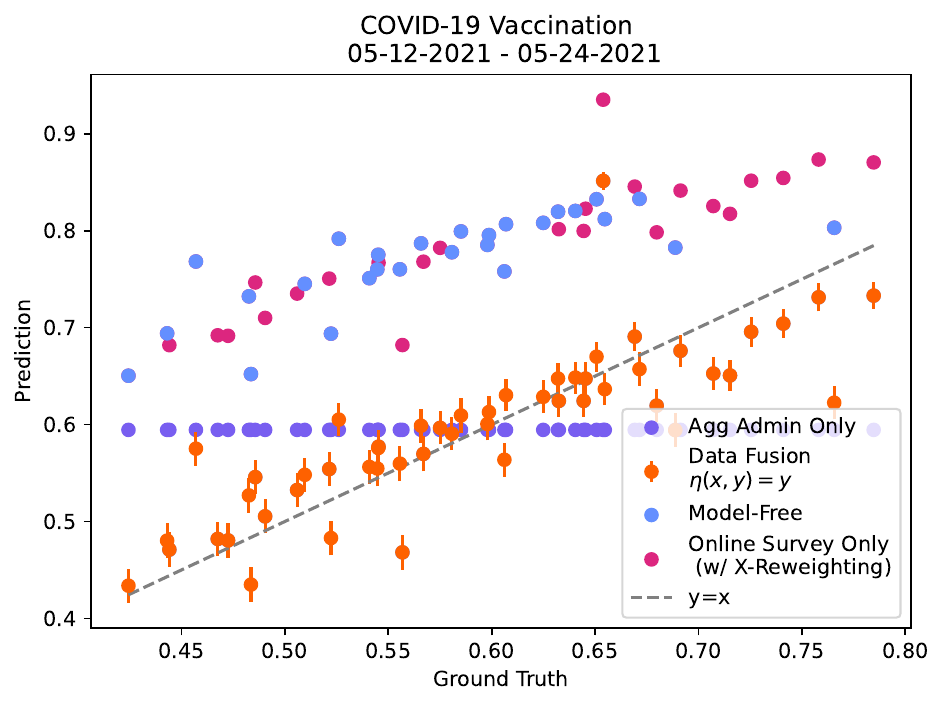}
\end{subfigure}%
\begin{subfigure}{0.48\textwidth}
    \centering
    \includegraphics[width=\linewidth]{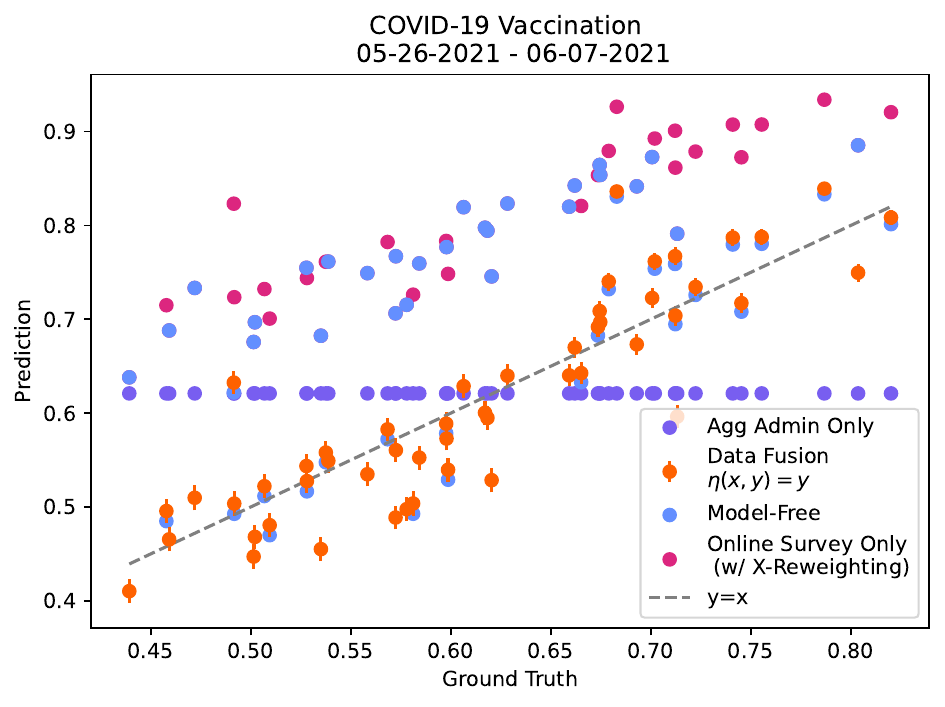}
\end{subfigure}
\end{figure}

\begin{figure}\ContinuedFloat
\begin{subfigure}{0.48\textwidth}
    \includegraphics[width=\linewidth]{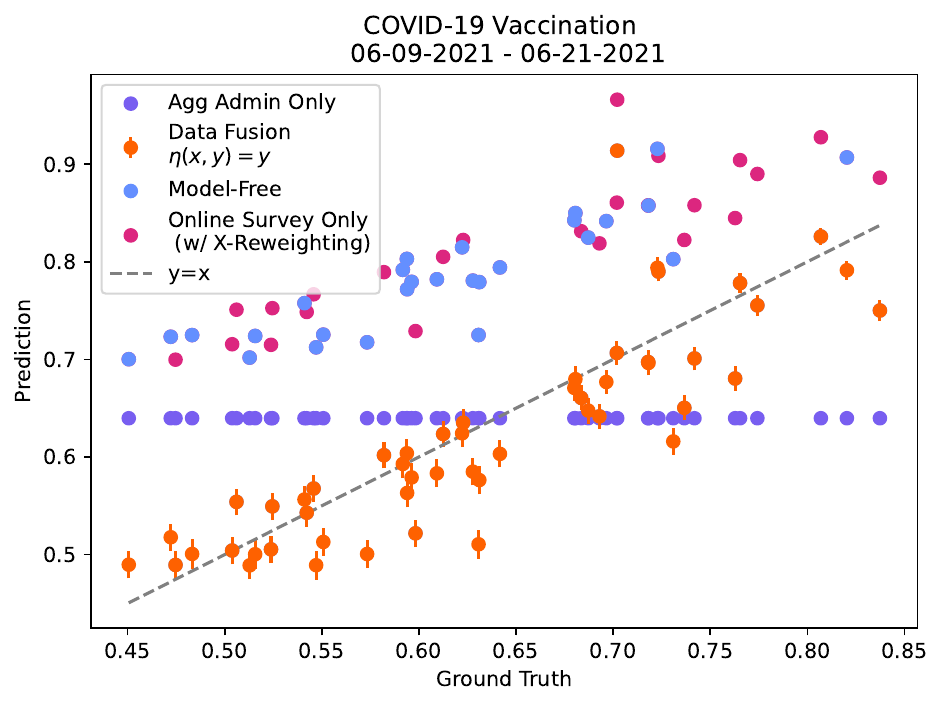}
\end{subfigure}%
    \begin{subfigure}{0.48\textwidth}
    \includegraphics[width=\linewidth]{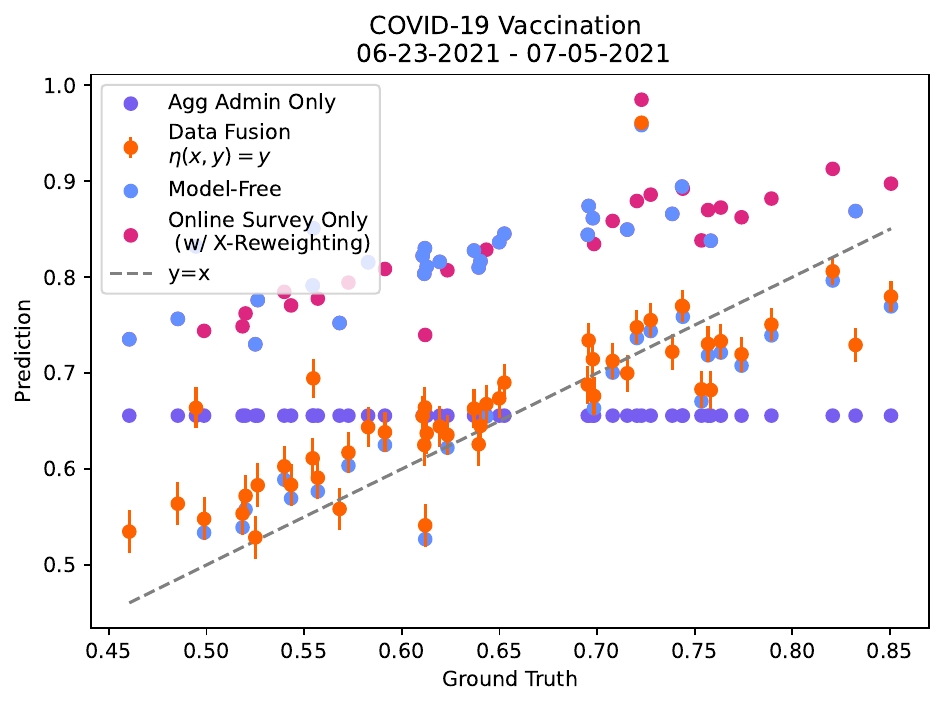}
\end{subfigure}
\begin{subfigure}{0.48\textwidth}
    \centering
    \includegraphics[width=\linewidth]{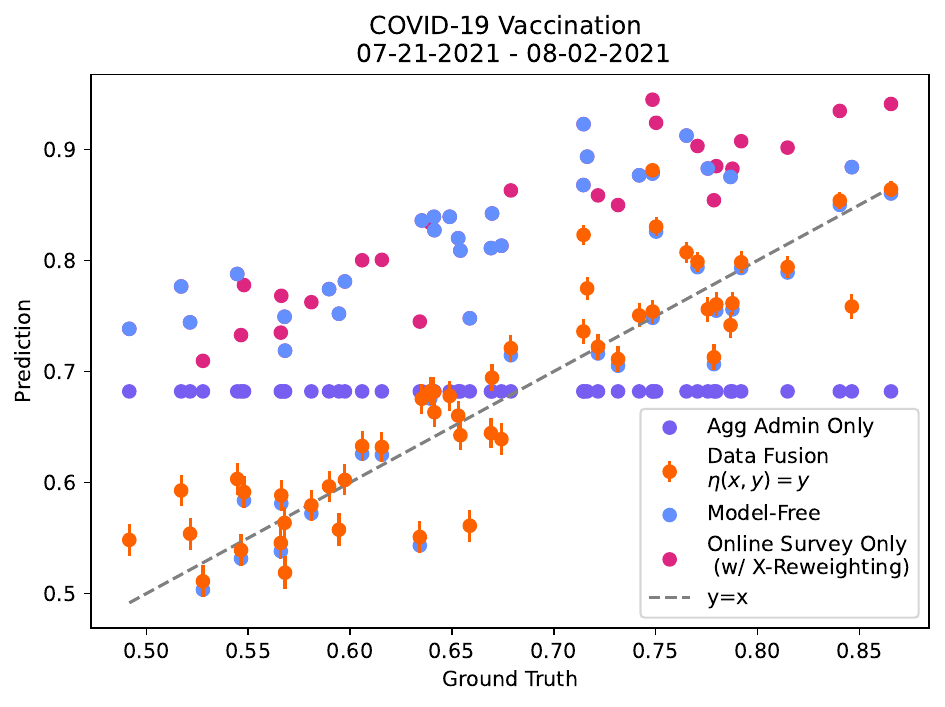}
\end{subfigure}%
\begin{subfigure}{0.48\textwidth}
    \centering
    \includegraphics[width=\linewidth]{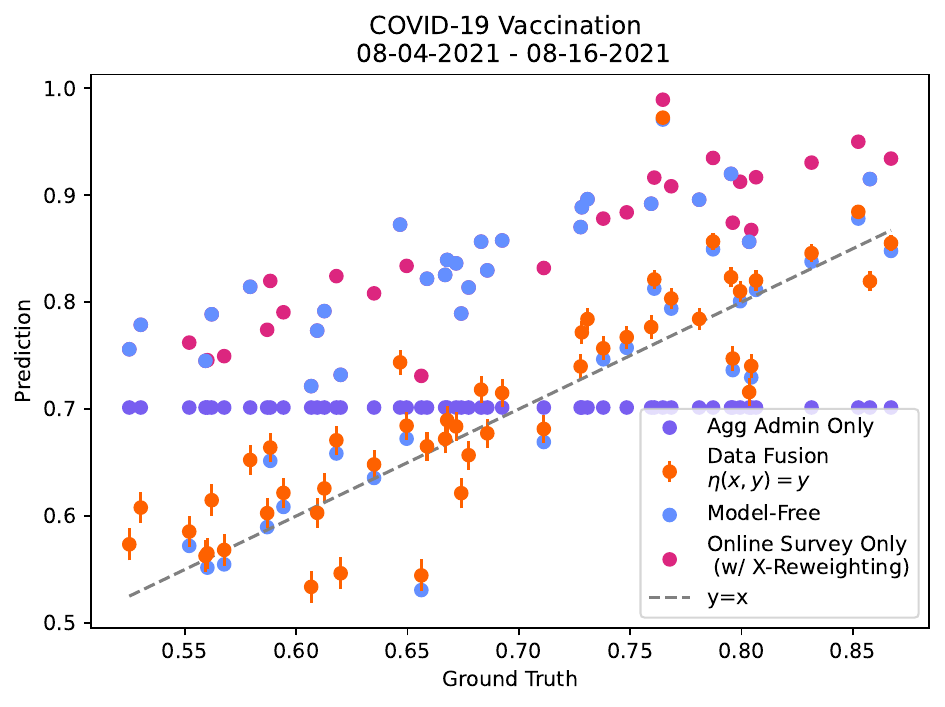}
\end{subfigure}
\begin{subfigure}{0.48\textwidth}
    \centering
    \includegraphics[width=\linewidth]{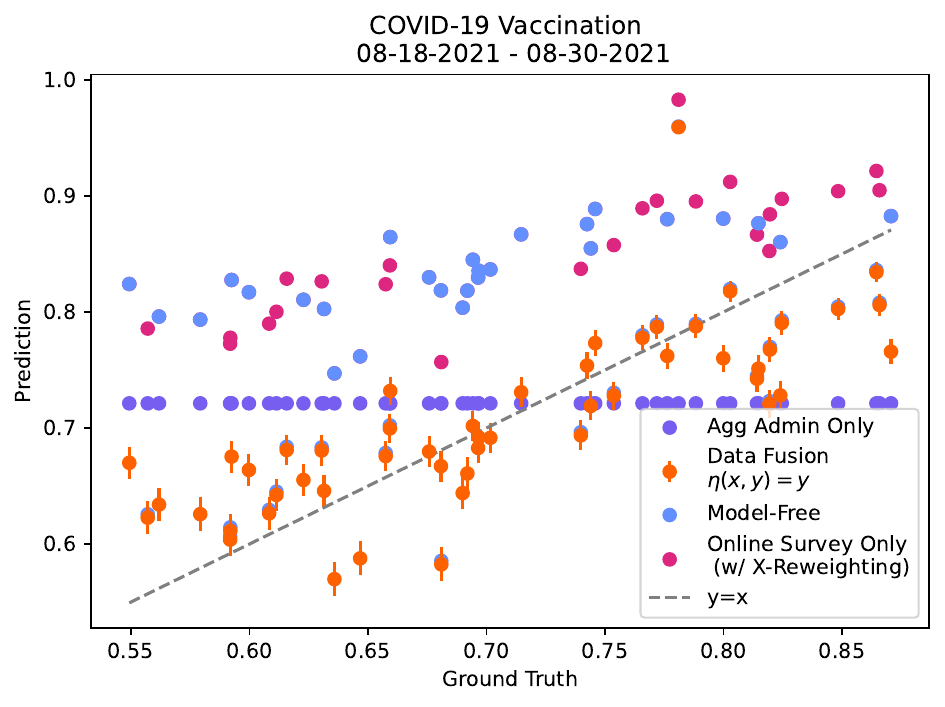}
\end{subfigure}%
\begin{subfigure}{0.48\textwidth}
    \centering
    \includegraphics[width=\linewidth]{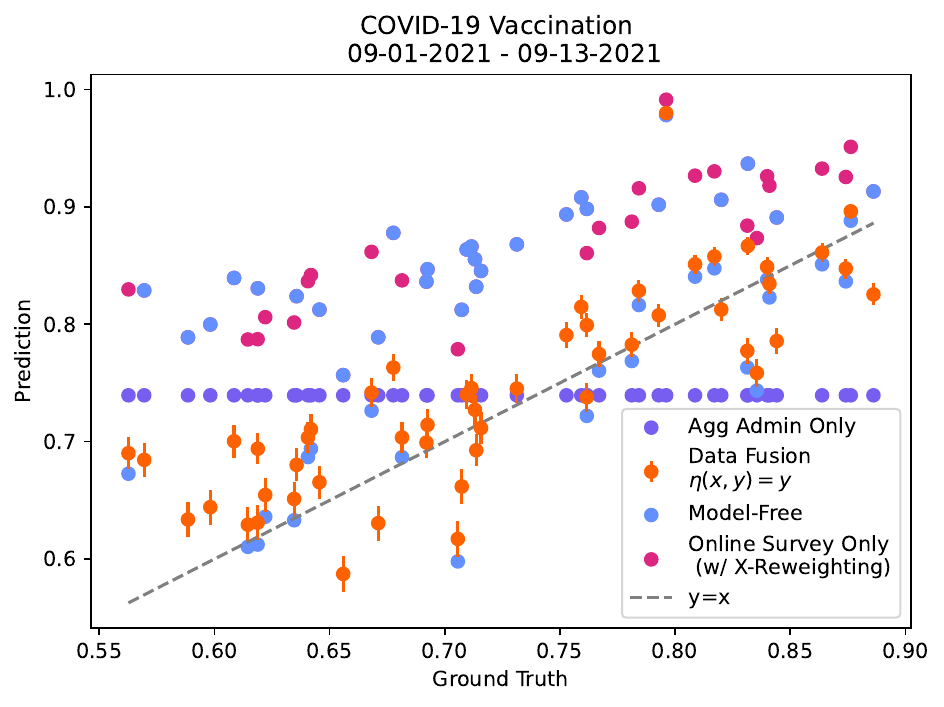}
\end{subfigure}
\end{figure}

\begin{figure}\ContinuedFloat
\begin{subfigure}{0.48\textwidth}
    \centering
    \includegraphics[width=\linewidth]{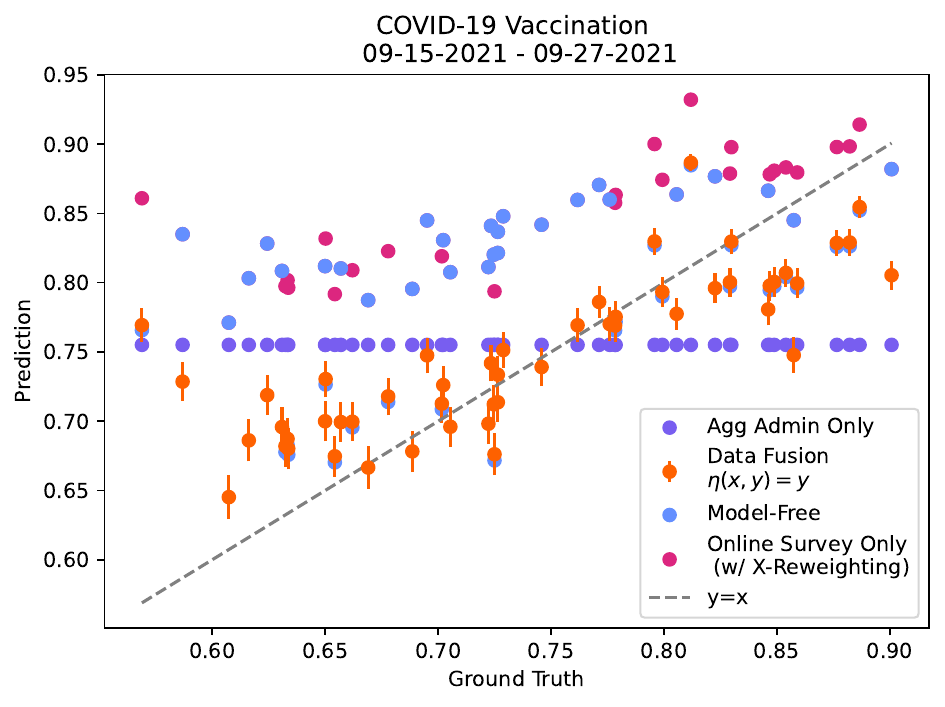}
\end{subfigure}%
\begin{subfigure}{0.48\textwidth}
    \centering
    \includegraphics[width=\linewidth]{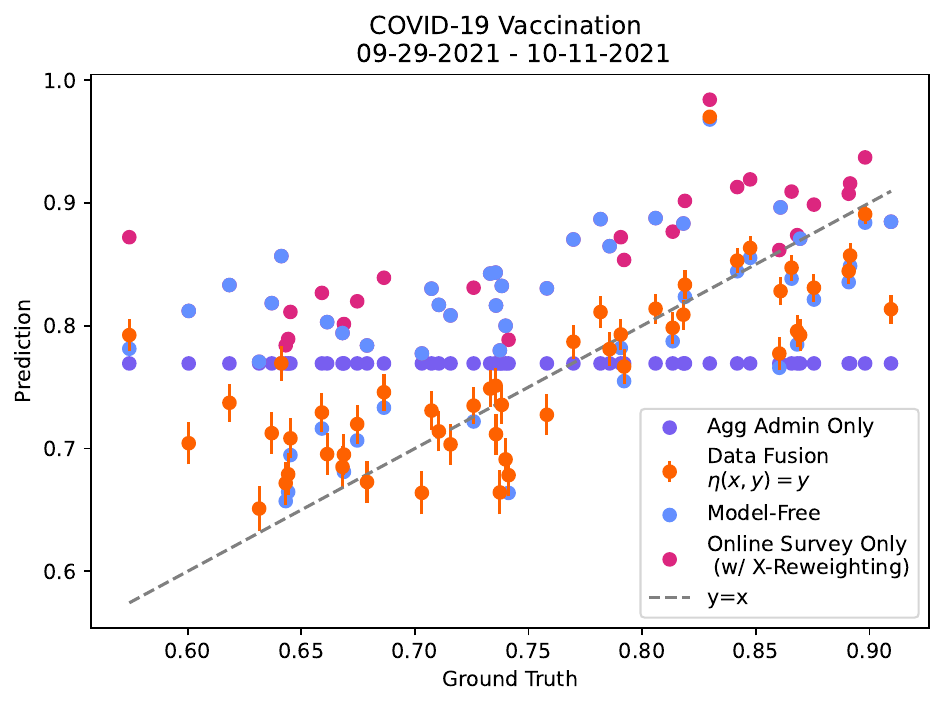}
\end{subfigure}%
\end{figure}

\begin{figure}
\caption{Medicaid Insurance coverage predictions over time.}
\begin{subfigure}{0.48\textwidth}
    \centering
    \includegraphics[width=\linewidth]{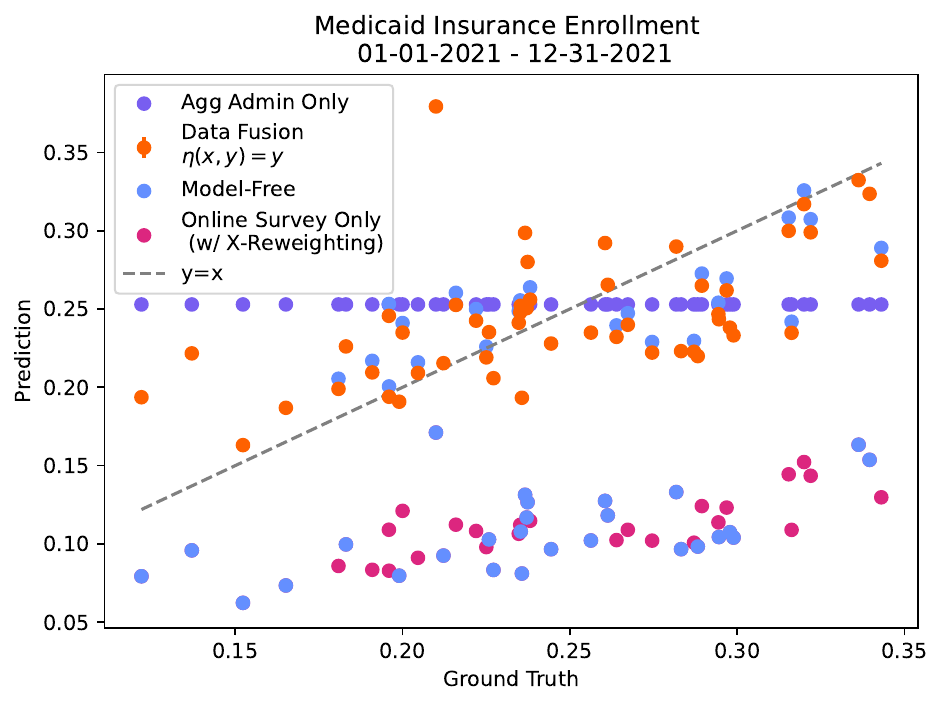}
\end{subfigure}%
\begin{subfigure}{0.48\textwidth}
    \centering
    \includegraphics[width=\linewidth]{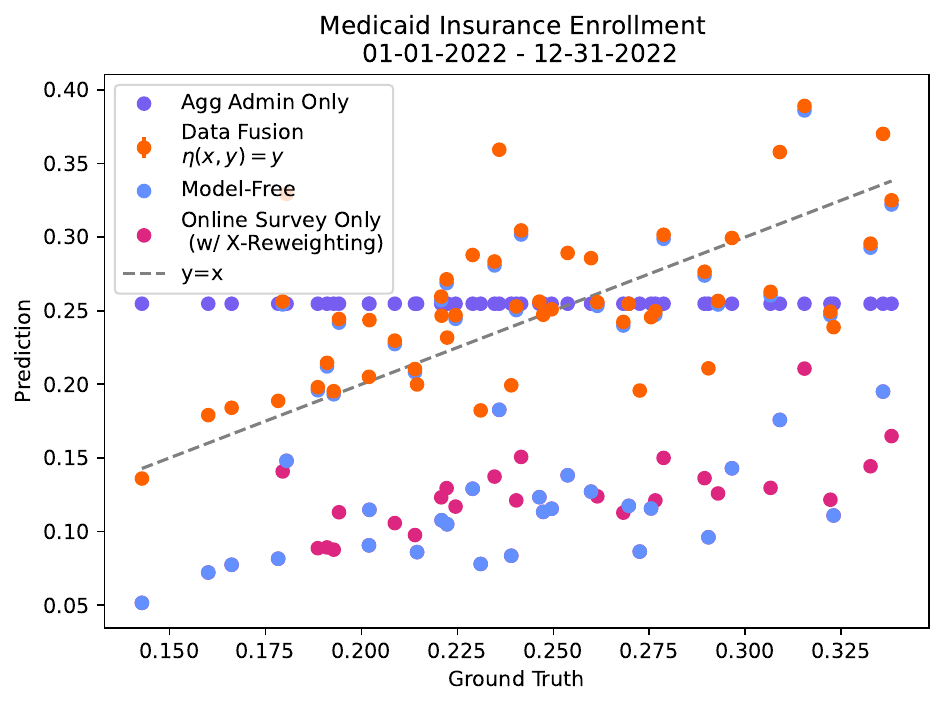}
\end{subfigure}
\end{figure}

\begin{figure}
\caption{SNAP participation over time.}
\begin{subfigure}{0.48\textwidth}
    \includegraphics[width=\linewidth]{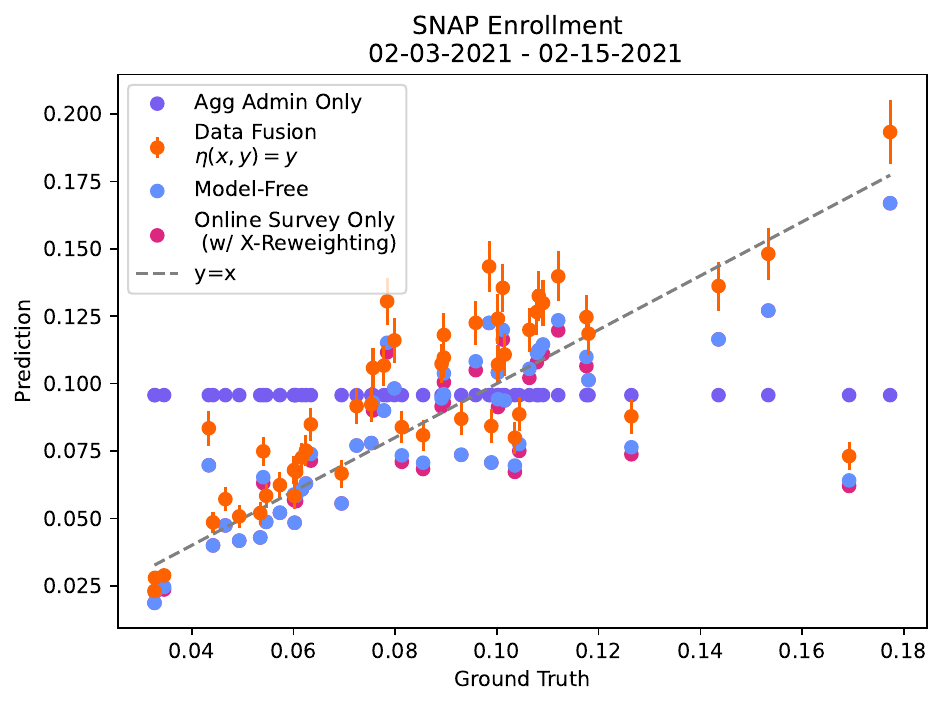}
\end{subfigure}%
\begin{subfigure}{0.48\textwidth}
    \includegraphics[width=\linewidth]{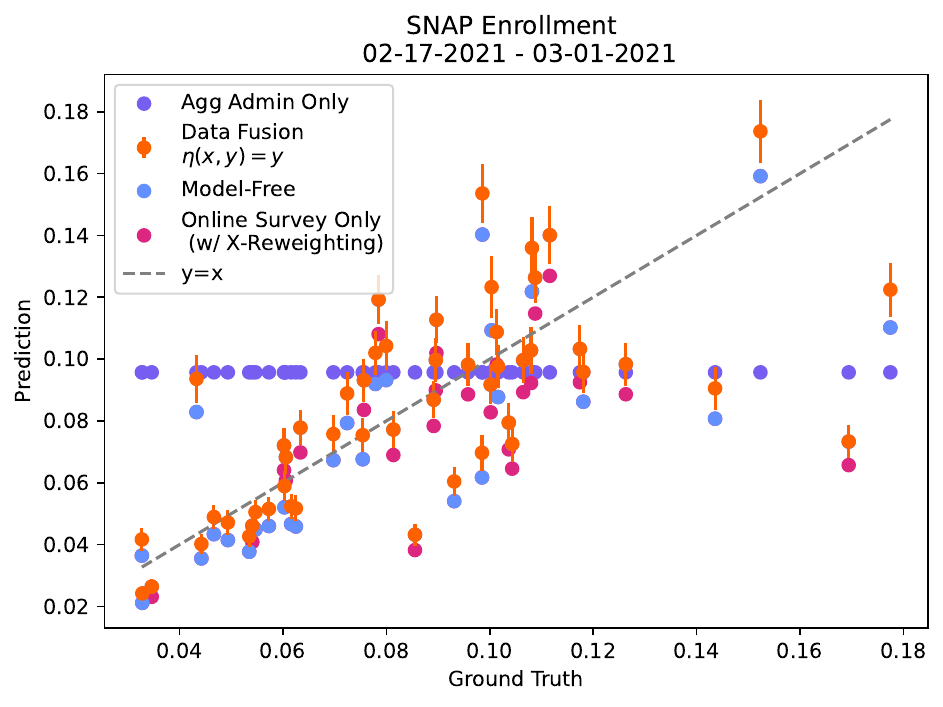}
\end{subfigure}
\begin{subfigure}{0.48\textwidth}
    \includegraphics[width=\linewidth]{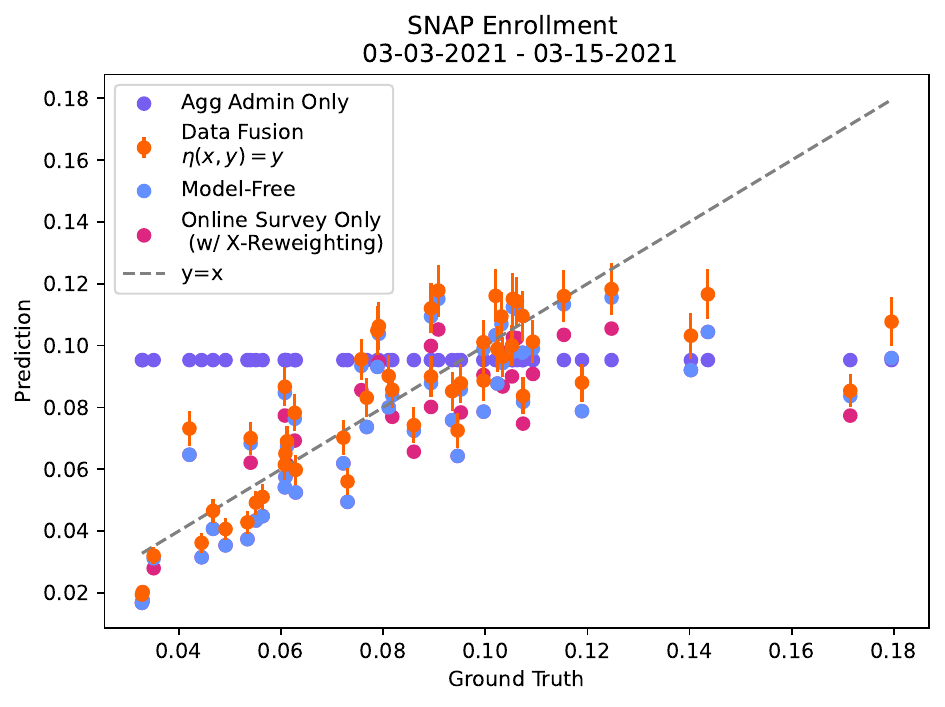}
\end{subfigure}%
\begin{subfigure}{0.48\textwidth}
    \includegraphics[width=\linewidth]{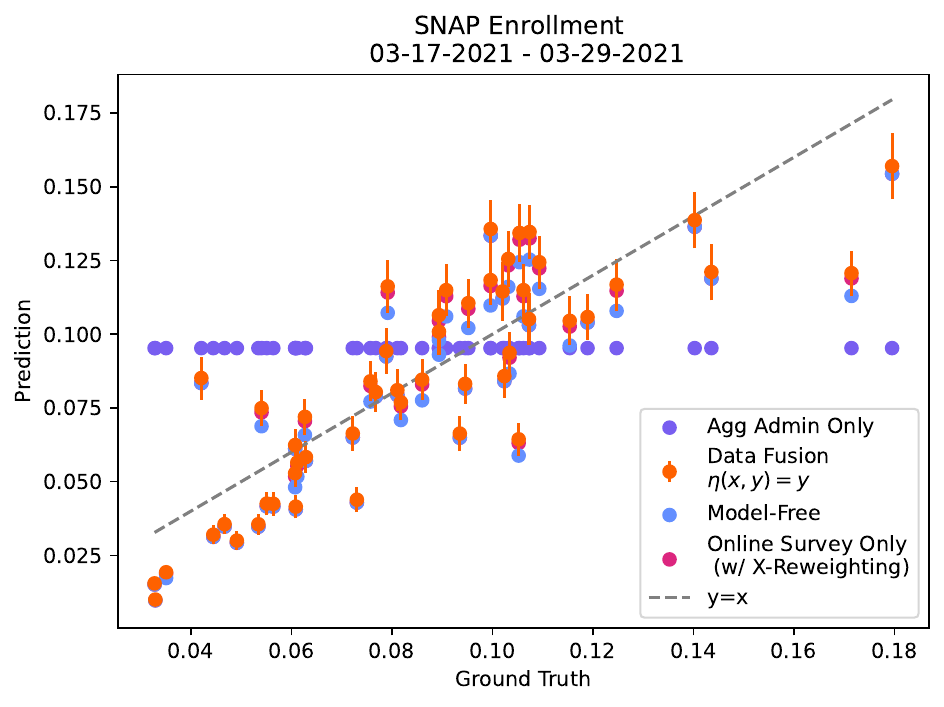}
\end{subfigure}
\begin{subfigure}{0.48\textwidth}
    \includegraphics[width=\linewidth]{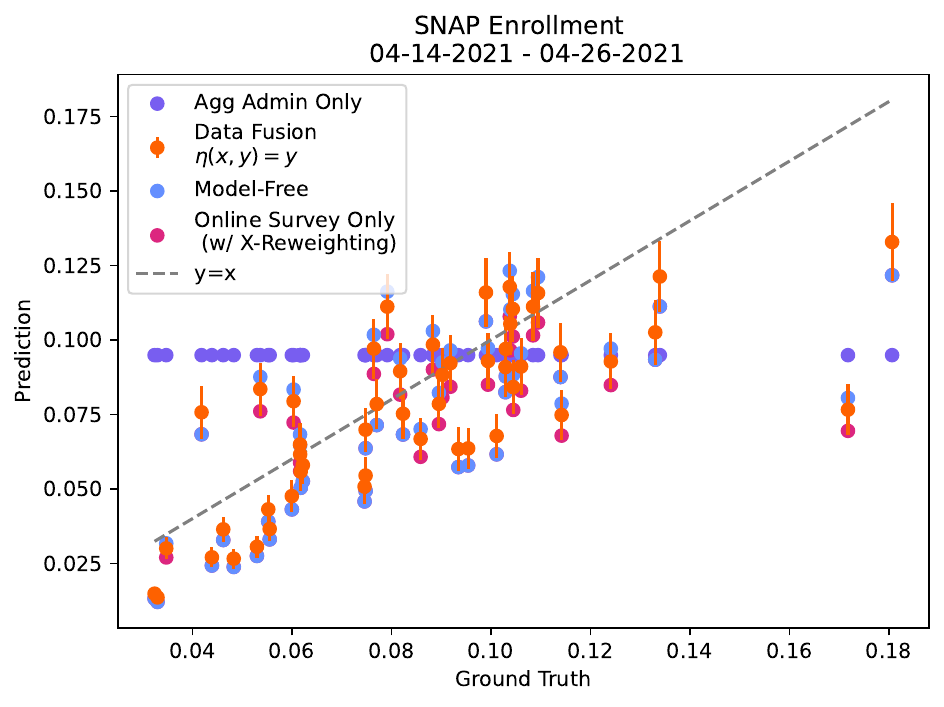}
\end{subfigure}%
\begin{subfigure}{0.48\textwidth}
\includegraphics[width=\linewidth]{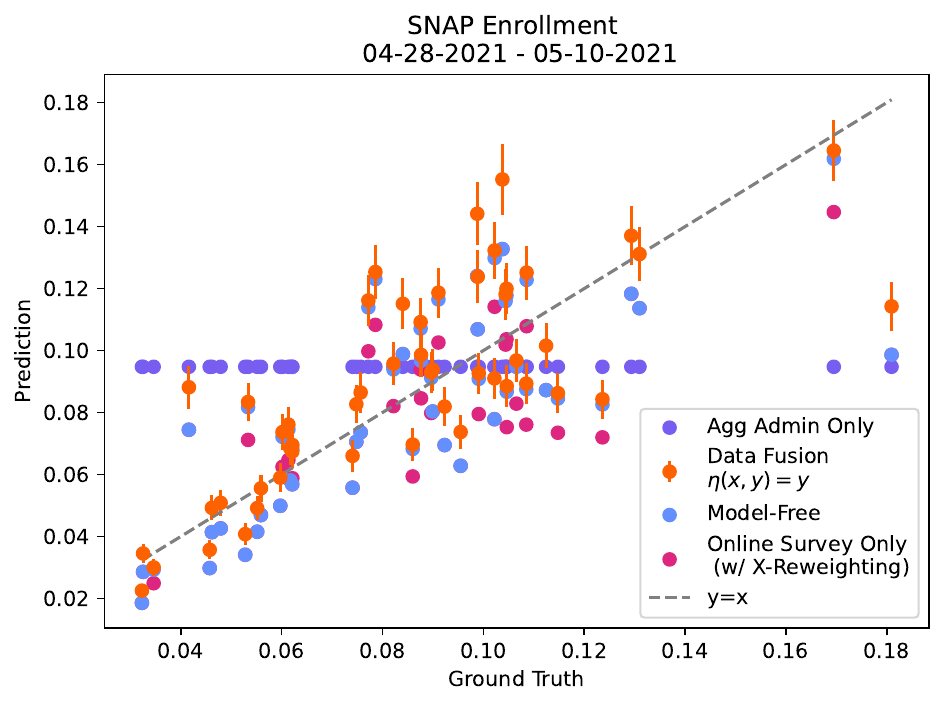}
\end{subfigure}
\end{figure}

\begin{figure}\ContinuedFloat
\begin{subfigure}{0.48\textwidth}
    \includegraphics[width=\linewidth]{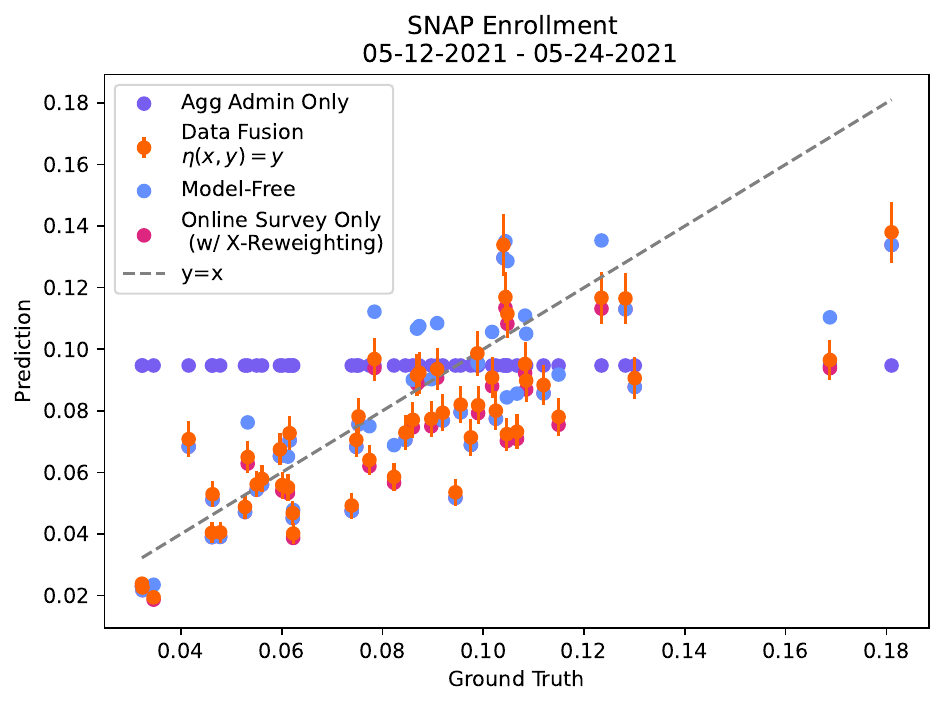}
\end{subfigure}%
\begin{subfigure}{0.48\textwidth}
    \centering
    \includegraphics[width=\linewidth]{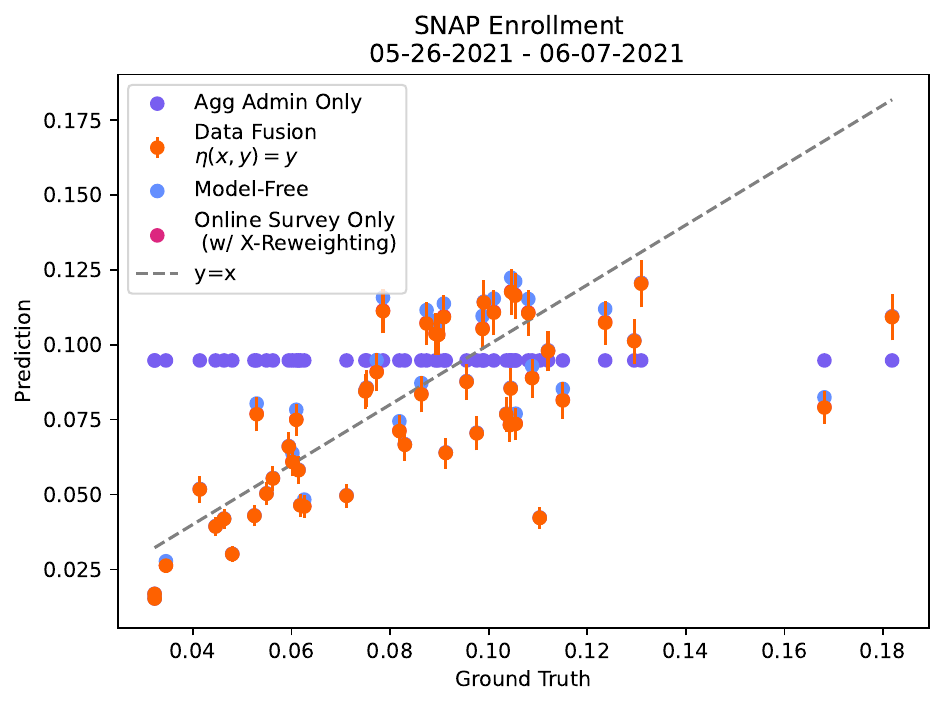}
\end{subfigure}
\begin{subfigure}{0.48\textwidth}
    \centering
    \includegraphics[width=\linewidth]{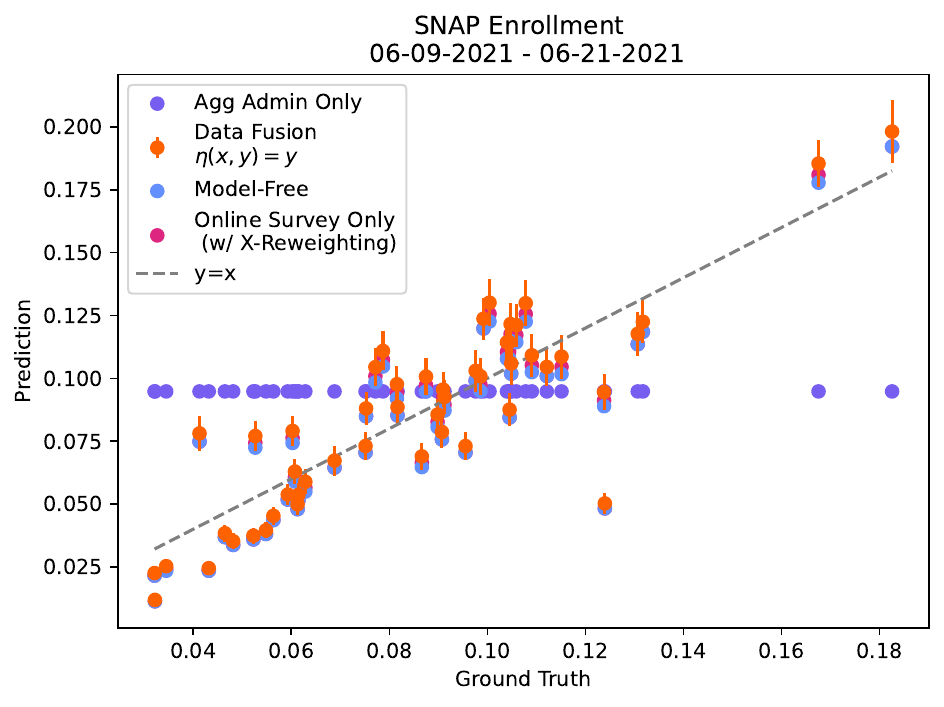}
\end{subfigure}%
\begin{subfigure}{0.48\textwidth}
    \centering
    \includegraphics[width=\linewidth]{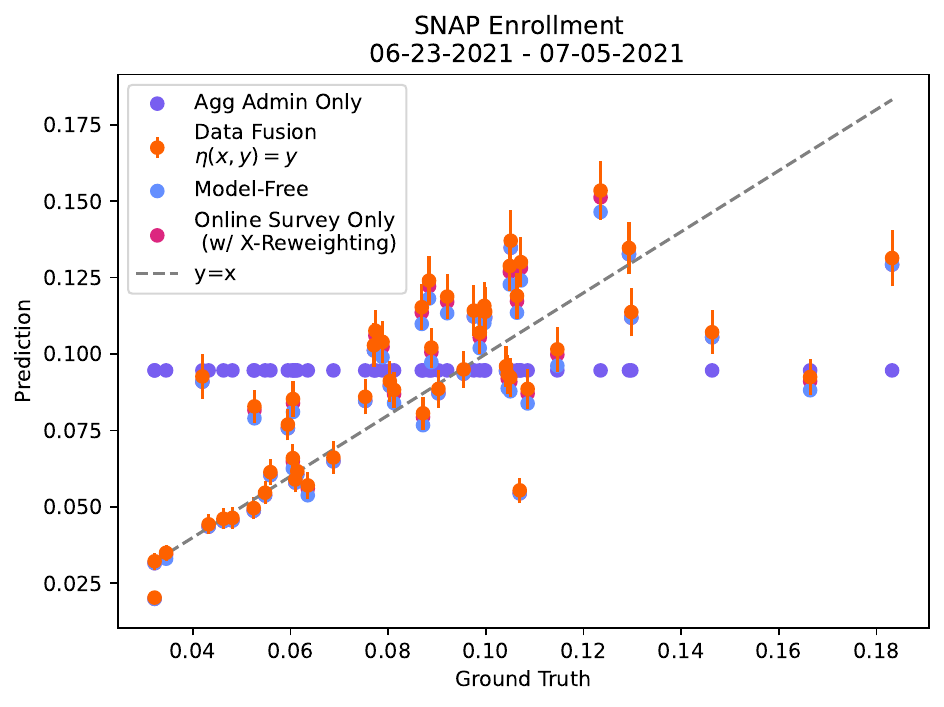}
\end{subfigure}
\begin{subfigure}{0.48\textwidth}
    \centering
    \includegraphics[width=\linewidth]{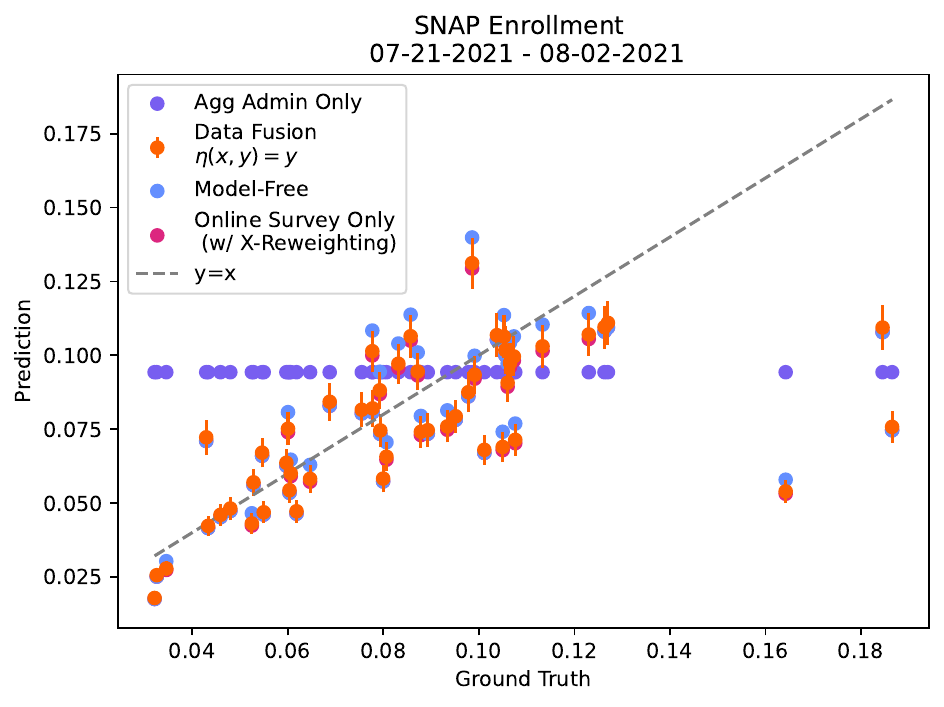}
\end{subfigure}%
\begin{subfigure}{0.48\textwidth}
    \centering
    \includegraphics[width=\linewidth]{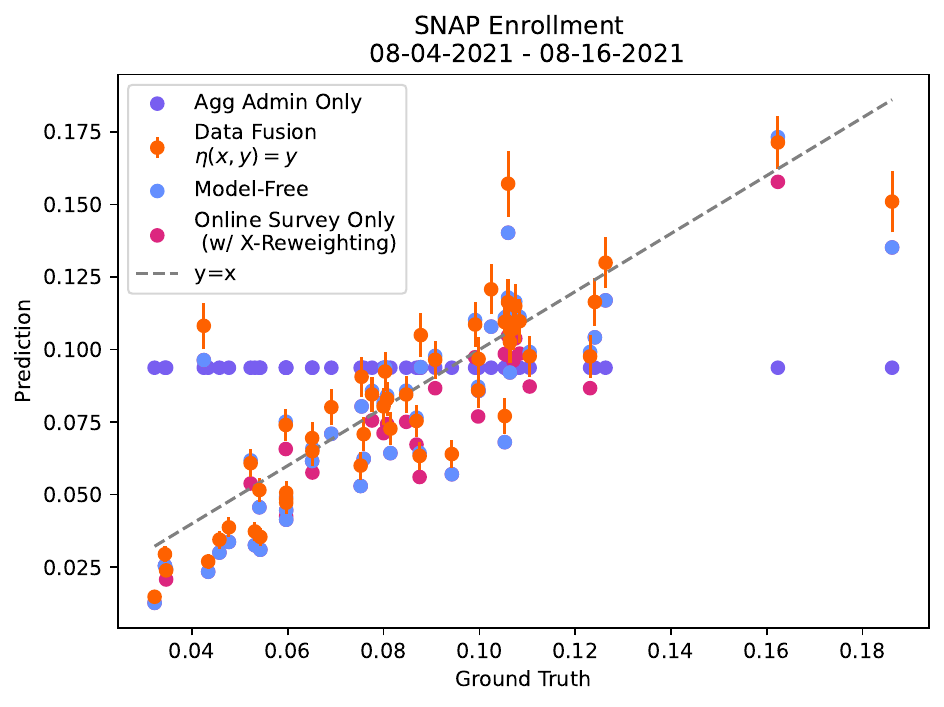}
\end{subfigure}
\end{figure}

\begin{figure}\ContinuedFloat
\begin{subfigure}{0.48\textwidth}
    \centering
    \includegraphics[width=\linewidth]{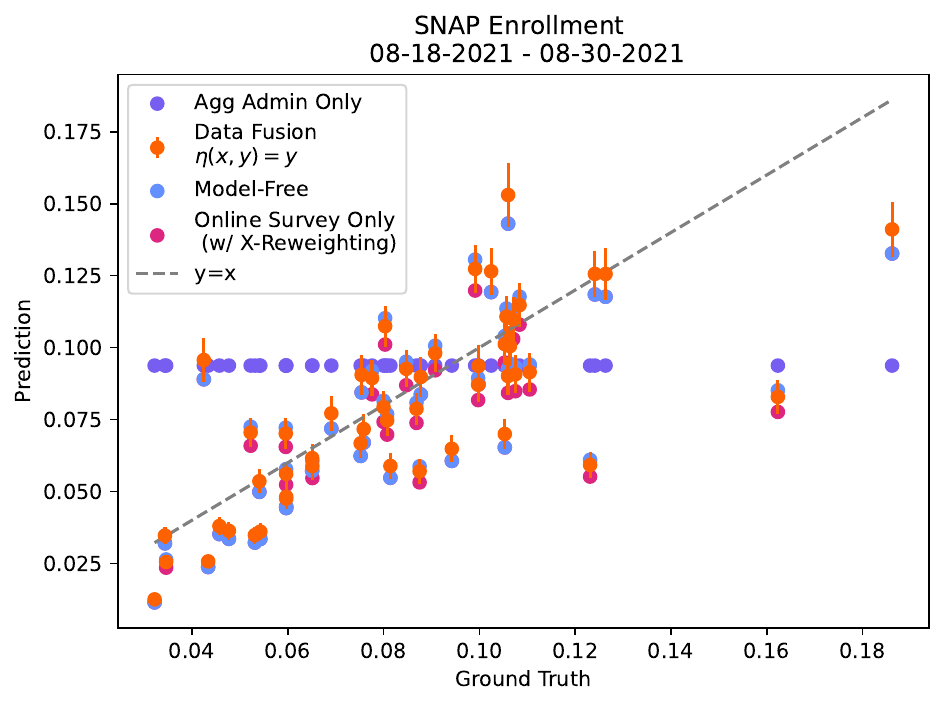}
\end{subfigure}%
\begin{subfigure}{0.48\textwidth}
    \centering
    \includegraphics[width=\linewidth]{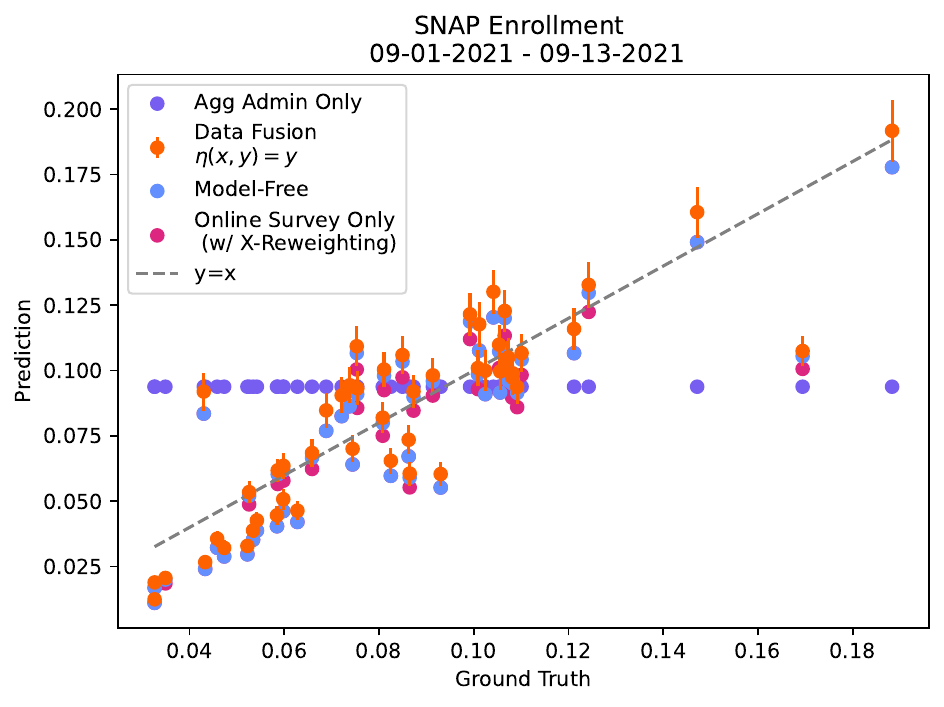}
\end{subfigure}
\begin{subfigure}{0.48\textwidth}
    \centering
    \includegraphics[width=\linewidth]{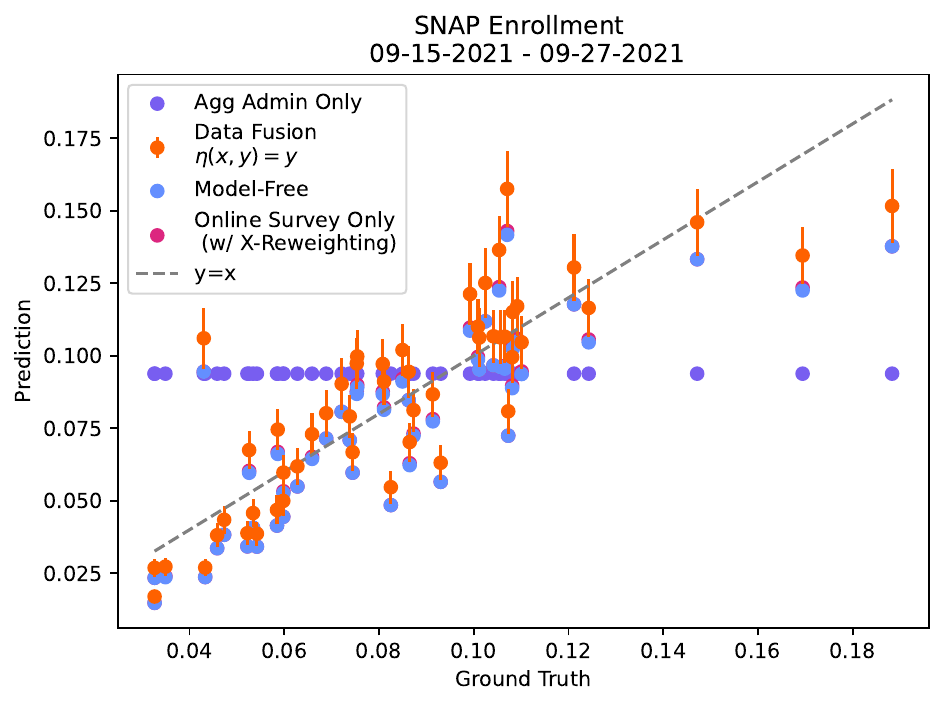}
\end{subfigure}%
\begin{subfigure}{0.48\textwidth}
    \centering
    \includegraphics[width=\linewidth]{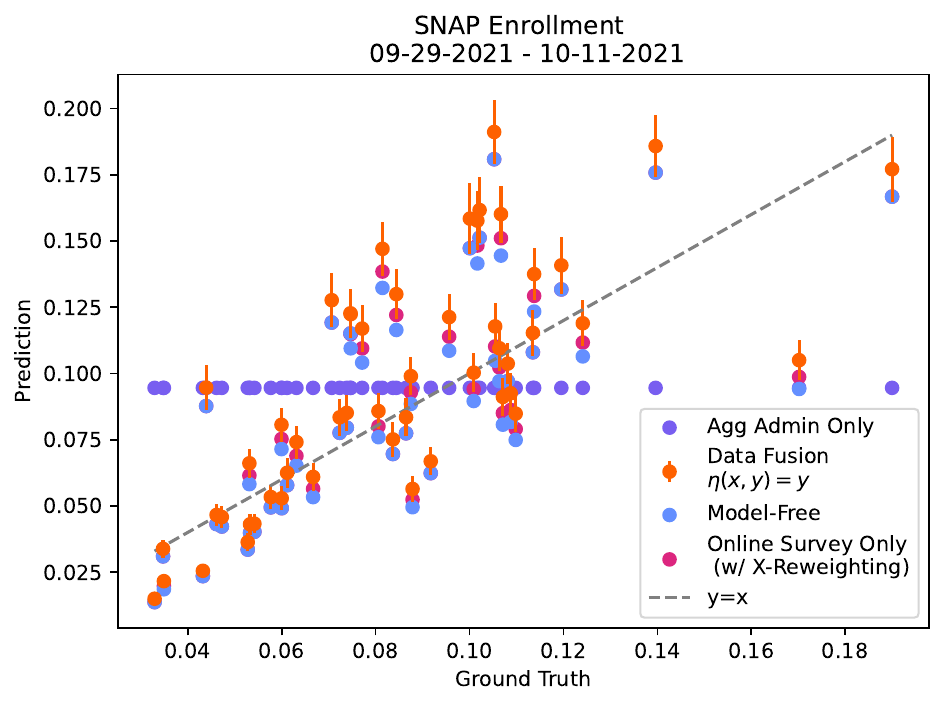}
\end{subfigure}
\begin{subfigure}{0.48\textwidth}
    \centering
    \includegraphics[width=\linewidth]{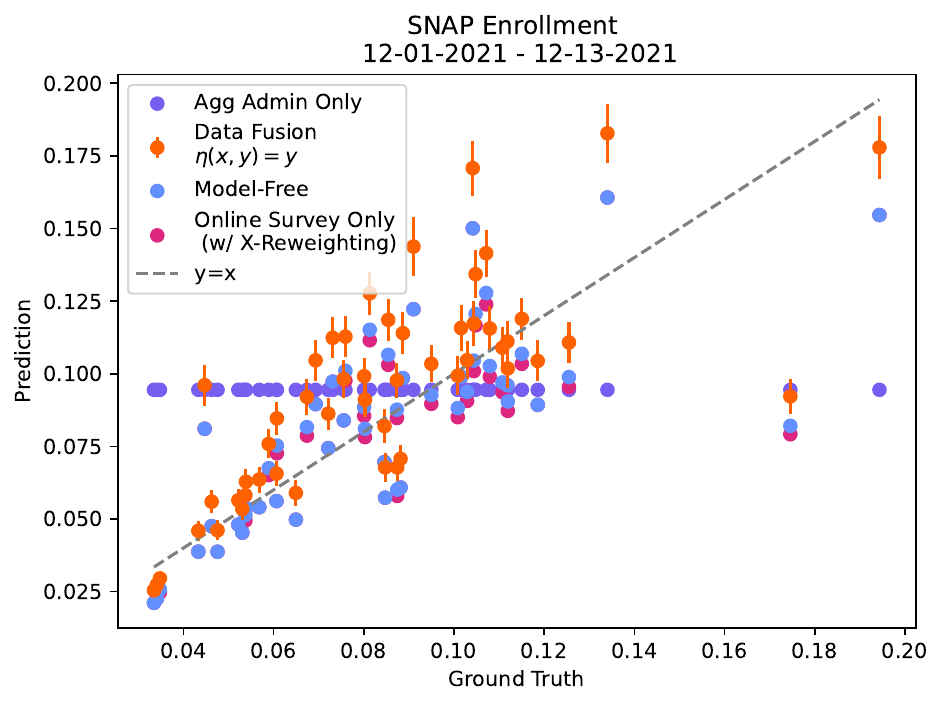}
\end{subfigure}%
\begin{subfigure}{0.48\textwidth}
    \centering
    \includegraphics[width=\linewidth]{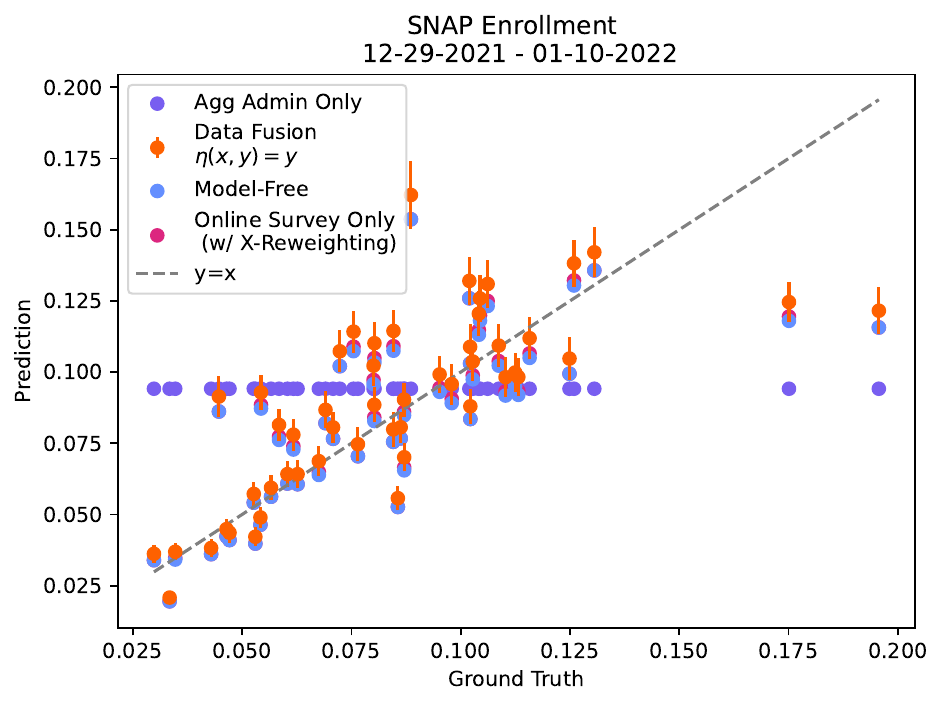}
\end{subfigure}
\end{figure}

\begin{figure}\ContinuedFloat
\begin{subfigure}{0.48\textwidth}
    \centering
    \includegraphics[width=\linewidth]{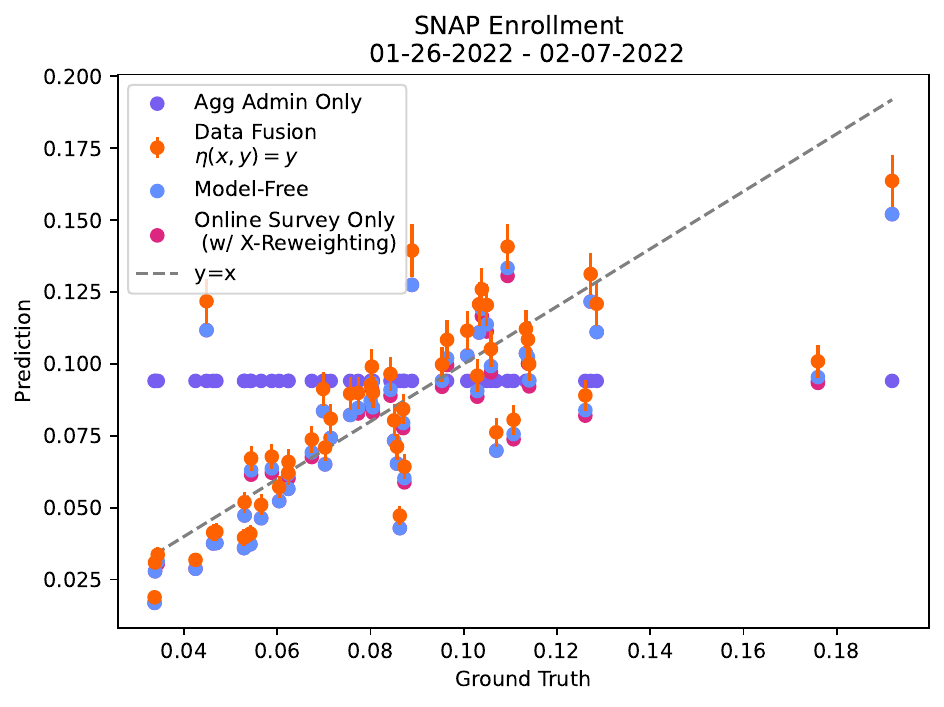}
\end{subfigure}%
\begin{subfigure}{0.48\textwidth}
    \centering
    \includegraphics[width=\linewidth]{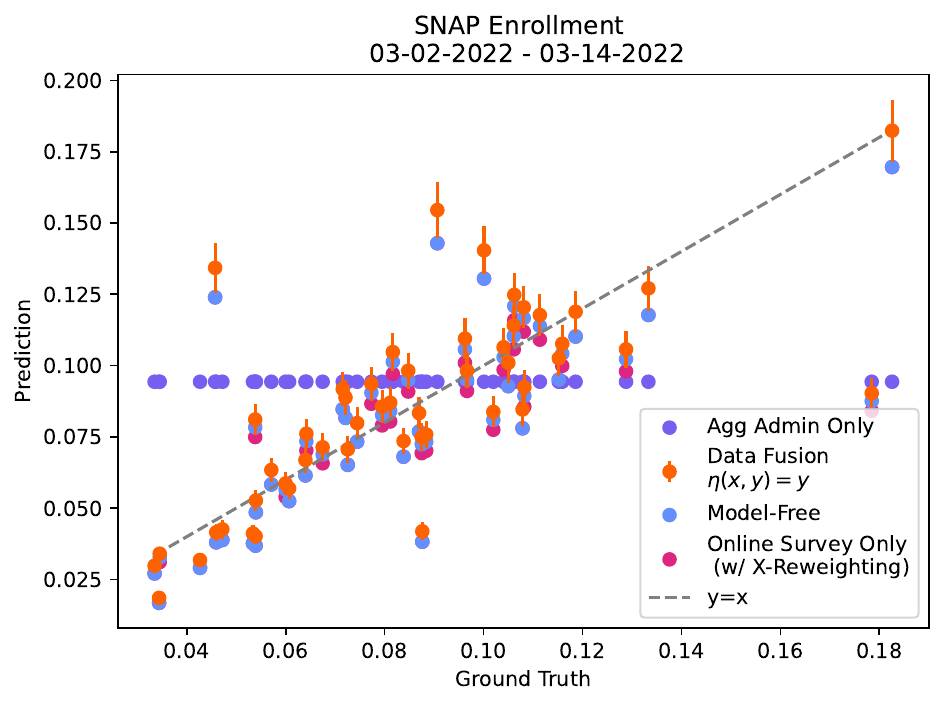}
\end{subfigure}
\begin{subfigure}{0.48\textwidth}
    \centering
    \includegraphics[width=\linewidth]{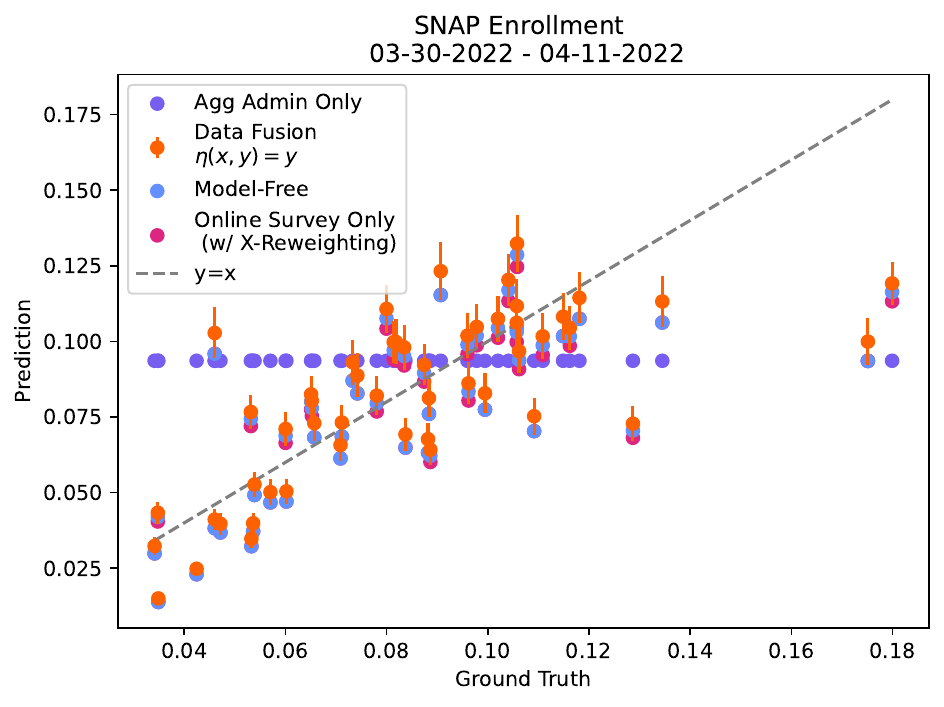}
\end{subfigure}%
\begin{subfigure}{0.48\textwidth}%
    \centering
    \includegraphics[width=\linewidth]{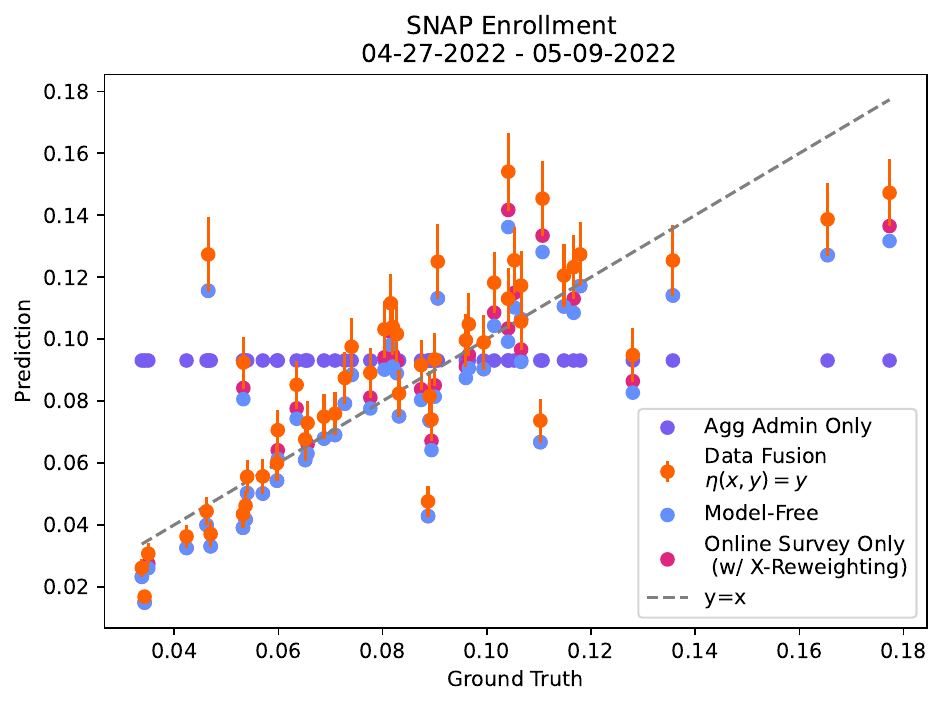}
\end{subfigure}
\begin{subfigure}{0.48\textwidth}
    \centering
    \includegraphics[width=\linewidth]{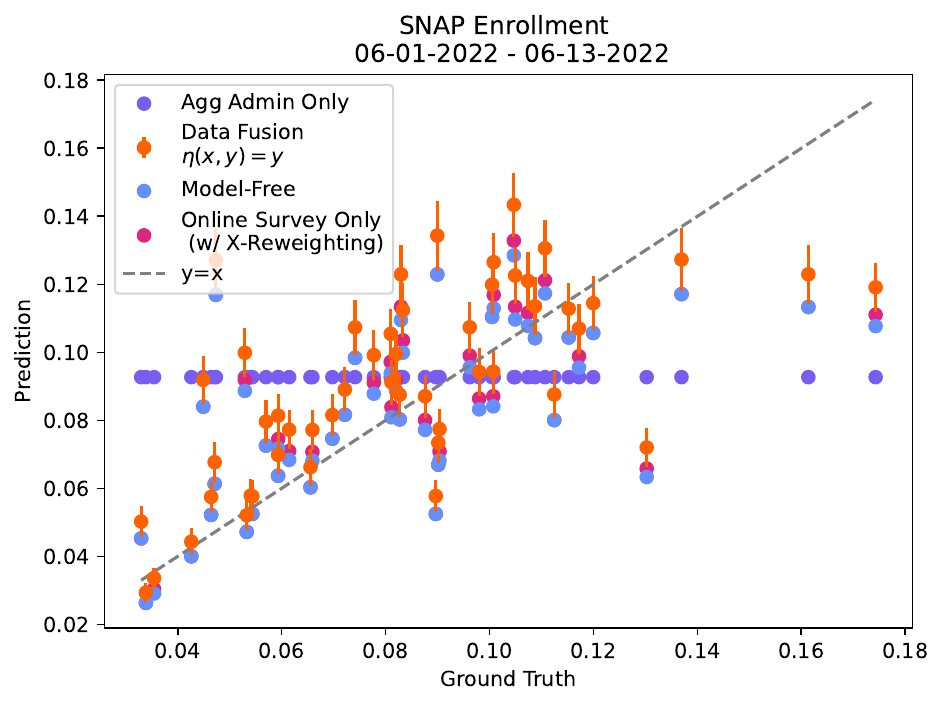}
\end{subfigure}%
\begin{subfigure}{0.48\textwidth}%
    \centering
    \includegraphics[width=\linewidth]{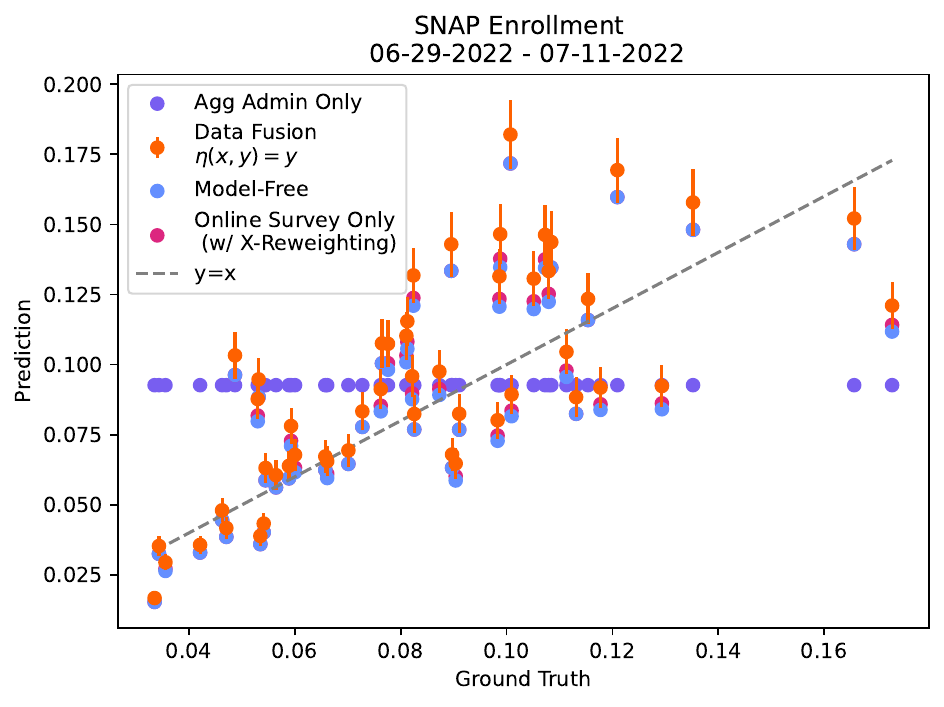}
\end{subfigure}
\end{figure}

\begin{figure}\ContinuedFloat
\begin{subfigure}{0.48\textwidth}
    \centering
    \includegraphics[width=\linewidth]{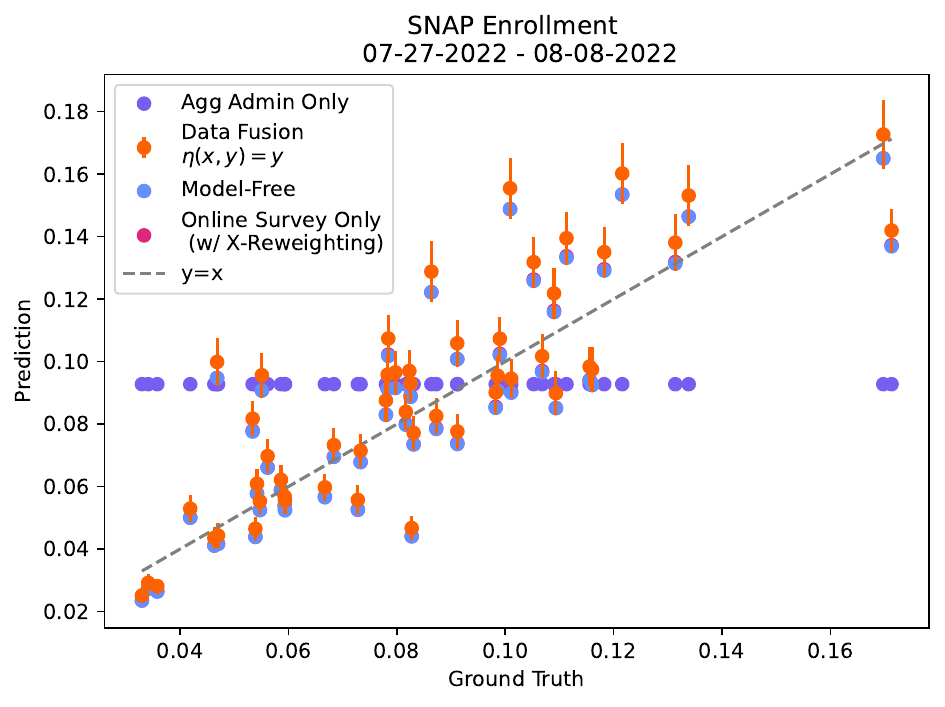}
\end{subfigure}%
\begin{subfigure}{0.48\textwidth}
    \centering
    \includegraphics[width=\linewidth]{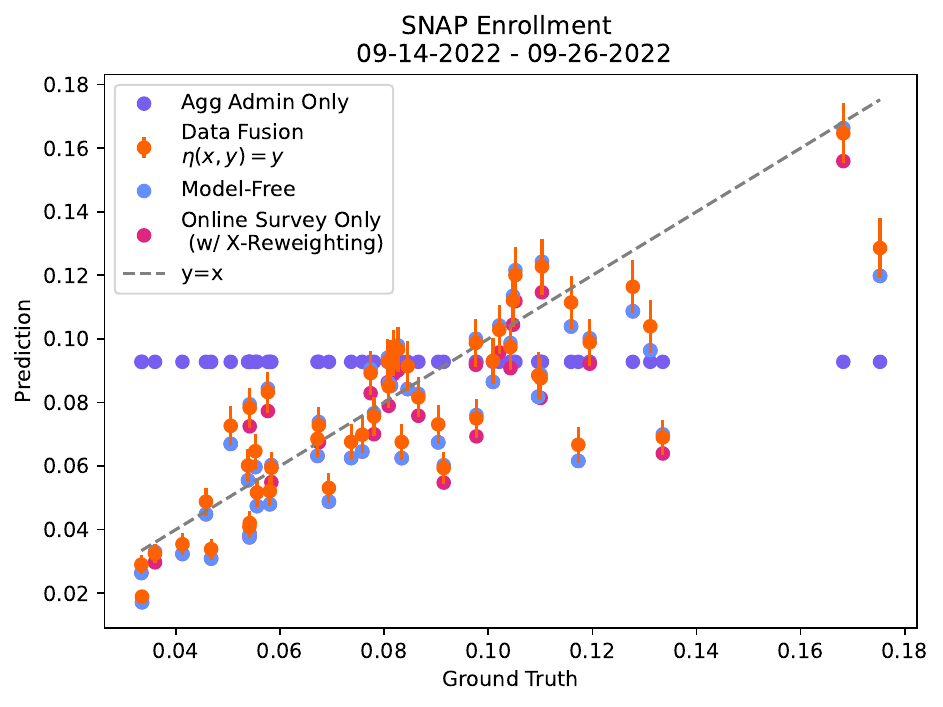}
\end{subfigure}
\begin{subfigure}{0.48\textwidth}
    \centering
    \includegraphics[width=\linewidth]{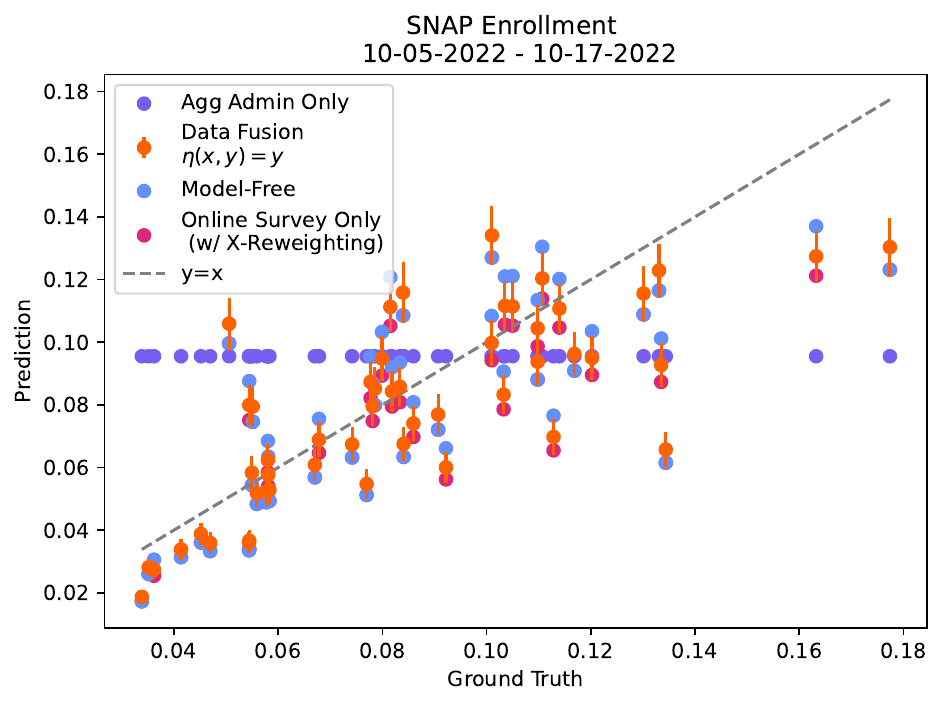}
\end{subfigure}%
\end{figure}

\end{section}

\end{document}